\documentclass[acmlarge,screen]{acmart}

\AtBeginDocument{%
  }

\usepackage{booktabs} % For formal tables

\usepackage{graphicx}
\usepackage{subfiles}
\usepackage{array}
\usepackage{amsmath}
\usepackage[ruled,linesnumbered]{algorithm2e}
\usepackage{algorithmic}
\usepackage{multirow}
\usepackage{subfigure}
\usepackage{caption}
\usepackage{flushend}
\usepackage{balance}
\usepackage{diagbox}
\usepackage{colortbl}
 \usepackage{enumitem}
 \usepackage{balance}
 \usepackage{makecell}
 \usepackage{float}
\usepackage{fancyhdr}
\usepackage{threeparttable}

\usepackage{url}
\usepackage{bm}
\usepackage{comment}
\usepackage{graphicx}
\usepackage{pifont}
\usepackage{bbding}

\usepackage{tcolorbox}

\usepackage{wrapfig}
\usepackage{subfigure}

\usepackage{listings}

\newcommand{\tool}{\textsc{VulRTex}}

\newcommand{\toolcompare}{\textsc{MemVul}}

\newcommand{\revise}[1]{{\color{red}[#1]}}

\newcommand{\firstrespto}[1]{
\fcolorbox{black}{black!15}{
\label{response:#1}
\bf
\scriptsize Referee-1 {#1}}~
}

\begin{document}

%%
%% The "title" command has an optional parameter,
%% allowing the author to define a "short title" to be used in page headers.

%\title{Security Wisdom of the Crowd: Automated Attack Tree Synthesis from Online Discussions}

% \title{{\tool}: An Automatic Approach to Synthesize Attack Trees from Crowd Security Posts}
% \title{{\tool}: Automatic Attack Tree Synthesis from Crowd Discussions to Enhance Security Knowledge Bases}

% \title{Towards Effective Identification of Vulnerability-Related Issue Reports with Rich-Text Reasoning}

\title{{\tool}: A Reasoning-Guided Approach to Identify Vulnerabilities from Rich-Text Issue Report}

%%
%% The "author" command and its associated commands are used to define
%% the authors and their affiliations.
%% Of note is the shared affiliation of the first two authors, and the
%% "authornote" and "authornotemark" commands
%% used to denote shared contribution to the research.
\author{Ziyou Jiang}
\orcid{0000-0003-1182-143X}
\affiliation{%
    \institution{State Key Laboratory of Intelligent Game}
  \state{Beijing}
  \country{China}
}
\affiliation{%
    \institution{Science and Technology on Integrated Information System Laboratory, Institute of Software Chinese Academy of Sciences}
  \state{Beijing}
  \country{China}
}
\affiliation{%
  \institution{University of Chinese Academy of Sciences}
  \state{Beijing}
  \country{China}
}
\email{ziyou2019@iscas.ac.cn}

\author{Mingyang Li}
\orcid{0009-0008-7936-5593}
\affiliation{%
    \institution{State Key Laboratory of Intelligent Game}
  \state{Beijing}
  \country{China}
}
\affiliation{%
    \institution{Science and Technology on Integrated Information System Laboratory, Institute of Software Chinese Academy of Sciences}
  \state{Beijing}
  \country{China}
}
\affiliation{%
  \institution{University of Chinese Academy of Sciences}
  \state{Beijing}
  \country{China}
}
\email{mingyang2017@iscas.ac.cn}
\authornote{Corresponding author.}

\author{Guowei Yang}
\orcid{0000-0002-1404-4560}
\affiliation{%
  \institution{School of Information Technology and Electrical Engineering, The University of Queensland}
  \city{Brisbane}
  \state{Queensland}
  \country{Australia}
}
\email{guowei.yang@uq.edu.au}

\author{Lin Shi}
\orcid{0000-0003-1476-7213}
\affiliation{%
  \institution{School of Software, Beihang University}
  \state{Beijing}
  \country{China}
}
\email{shilin@buaa.edu.cn}

\author{Qing Wang$^{*}$}
\orcid{0000-0002-2618-5694}
\affiliation{%
    \institution{State Key Laboratory of Intelligent Game}
  \state{Beijing}
  \country{China}
}
\affiliation{%
    \institution{Science and Technology on Integrated Information System Laboratory, Institute of Software Chinese Academy of Sciences}
  \state{Beijing}
  \country{China}
}
\affiliation{%
  \institution{University of Chinese Academy of Sciences}
  \state{Beijing}
  \country{China}
}
\email{wq@iscas.ac.cn}

% \date{\today}

%%
%% The abstract is a short summary of the work to be presented in the
%% article.
\begin{abstract}
Software vulnerabilities exist in open-source software (OSS), and the developers who discover these vulnerabilities may submit issue reports (IRs) to describe their details.
% However, some IR authors who lack knowledge of software security cannot accurately identify these vulnerabilities, 
Security practitioners need to spend a lot of time manually identifying vulnerability-related IRs from the community, and the time gap may be exploited by attackers to harm the system.
% and disclosing them to the security community.
% \revise{However, some developers who lack security experience may produce vulnerabilities during their coding process, thus resulting in millions of losses in today's business.
% If the project users discover these vulnerabilities, they may submit the OSS issue reports (IRs) with rich-text information, and wait for developers to fix these vulnerabilities.}
Previously, researchers have proposed automatic approaches to facilitate identifying these vulnerability-related IRs, 
but these works focus on textual descriptions but lack the comprehensive analysis of IR's rich-text information.
% on IRs with rich-text information.
% the comprehensive analysis of rich-text information, 
% XXXX
% leaving a large margin for performance improvement in vulnerability identification.
% Previous researchers have proposed the {\toolcompare} to automatically detect the vulnerabilities from vulnerability-related IRs. 
% \lin{bad rationale}
% % Due to the lack of security expertise, 
% Due to the lack of security expertise, some of these vulnerabilities are . 
% and the improper usage of OSS may produce some software vulnerabilities. 
% To disclose these vulnerabilities to the public, security practitioners need to track and discuss their details, where the time lag between IR creation and vulnerability disclosure may be utilized by attackers to deploy the exploits.
% Therefore, we utilize the reasoning ability of the Large Language Model (LLM) to analyze the rich-text information and introduce the LLM-Agent in IR vulnerability detection. 

In this paper, we propose {\tool}, a reasoning-guided approach to identify vulnerability-related IRs with their rich-text information.
In particular, {\tool} first utilizes the reasoning ability of the Large Language Model (LLM) to prepare the {Vulnerability Reasoning Database} with historical IRs.
Then, it retrieves the relevant cases from the prepared reasoning database to generate reasoning guidance, which guides LLM to identify vulnerabilities by reasoning analysis on target IRs' rich-text information. 
% {\tool} detects the vulnerability of New-IR with the generated guidance prompt.

To evaluate the performance of {\tool}, we conduct experiments on 973,572 IRs,
% \lin{Only positive data, no negative IRs that are normal?}
and the results show that {\tool} achieves the highest performance in identifying the vulnerability-related IRs and predicting CWE-IDs when the dataset is imbalanced, outperforming the best baseline with +11.0\% F1, +20.2\% AUPRC, and +10.5\% Macro-F1, and 2x lower time cost than baseline reasoning approaches.
Furthermore, {\tool} has been applied to identify 30 emerging vulnerabilities across 10 representative OSS projects in 2024's GitHub IRs, 
and 11 of them are successfully assigned CVE-IDs, which illustrates {\tool}'s practicability.
% \ziyou{Case-based Reasoning?}
\end{abstract}

\maketitle

\section{Introduction}\label{sec:intro}

Software vulnerabilities widely exist in open-source software (OSS) projects.
When developers discover the vulnerabilities, they may submit the IRs to describe the details of these vulnerabilities.
%In the OSS, software vulnerabilities widely exist in the projects. When developers discover the vulnerabilities, they may submit the IRs to describe the details of these vulnerabilities.
To identify these vulnerability-related IRs,
security practitioners usually spend a lot of time manually analyzing their contents~\cite{householder2017cert}.
They often track these IRs with issue-tracking systems~\cite{bugzilla,jira}, then classify these vulnerability-related IRs with {Common Weakness Enumeration (CWE)}~\cite{CWE}, which 
categorizes the types of vulnerabilities with their causes, behaviors, and consequences.
% and privately report the discovered vulnerabilities to the OSS maintainers.
Some of these identified vulnerabilities will be disclosed in the Common Vulnerabilities and Exposure (CVE)~\cite{CVE}, a security database that provides a standardized method to catalog publicly disclosed vulnerabilities, and to alert downstream users in the supply chain to the security risks in the project~\cite{iso_disclosure}.
% After the vulnerability identification, security practitioners will apply for an ID from the 
% \lin{any evidence to support the statement? are vuls reported following this rationale?} 
% to disclose this vulnerability, which is the
However, the manual identification of vulnerability is tedious and time-consuming. The time interval between the creation of vulnerability IRs and the vulnerability disclosure 
can be exploited by attackers (e.g., zero-day attacks~\cite{DBLP:conf/ccs/BilgeD12}) to harm the system, resulting in millions of dollars of losses in today’s businesses~\cite{article}.

%provides attackers with the opportunity to deploy the exploits (e.g., zero-day attacks~\cite{DBLP:conf/ccs/BilgeD12}) to harm the system, resulting in millions of dollars of losses {in today’s businesses}.~\cite{article}.}

To alleviate this risk at an early stage, researchers have proposed automatic approaches to facilitate the identification of vulnerability-related IRs with their textual description~\cite{DBLP:conf/msr/GegickRX10,DBLP:journals/sqj/OyetoyanM21,DBLP:conf/sigsoft/PanZC0BHLH22}.
% lack the comprehensive analysis of rich-text information, leaving a large margin for performance improvement in vulnerability identification.
% To analyze how project users describe vulnerabilities through IRs, we manually investigate these vulnerability-related IRs collected from GitHub.
However, we find that 39.1\% of vulnerability-related IRs only have a few-text information in our manual analysis, and they utilize \textbf{rich-text information} to describe the vulnerabilities, e.g., page screenshots, video streams, music files, code snippets, etc., which implicitly indicate how the vulnerabilities are triggered, thus helping security practitioners identify vulnerability-related IRs and analyze their CWE-IDs.
These previous works focus on textual descriptions but lack a comprehensive analysis of IR’s rich-text information, so their practical usage is limited.
Among them, we find that over 95\% of rich-text IRs utilize page screenshots and code snippets to illustrate how the vulnerabilities are triggered, and their details and contributions are shown as follows:

\begin{itemize}[leftmargin=*]
    \item \textbf{Page Screenshots:} Some developers and users will use page screenshots to display special states during the program's runtime, thereby reflecting possible vulnerabilities in the system. By capturing the specific element in the page screenshots or the transitions between different pages, we may identify which type of vulnerability the project will encounter. 
    \item \textbf{Code Snippets:} Some users will provide the simple Proof-of-Concept (PoC) with code snippets to validate the existence of vulnerabilities, and developers will show the vulnerability-related warnings or bug reports with specific code lines in the projects.
\end{itemize}

To intuitively illustrate how the relationships between these rich-text elements reflect the vulnerabilities, we will introduce them in the following Section \ref{sec:preliminary}.
From the previous investigations, we focus on exploring how the vulnerabilities are triggered by analyzing the rich-text elements of the page screenshots and code snippets.
% First challenge: lacks the relationships between rich-text elements->CoT&Agent
% Second challenge: illusions->Correct
% Third challenge: time aspect->CBR's contribution
However, since some information is implicitly described by IR's texts and rich-text elements, it is challenging to identify vulnerability and predict the CWE-IDs: 

\textbf{Challenge-1: Difficulty in analyzing the triggering process of vulnerability.}
The first challenge comes from how to accurately understand the textual semantics of the rich-text elements and construct the relationships between rich-text elements that describe the triggering process of vulnerabilities.
To address it, we utilize the text understanding and reasoning ability of LLMs to construct the relationships between these rich-text elements.

\textbf{Challenge-2: Factual errors in LLM's reasoning process.}
The second challenge comes from the factual errors in the reasoning graphs that are inconsistent with the real-world vulnerability triggering process, mainly because the pre-trained data may have some flaws and be outdated.
To address it, we incorporate a module to correct the factual errors in the LLM's reasoning steps.

\textbf{Challenge-3: Heavy time cost in LLM's reasoning process.}
The third challenge comes from heavy time costs.
The security practitioners who manage vulnerability identification need to deal with thousands of IRs. If the efficiency of the automated approach is lower than the manual analysis, its value of practical usage is limited.
For each target IR that needs to identify vulnerabilities, LLM needs to fully explore all the relationships between page screenshots and code snippets to reason how the pages are redirected and determine the vulnerability's type.
{To improve the effectiveness of reasoning, we utilize the LLM's \textbf{Retrieval Augmented Generation (RAG)} in the reasoning process~\cite{DBLP:conf/nips/LewisPPPKGKLYR020}, where we can retrieve some relevant cases in the history, then use these cases to guide the reasoning of the target IRs~\cite{DBLP:journals/corr/abs-2402-17453}.}
In practical analysis,
we find that vulnerability-related IRs with the same CWE-ID may have commonalities in describing the vulnerabilities.
For example, both \textit{Fuel-CMS/issues/536}~\cite{ir_multimodal_compare} and \textit{PhpldapAdmin/issues/130}~\cite{ir_multimodal_example} belong to the CWE-79, which utilizes the redirection of page screenshots to show how the attackers set XSS payloads to attack the project.
We can treat the first IR as a retrieved case for the second one, which incorporates how the vulnerability is triggered and can guide the vulnerability identification of the second IR.

{
}
% \textbf{LLM} XXX, which outperforms other Natural Language Process (NLP) models on most few-shot tasks~\cite{LLMBackground,DBLP:journals/corr/abs-2103-10385}.
% With the development of modality alignment~\cite{pmlr-v139-radford21a} and task unity~\cite{pmlr-v162-wang22al}, 
% researchers have proposed the Multimodal LLM (MLLM), such as GPT-4V~\cite{DBLP:journals/corr/abs-2303-08774}, to analyze rich-text information in the documents~\cite{DBLP:journals/corr/abs-2306-13549}.
% \ding{183} \textbf{Misleading of Noise Data:} some page screenshots and code snippets in IR do not indicate the security issues in the OSS, so the introduction of such noisy rich-text information will affect LLM's analysis of vulnerabilities.

% \ding{182} \textbf{Limitation on MLLM Multimodal Interface}, where the latest interface of GPT-4V (i.e., \texttt{gpt-4-vision-preview}) has only 100 calls per day~\cite{gpt4vcalllimit}, so it is difficult to be applied on a large number of vulnerability-related IRs from GitHub;
% \ding{182} \textbf{Lacking the Background Knowledge of IR Vulnerability Detection}, where existing LLMs have not been fine-tuned on security datasets, resulting in fluctuations when detecting IR vulnerabilities.

% \lin{what is the relationship between last para and this para?}
In this paper, we propose {\tool}, which is a reasoning-guided approach to {identify} vulnerabilities from IRs with rich-text information.
%Specifically, it utilizes the VulAgent to prepare the \textit{Vulnerability Reasoning Database} from historical IRs, 
Specifically, it prepares the \textbf{Vulnerability Reasoning Database} from historical IRs with LLM, 
% where each record incorporates the content of \textbf{\textit{Historical IR},} the \textbf{\textit{CWE label}} that reflects the type of vulnerability, 
which treats the LLM as an agent~\cite{DBLP:conf/iclr/YaoZYDSN023} that interacts with external tools to understand the semantic information of screenshots and code snippets, and utilize the reasoning ability of LLM~\cite{DBLP:conf/nips/Wei0SBIXCLZ22} 
% (e.g., the chain-of-thought, CoT) 
to construct the relationships between rich-text elements.
The reasoning database contains reasoning graphs that describe how to explore rich-text information in historical IRs to identify whether they contain vulnerabilities.
% The LLM-agent mechanism asks LLM as the agent to apply various rich-text data analysis tools~\cite{DBLP:conf/iclr/YaoZYDSN023}, and LLM will think step-by-step to obtain the semantic information of the vulnerability-related IR.
Then, {\tool} incorporates a novel RAG method to retrieve relevant reasoning graphs from the reasoning database that have a similar vulnerability triggering logic to the target IR.
Finally, with the retrieved reasoning graphs, {\tool} generates the guidance prompt to guide LLM to identify vulnerabilities.
% which is a specific prompt to guide the LLM on how to analyze the rich-text information of target IR based on the background knowledge. 
% , which is used to determine whether it contains the vulnerability and output a summary to describe the vulnerability.

To evaluate the performance of {\tool}, 
we conduct experiments on 973,572 IRs with 4,002 vulnerability-related IRs.
We compare {\tool} with three types of baselines, and the results show that {\tool} achieves the best performance in identifying the vulnerability-related IRs when the dataset is imbalanced, outperforming baselines with +11.0\% F1 and +20.2\% AUPRC, with over 2x lower time cost than the baseline reasoning approaches.  
{\tool} also achieves the best performance in CWE-ID prediction, outperforming the best baseline with +10.5\%.
Furthermore, {\tool} has been successfully applied on newly tracked IRs across 10 representative projects outside the original dataset after {2024}. 
Among them, 30 emerging vulnerabilities are identified by {\tool}, 11 of them (36.7\%) are assigned CVE-IDs, and 9 (30.0\%) will potentially be disclosed, which further illustrates its practicality.
% has been successfully applied to detect the 20 emerging vulnerabilities from 173 tracked IRs\lin{again}, and 12 of them are finally assigned a CVE-ID.
The major contributions are summarized as follows:

\begin{itemize}[leftmargin=*]
    \item 
    \textbf{Technique}: {\tool}, an automated approach to {identify} the vulnerability-related IRs. To the best of our knowledge, this is the first work on introducing LLM's reasoning ability to identify rich-text vulnerability-related IRs, as well as using the thought of RAG to reduce the time cost of {\tool} in practical usage.
    \item 
    \textbf{Evaluation}: An experimental evaluation of {\tool}, which shows that {\tool} outperforms all baselines on identifying the vulnerabilities, {and the application study on OSS projects further demonstrates its usefulness in practice.}
    \item \textbf{Data}:
    {We release the datasets and source code~\cite{model_data} to facilitate the replication and {the application of {\tool} in the more extensive contexts.}}
\end{itemize}

% {In the rest of the paper, Section \ref{sec:preliminary} illustrates the motivation example. Section \ref{sec:approach} presents the details of {\tool}. Section \ref{sec:exp} sets up the experiments. Section \ref{sec:results} describes the experimental results. Section \ref{sec:discussion} presents the discussion and threats to validity. Section \ref{sec:rw} discusses the related works, and Section \ref{sec:conclusion} concludes our paper.}

\section{Motivation Example}\label{sec:preliminary}

Recent researchers have conducted a preliminary study on 1,221,677 GitHub IRs in the GHArchive~\cite{GHArchive}, which is a massive IR dataset, archiving the original information of IRs since 2015.
Among these IRs, 3,937 of them contain vulnerabilities and have been assigned a CVE-ID, and 3,886 (98.7\%) of these vulnerability-related IRs were created earlier than the vulnerability disclosure~\cite{DBLP:conf/sigsoft/PanZC0BHLH22}.
Moreover, many developers utilize rich-text information to describe the details of vulnerabilities.
To analyze the proportion of these vulnerability-related IRs with rich-text information, we manually inspect these 3,886 IRs created earlier than vulnerability disclosure.
We find that 1,520 (39.1\%) contain rich-text elements that relate to the details of vulnerabilities, and over 95\% of them are page screenshots and code snippets.
Therefore, mining the rich-text information
may improve the performance of identifying vulnerability-related IRs, thereby reducing business losses.

\begin{figure}[t]
\centering
\includegraphics[width=\columnwidth]{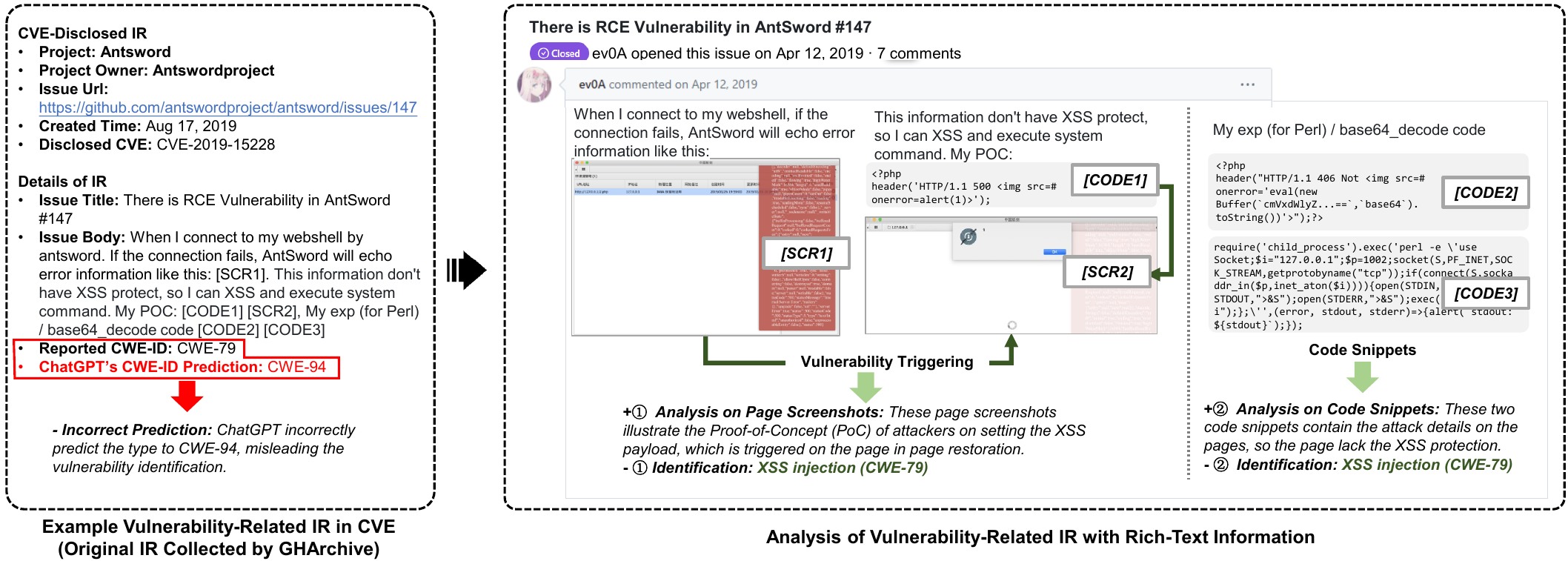}
\vspace{-0.6cm}
\caption{{The vulnerability-related IR with rich-text information, which has been assigned CVE-2019-15228.}
}
\label{fig:motivation_example}
\vspace{-0.2cm}
\end{figure}

Fig. \ref{fig:motivation_example} shows an example of vulnerability-related IR that contains rich-text information (\textit{antswordproject/antsword/issues/147}). 
{The author of the project {AntSword} reports an IR \#147 that may have a Remote Code Execution (RCE) vulnerability.
This vulnerability relates to multiple vulnerability types in CWE (CWE-79, CWE-94, and CWE-119, etc.), and 
security practitioners analyze IR's content to determine that this project may encounter the ``XSS Injection'' (CWE-79)~\cite{cwe_79}.}
Different from the other IRs, the author utilizes page screenshots and code snippets to describe this IR, 
which makes traditional approaches fail to identify it.
Besides, we also utilize the widely-used LLM, i.e., ChatGPT~\cite{LLMBackground}, to identify the vulnerability from its textual description, and find that ChatGPT incorrectly predicts the type to CWE-94~\cite{cwe_94}.

To analyze how the rich-text information affects the vulnerability identification,
we obtain the original page of this IR.
We can see that, the author utilizes two screenshots and three code snippets to describe the triggering process of vulnerability.
The process of this attack is as follows: 
\ding{182} On the main page of this project \textbf{\textit{[SCR1]}}, the attacker may set a field as XSS payload with the \texttt{PHP} script, and attack the project that incorrectly neutralizes user-controllable input.
The PoC of this attack is shown in \textbf{\textit{[CODE1]}}. 
\ding{183} If the administrator wants to restore this page, the vulnerable script will be executed, the system command will execute and the XSS payload will be triggered on the page.
The page with triggered vulnerability is shown in \textbf{\textit{[SCR2]}}.
To analyze the details of this vulnerability, 
we can also refer to the code snippets in \textit{\textbf{[CODE2]}} and \textit{\textbf{[CODE3]}}, which illustrate how the attacker utilizes the \texttt{PHP} script to call the system command and perform the XSS injection to the system.

From the above descriptions, we can construct a \textbf{{reasoning graph}} to illustrate how the developers describe the details of this vulnerability. The first path \textbf{\textit{[SCR1]}}$\rightarrow$\textbf{\textit{[CODE1]}}$\rightarrow$\textbf{\textit{[SCR2]}} indicates how the vulnerability CWE-79 is triggered, and the second path \textbf{\textit{[CODE2]}}$\rightarrow$\textbf{\textit{[CODE3]}} indicates the content of \texttt{PHP} attack script.
{Based on this reasoning graph, we can determine whether this IR contains vulnerabilities and its relevant CWE-ID.}
{Therefore, we believe that rich-text information can help identify the vulnerability-related IR.}
\section{Approach}\label{sec:approach}
\begin{figure*}[t]
\centering
\includegraphics[width=\textwidth]{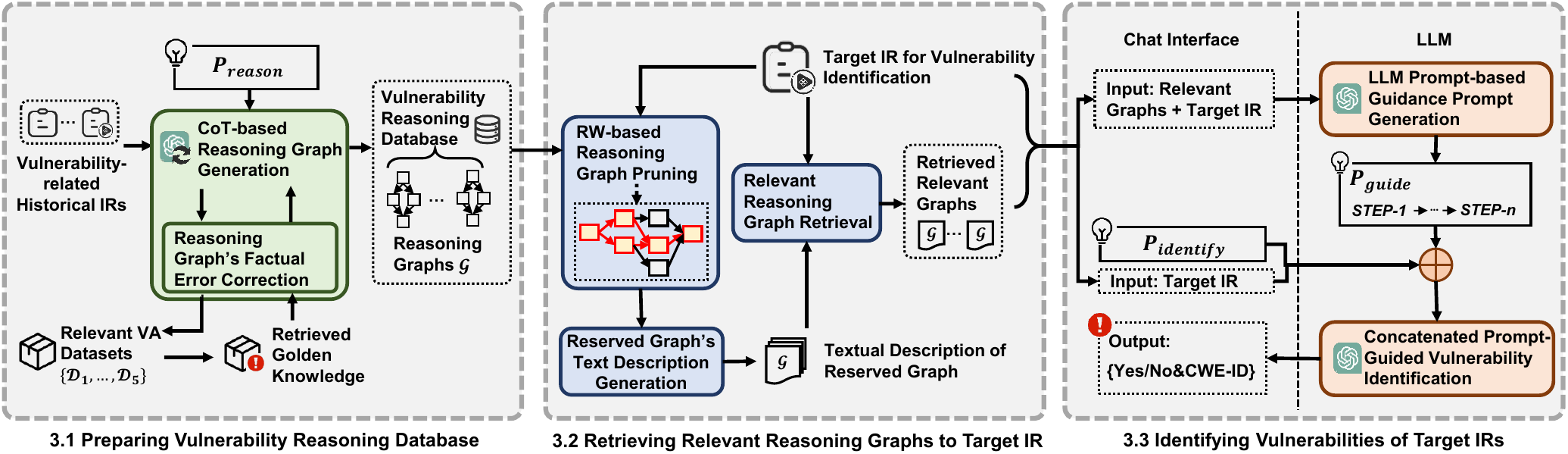}
\vspace{-0.7cm}
\caption{{The overview of our approach {\tool}.}}
\vspace{-0.3cm}
\label{fig:model}
\end{figure*}

{In this section, we introduce the details of {\tool}, and the overall framework is illustrated in Fig. \ref{fig:model}.
First, {\tool} prepares the vulnerability reasoning database with historical IRs, which contains reasoning graphs $\mathcal{G}$ that describe how the vulnerabilities are triggered based on the rich-text information in historical IRs. 
{Second, to identify the vulnerability from the target IR, {\tool} does not need to reuse the LLMs to explore the IR's rich-text elements. Instead, it retrieves its relevant reasoning graphs from the database to generate prompts to guide LLM to identify vulnerability-related IRs.}
{Afterwards, we utilize the guidance prompts to identify vulnerabilities from target IRs.}}

\subsection{Preparing the Vulnerability Reasoning Database}\label{sec:preparing_vul_database}

In Section \ref{sec:preliminary}, we discussed that the relationship between rich-text elements that describe the details of vulnerabilities is complex, which can be formulated as a reasoning graph that illustrates steps of exploring rich-text information.
Therefore, we prepare the vulnerability reasoning database with reasoning graphs $\mathcal{G}$ from the historical IRs, which treats the LLM as an agent, asking it to utilize tools and reasoning ability to construct relationships between rich-text elements.
To implement this logic, we design a specific prompt~\cite{DBLP:journals/corr/abs-2207-00747} to control LLM to reason the IR's rich-text information step by step and identify the vulnerabilities.

Moreover, In Huang et al's survey~\cite{DBLP:journals/corr/abs-2311-05232}, they indicate that the pre-trained data, training, and decoding strategies of LLMs have flaws that result in the inconsistent with real-world facts, which is called LLM's factual error~\cite{DBLP:journals/nca/ZhuLSZ23}.
This factual error is a typical LLM hallucination that exists in the outputs and affects the LLM's practical usage.
To bridge this gap,
we utilize the following two processes to generate the vulnerability reasoning database:

\begin{itemize}[leftmargin=*]
    \item \textbf{Reasoning Graph Generation:} To realize the LLM reasoning and reduce the time and memory cost, we build a simple \textbf{Chain-of-Thought (CoT)} framework, and the generated graphs only contain two types of nodes, i.e., \textbf{Observation} and \textbf{Action}, and we utilize a specific prompt to guide the historical IR's reasoning graph generation,
    We also utilize the \textbf{inclusion relationship} to analyze texts, screenshots, and code snippets in sequence, which can effectively reduce the nodes and infinite loops in the reasoning graphs.
    \item \textbf{Factual Error Correction:} 
    In each LLM reasoning step, we introduce the external vulnerability awareness (VA) datasets~\cite{DBLP:journals/tosem/YuHLLWX22,DBLP:conf/kbse/JiangSYW23} to check and correct the factual errors, which contain the latest vulnerabilities released in the security community and have been perceived and reviewed by experienced security practitioners. 
    This external knowledge is the \textbf{golden knowledge} and can be used to reduce the inconsistency in the generated reasoning graph's nodes and edges.
\end{itemize}

\subsubsection{The Definition of CoT-based LLM Reasoning}
Given the timestamp $t$, the CoT-based LLM reasoning incorporates an observation $O_t$ that reflects the result of vulnerability-related IR identification at this timestamp. Then, LLM will conduct an action $A_t$ based on the current observation, which controls the reasoning steps. The policy $\pi(A_t|C_t)=C_t\mapsto A_t$ specifies the way of searching actions, where the $C_t$ is the context sequence of the reasoning steps:
\begin{equation}
    C_t=(O_1, A_1, O_2, A_2,..., O_{t-1}, A_{t-1}, O_t)
\end{equation}
where policy $\pi(A_t|C_t)$ is determined by the LLM. If the observation $O_t$ indicates that the vulnerability is still unidentified, the LLM will search for an action $A_t$ that can analyze other rich-text information to help the identification; otherwise, the LLM will terminate the reasoning if the vulnerability is identified.

In each action $A_t\in Tools$, it is selected from a set of tools. These tools can effectively \firstrespto{Q2} \revise{parse} the texts in screenshots and understand the semantic information of the code snippets. 
Recently, OpenAI has released tools such as GPT-4o and o1, which can be used to analyze the semantics of multimodal text.
However, these models require high time, space, and financial costs, which makes them difficult to use for preparing the reasoning database, and we will not use them in the {\tool}.
We also incorporate a new inner tool \textbf{AgentTerminator} to stop the reasoning when the LLM cannot obtain the new knowledge to continue the reasoning steps.

The reasoning graph $\mathcal{G}$ includes all the observations and actions that identify vulnerability-related IRs.
At the timestamp $t\in\{1,...,T\}$, the paths of $\mathcal{G}$ is formulated as follows:

\begin{equation}
   C_{t,i}=(O_1,A_{1,i},O_{2,i},...,O_{t,i}), \mathcal{G}=\{C_{T,1},...,C_{T,m}\}
\end{equation}
where $C_{t,i}$ is the $i_{th}$ path in the graph.
Finally, the graph can be formulated as $\mathcal{G}=(\mathcal{V},\mathcal{E})$, where $\mathcal{V}=\{O_1, O_2,..., O_n\}$ indicates the observations, and $\mathcal{E}=\{A_1, A_2,..., A_n\}$ indicates the actions that control the steps of reasoning. 
As is shown in Fig. \ref{fig:model_example}, the reasoning graph $\mathcal{G}$ has four paths, seven observations, and 10 actions to identify the vulnerabilities.

\subsubsection{CoT-based Reasoning Graph Generation}

\begin{figure}[t]
\centering
\includegraphics[width=\columnwidth]{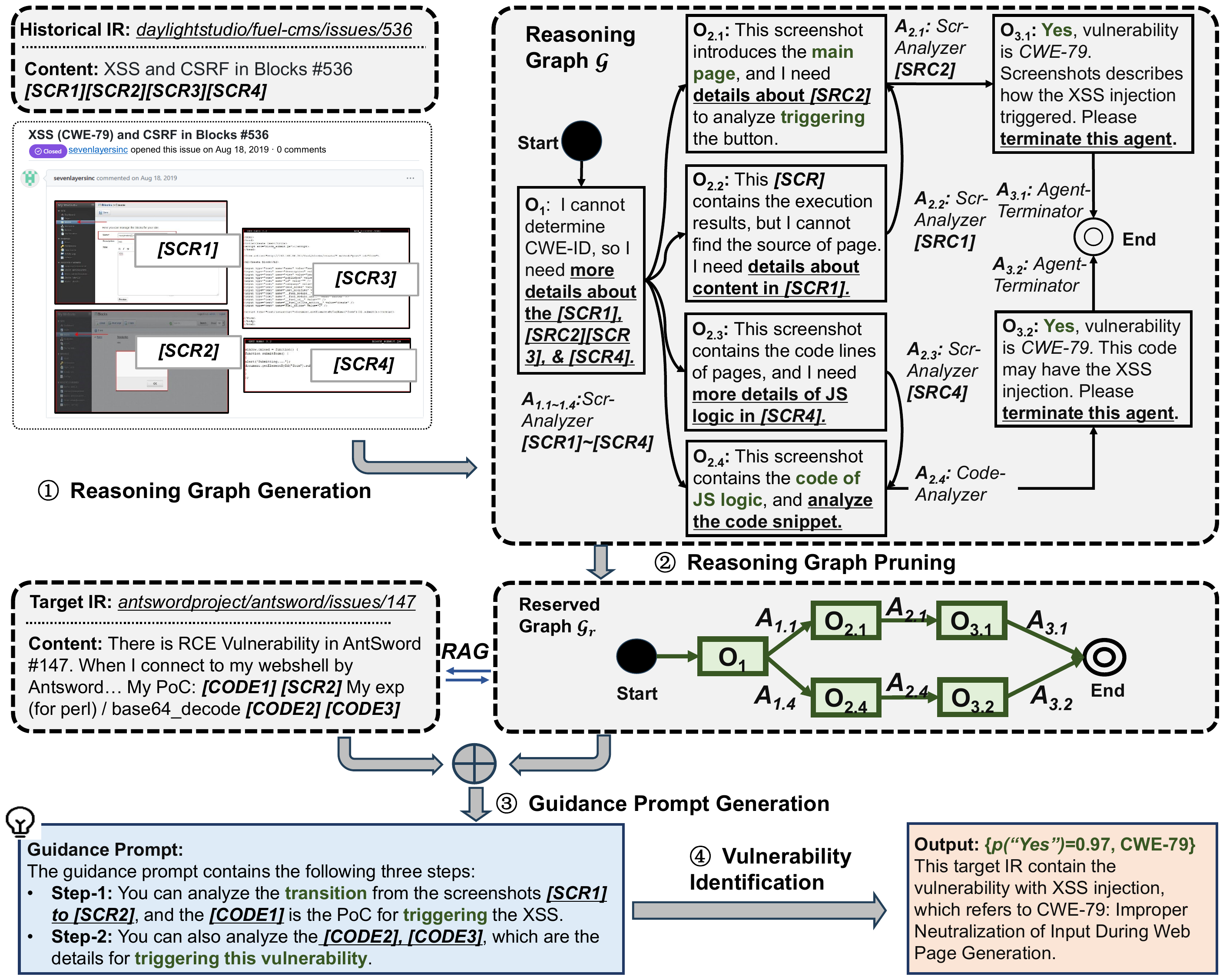}
\vspace{-0.5cm}
\caption{The steps of identifying target IR's (i.e., Fig. \ref{fig:motivation_example}'s example) vulnerability with rich-text information.}
\vspace{-0.2cm}
\label{fig:model_example}
\end{figure}

The historical IR incorporates the $\{Title, Body\}$, where $Title$ summarizes the main topic of this IR, and
$Body$ contains a set of sentences with rich-text information that describes the details of IR.
The CoT-based reasoning incorporates a specific prompt to ask the LLM to analyze this content and generate the reasoning graph $\mathcal{G}$. 
The following prompt $P_{reason}$ specifies how to find the vulnerabilities step by step:
\begin{center}
\vspace{-0.2cm}
\small
\begin{tcolorbox}[colback=white,%gray background
                  colframe=black,% black frame colour
                  width=\columnwidth,% Use 8cm total width,
                  arc=1mm, auto outer arc,
                  boxrule=0.4pt,
                  left=0.1pt,
                  right=0.1pt,
                  top=0.1pt,
                  bottom=0.1pt,
                colbacktitle=white!80!gray, coltitle=black, %标题框的背景和线条颜色 
                 title={{\textbf{$P_{reason}$: Prompt for Generating Reasoning Graph}}}%标题
                 ]
{Please think \textbf{step by step}. For each step, you need to select {\textbf{multiple}} rich-text elements that relate to the vulnerability. Then, you should identify whether this IR contains the vulnerability and output an ``\textbf{Observation}'' based on context information. 
It would be best to choose the ``\textbf{Action}'' from ``\textbf{Tools}'' to control the reasoning into the next ``Observation'' after thinking.}

\textit{- \textbf{Input:} The content of historical IR and the rich-text information.}

\textit{{- \textbf{Input:} The definition of ``Observation'' and ``Action''.}}

\textit{$^*$ \textbf{Note of Inclusion Relationship:} We suggest you first analyze the text, then explore the page screenshot \textbf{[SCR]}, and finally analyze the code snippets \textbf{[CODE]}.}
\end{tcolorbox}
\end{center}
where the prompt controls the graph generation by asking the LLM to think \textit{step by step}.
In each reasoning step, we ask the LLM to select \textbf{multiple} rich-text elements to analyze whether the IR contains vulnerabilities, as is shown in Fig. \ref{fig:model_example}'s observation $O_1$ (i.e., \textit{\textbf{[SCR1]}}$\sim$\textit{\textbf{[SCR4]}}, and edges are $\{(O_1\rightarrow O_{2.1}),..., (O_1\rightarrow O_{2.4})$\}).
To analyze each of the selected elements, {\tool} will continuously conduct multiple actions based on the observation until it terminates.
Therefore, any node that represents the observations may be connected with multiple edges that represent the actions, so the output is a reasoning graph rather than a single-directional path.

In addition, we also define the observation and action in this prompt, as well as the tool list for the action selection. The details of these three tools are shown as follows:

\begin{itemize}[leftmargin=*]
    \item \textbf{ScrAnalyzer:} We utilize Tencent Cloud's OCR~\cite{tecentOCR} to analyze all the page elements in the screenshots \textit{\textbf{[SCR]}}. Compared with other OCRs, Tencent Cloud can analyze more page elements with high accuracy and time efficiency.
    \item \textbf{CodeAnalyzer:} We utilize the CAST~\cite{DBLP:conf/emnlp/Shi0D0HZS21} to generate the description of code \textit{\textbf{[CODE]}}, which is a novel method to analyze codes with Abstract Syntax Trees (AST)~\cite{DBLP:conf/icse/ZhangWZ0WL19}.
    \item \textbf{AgentTerminator}: We ask the LLM to terminate the agent.
\end{itemize}

Moreover, we find that IRs allow the \textbf{inclusion relationship} \textit{\textbf{Text}}$\stackrel{Inc}{\longrightarrow}$\textit{\textbf{[SCR]}}$\stackrel{Inc}{\longrightarrow}$\textit{\textbf{[CODE]}} among rich-text information.
This relationship comes from our investigation of issues from GitHub, GitLab, and other OSS platforms, which means IR's text contains screenshots and code snippets, and some screenshots contain code.
Therefore, we suggest analyzing the comment, screenshot, and code in order based on the greedy search. 
Our manual analysis found that the reasoning graphs generated by the {\tool} have 3 fewer nodes on average with the inclusion relationship, which means the {\tool} can explore more elements with fewer iterations.
We iteratively generate the observations and actions, until the reasoning graph is determined.

\subsubsection{Reasoning Graph's Factual Error Correction}\label{sec:external_knowledge}
{As illustrated in the previous section, the LLM factual errors may affect the reliability of reasoning graphs. 
We select the five representative VA datasets $\mathcal{D}$ in Table \ref{tab:va_dataset} to correct the factual errors,
{To obtain more comprehensive and up-to-date external knowledge, we utilize the following three criteria to select the VA dataset:}}

% Please add the following required packages to your document preamble:
% \usepackage{multirow}
\begin{wraptable}{r}{0.5\textwidth}
    \centering
    \tiny
 \renewcommand{\arraystretch}{1}
 \vspace{-0.3cm}
\caption{The statistics of VA datasets, which are treated as golden knowledge for factual error correction.}
\vspace{-0.2cm}
\resizebox{0.5\columnwidth}{!}{
\begin{tabular}{llccl}
\toprule
\textbf{Id} & \textbf{VA Dataset}         & \textbf{Updated}  & \textbf{Lang} & \multicolumn{1}{c}{\textbf{Size}}                            \\
\midrule
$\mathcal{D}_{1}$ & {KB}~\cite{DBLP:conf/msr/PontaPSBD19}              & 2023      &  All & 1.2K \\
$\mathcal{D}_{2}$ & {BigVul}~\cite{DBLP:conf/msr/FanL0N20}            & 2023      &  C/C++ & 4.6K                     \\
$\mathcal{D}_{3}$ & {OWASP}~\cite{OWASP}           & 2024    &  C\#/Python   & 21K            \\
$\mathcal{D}_{4}$ & {Debian}~\cite{debian_dataset} & 2024   &   C/C++   & 3.3K                              \\
$\mathcal{D}_{5}$ & {VDISC}~\cite{DBLP:conf/icmla/RussellKHLHOEM18}           & 2024    &  C/C++/Java   & 1.2M     \\
\bottomrule
\end{tabular}}
\vspace{-0.2cm}
\label{tab:va_dataset}
\end{wraptable}

\begin{itemize}[leftmargin=*]
    \item \textbf{Criterion-1: Inclusion of Recent Vulnerabilities:} The selected VA datasets have been recently updated and maintained, and the latest vulnerabilities have been included in these datasets.
    \item \textbf{Criterion-2: Number of Vulnerabilities:} The vulnerabilities in the VA dataset cover a large number of projects, and the number of vulnerabilities is considerable. For example, VDISC covers the maximum number of projects (over 1K distinct projects) with 1.2M unique vulnerabilities~\cite{DBLP:conf/msr/PontaPSBD19}.
    \item \textbf{Criterion-3: Usage in Security Communities:} The selected VA datasets contain multiple programming languages and have been widely used in security communities. For example, the KB, BigVul, and Debian datasets are referenced by over 100 works as benchmark datasets.
\end{itemize}

In the path $C_t$ at timestamp $t$, 
we generate plain texts to describe these paths based on reasoning. For example, the generated \textbf{text description} of path $(O_1, A_1, O_2)$ is ``\textit{\textbf{from the observation $O_1$, we ask LLM to take the action $A_1$, and the next operation is $O_2$}}''. 
Then, we aim to retrieve the golden knowledge to correct the factual errors with TF-IDF~\cite{tfidf} text similarity.
The TF-IDF utilizes the term frequency to calculate the text similarities, which has high efficiency and can focus on vulnerability-related keywords in similarity calculation, e.g., ``XSS'' and ``CSRF'', etc.
{Based on these five selected VA datasets, we extract the golden knowledge $Know_t$ in these datasets that have TF-IDF similarities higher than the threshold $\theta_{sim}$:
\begin{equation}   
Know_t|sim_{Know_t\in\mathcal{D}}(Know_t, C_t)>\theta_{sim}, \text{where TF-IDF}\rightarrow sim(\cdot,\cdot)
\end{equation}
where $sim(\cdot,\cdot)$ is the function for TF-IDF similarity.} 
{The TF-IDF is based on the term frequency in the texts, so we calculate the term frequency in all the rich-text information in IRs, and the screenshots are parsed to text with OCR.}
{Since we utilize the TF-IDF similarity to retrieve the external knowledge from the VA dataset, it might introduce some noisy data that cannot be treated as the golden knowledge for the corresponding IR.
To address it, we carefully tune the parameter $\theta_{sim}$'s value from $[0,1]$ to make sure that all the IRs with factual errors can find the matched golden knowledge in the retrieved knowledge in $Know_t\in\mathcal{D}$.
The LLMs can utilize their reasoning ability to analyze the retrieved $Know_t$ and correct the errors in the nodes and edges of reasoning graphs (details of this threat and its alleviation are shown in Section \ref{sec:threat}).}
We concatenate knowledge and path $Know_t\oplus C_t$ and feed it to the LLM. 
After these three steps, we output the vulnerability reasoning database, where each record is the $\mathcal{G}\in Reason\_Base$.

\subsection{Retrieving the Relevant Reasoning Graphs to Target IR}\label{sec:retrieval}
As illustrated in the previous sections, it is time-consuming to reconduct the reasoning to identify vulnerability-related target IRs.
Therefore, we introduce the thought of RAG to retrieve the necessary knowledge and generate guidance for identifying the target IR's vulnerability. This incorporates two steps to improve the quality of the generated guidance prompt, i.e., external knowledge retrieval and text generation.
Previous work indicates that the generation results of RAG are diversified based on these two steps~\cite{DBLP:conf/emnlp/JiangXGSLDYCN23}.
However, existing RAG approaches cannot be directly applied to retrieve relevant reasoning graphs from the database, due to the following challenges:

\textbf{Challenge-1: Heavy time cost in reasoning graph retrieval.} In Fig. \ref{fig:model_example}'s example we can see that the reasoning graph contains massive exploration paths, and directly sampling the relevant records based on the complete graph may have massive time and memory costs. We have applied some previous RAG approaches to this task, such as AutoRAG~\cite{DBLP:journals/corr/abs-2410-20878}, REALM~\cite{DBLP:journals/corr/abs-2002-08909}, and MemoRAG~\cite{DBLP:journals/corr/abs-2409-05591}, etc. (note that, we investigate the RAGs that have open-sourced dataset and models, and released mature tools), need to take  $\geq 5$ min/per target IR to retrieve the relevant graphs, which cannot be directly applied in the reasoning graph retrieval.

\textbf{Challenge-2: Low relevance in retrieved graphs after pruning.} To improve the efficiency of the graph retrieval, we need to prune the reasoning graphs and find a short path to indicate the vulnerability triggering. 
However, the vulnerability triggering logic is quite complex, and the traditional methods to prune the reasoning graphs, such as DFS~\cite{DBLP:journals/corr/abs-2304-08354}, CART~\cite{DBLP:journals/imcs/MathieuT23} and PageRank~\cite{Page1999Page}, lack the background knowledge of the vulnerability triggering.
Even though these graph pruning methods traverse the whole reasoning graph to determine what nodes and edges will be pruned, the retrieved graphs may not be relevant to the target IRs.

To address these two challenges, we introduce the \textbf{Random-Walking}~\cite{DBLP:conf/kdd/PerozziAS14} method to prune the reasoning graph, which randomly selects paths to represent the $\mathcal{G}$ based on the sampling probabilities, thus reducing the cost of relevant record sampling.
We introduce two TF-IDF similarities in the record sampling. 
\ding{182} The first similarity constructs the weights that measure the similarity between the rich-text information in target IR and the texts of $\mathcal{G}$'s nodes, and the sampling probabilities are calculated with these weights. Nodes with higher probabilities are more likely to be reserved.
\ding{183} The second similarity selects the relevant graphs by calculating the similarity between the target IRs and the text description of pruned $\mathcal{G}$.

\begin{algorithm}[t]
\footnotesize
	%\SetAlgoNoLine %控制有无竖线
	\caption{Process of $\mathcal{G}$'s random-walking.} 
 \label{alg:rw}
	\KwIn{The target IR $TarIR$; reasoning graph $\mathcal{G}=(\mathcal{V},\mathcal{E})$, where $\mathcal{V}=\{O_1, O_2,..., O_n\}$, $\mathcal{E}=\{A_1, A_2,..., A_n\}$.} 
	\KwOut{The reserved reasoning graph after pruning $\mathcal{G}_{r}=(\mathcal{V}_{r}, \mathcal{E}_{r})$, where \firstrespto{Q2}\revise{$\mathcal{V}_{r}\subset\mathcal{V}$, $\mathcal{E}_{r}\subset\mathcal{E}$.}}
    $\tilde{\bm{M}}=\mathbf{0}$\tcp*{Initialize the adjacency matrix.}\label{line:1}
    \For {$O_i\in\mathcal{V}$, $O_j\in\mathcal{V}$, and $O_i\stackrel{A_i}{\longrightarrow}O_j$}{
        $\tilde{\bm{M}}_{i,j}=sim(O_j, TarIR)-sim(O_i, TarIR)$, where TF-IDF$\rightarrow sim(\cdot,\cdot)$\tcp*{\ding{182} Construct the adjacency matrix $\tilde{\bm{M}}_{i,j}$ of the reasoning graph $\mathcal{G}$.}\label{line:4}
        
         $deg(O_i)=\sum_{O_j\in\mathcal{V}}\tilde{\bm{M}}_{i,j}$, $deg(O_j)=\sum_{O_i\in\mathcal{V}}\tilde{\bm{M}}_{j,i}$ \tcp*{\ding{183} Calculate the sampling probability.}\label{line:5}
        $p(O_i,O_i)=\left(\frac{1}{deg(O_i)}+\frac{1}{deg(O_j)}\right)/\sum_{(O_i^{'},O_j^{'})}\left(\frac{1}{deg(O_i^{'})}+\frac{1}{deg(O_j^{'})}\right)$\;\label{line:8}
    }
    % \For{$O_i,O_j\in|\tilde{\bm{M}}|$}{
        
    % }
    
    $\mathcal{V}_r\cup\{O_1\}$, $\mathcal{V}_{walk}=\{O_1\}$\tcp*{\ding{184} Random-walking with sampling probability.}\label{line:9}
    \For{$O_i\in \mathcal{V}_r\cup\{O_1\}\backslash\mathcal{V}_{walk}$ and $O_j\in\mathcal{V}\backslash\mathcal{V}_r$}{
        Select $O_j$ for the node $O_i$ with the probability $p(O_i,O_j)$\;
        $\mathcal{V}_{walk}\cup\{O_i\}$,         $\mathcal{V}_r\cup\{O_j\}$, $\mathcal{E}_r\cup\{A_i|O_i\stackrel{A_i}{\longrightarrow}O_j\}$\;
        \If{$A_i=AgentTerminator$}{
        \textbf{break}\tcp*{Loop terminates.}
        }
    }\label{line:16}
    % \SetKwProg{Fn}{Function}{}{}
    % \Fn{AAA}{123} \tcc*[r]{AAA}
    return $\mathcal{G}_r$;
\end{algorithm}
% \vspace{-0.3cm}

\subsubsection{Random-Walking-based Reasoning Graph Pruning}
{We randomly prune the $\mathcal{G}$ to a few reserved nodes and edges based on edge weights.
Compared with the traditional graph pruning methods~\cite{DBLP:journals/imcs/MathieuT23,Page1999Page}, it does not require the traversal of the whole reasoning graph, which has high efficiency in the relevant record sampling.}

We propose the Algorithm \ref{alg:rw} to reserve the $\mathcal{G}_r=(\mathcal{V}_r,\mathcal{E}_r)$ to represent the reasoning graphs.
In this algorithm, from line \ref{line:1} to \ref{line:4}, we construct the adjacency matrix $\tilde{\bm{M}}$ of the graph with the target IR, where $\tilde{\bm{M}}_{i,j}$ measures the importance of the reasoning $O_i\stackrel{A_i}{\longrightarrow}O_j$.
We utilize the increment of TF-IDF similarity to measure the importance of reasoning. 
% The higher $\tilde{\bm{M}}_{i,j}$ indicates the more important $A_i$.
From line \ref{line:5} to \ref{line:8}, we calculate the sampling probability $p(O_i,O_j)$ with the GraphSaint edge probability~\cite{DBLP:conf/iclr/ZengZSKP20}. 
This probability is calculated with the node degrees, which have unbiasedness in random-walking.
From line \ref{line:9} to \ref{line:16}, we random walk the $\mathcal{G}$ with the sampling probability $p(O_i,O_j)$. We first sample all observations adjacent to $O_1$, and then we sample other nodes based on the visited observations until the action is terminated.
All nodes and edges will not be repeatedly explored.
Finally, we reserve the graph $\mathcal{G}_r$ after the random-walking algorithm.

\subsubsection{Reserved Graph's Text Description Generation}

For each record in the vulnerability reasoning database, we first prune the graph and reserve the subgraph $\mathcal{G}_r$.
Second, we extract the paths from $\mathcal{G}_r$ that can achieve the termination.
For example, $\mathcal{G}_r$ in Fig. \ref{fig:model_example} has two paths: $(O_1,A_{1.1},O_{2.1},A_{2.1},O_{3.1})$ and $(O_1,A_{1.4},O_{2.4},A_{2.4},O_{3.2})$.
The textual description of $\mathcal{G}_r$ is the concatenation of all the paths' textual descriptions, as is described in Section \ref{sec:external_knowledge}.

\subsubsection{Relevant Reasoning Graph Retrieval}

{Finally, we utilize the TF-IDF text similarity to select the relevant graphs with values higher than the threshold $\theta_{sim}$.
\begin{equation}
\mathcal{G}^{Tar}_{r}|sim_{\mathcal{G}^{Tar}_{r}\in Reason\_Base}(TarIR,\mathcal{G}^{Tar}_{r})>\theta_{sim}, \text{where TF-IDF}\rightarrow sim(\cdot,\cdot)
\end{equation}
where $\mathcal{G}^{Tar}_{r}$ is the textual description of selected reasoning graphs, and $TarIR$ indicates all the rich-text information in target IRs, which are parsed to text by OCR and CAST tools.}

\subsection{Identifying the Vulnerability of Target IR}\label{sec:vul_identification}

Since manually designed prompts are not appropriate for some specific tasks, researchers utilize LLM to generate specialized guidance prompts, thus improving their performances~\cite{DBLP:conf/iclr/ZhouMHPPCB23}.
Given the target IR and its relevant reasoning graphs $\mathcal{G}^{Tar}_{r}$, we utilize the LLM to generate guidance prompts for different target IRs.
We ask the generated guidance prompt $(\mathcal{G}^{Tar}_{r}, TarIR)\mapsto P_{guide}$ to include several steps, which describe the instructions for how to analyze target IR.
Fig. \ref{fig:model_example} shows the example of the guidance prompt that includes the three steps to analyze the rich-text information in target IR.
In these steps, the guidance prompt refers to the reserved graph. It finds the similarity between the target IR and relevant record in page redirection logic and asks the LLM to analyze the {main page} and the {XSS-triggered page} to identify vulnerabilities.

After generating the guidance prompt, we concatenate it with the prompt for identifying vulnerability $P_{identify}$. 
The following is the concatenated prompt $P_{guide}\oplus P_{identify}$:

\begin{center}
\vspace{-0.2cm}
\small
\begin{tcolorbox}[colback=white,%gray background
                  colframe=black,% black frame colour
                  width=\columnwidth,% Use 8cm total width,
                  arc=1mm, auto outer arc,
                  boxrule=0.4pt,
                  left=0.1pt,
                  right=0.1pt,
                  top=0.1pt,
                  bottom=0.1pt,
                colbacktitle=white!80!gray, coltitle=black, %标题框的背景和线条颜色 
                title={{\textbf{Concatenated Prompt for Vulnerability Identification}}}%标题
                 ]
$P_{identify}$: Please identify whether the following IR contains the vulnerability, and predict the type (CWE-ID) of the vulnerability.
\textit{This is a classification task, so please directly output whether the IR contains the vulnerability with ``Yes, No''.}
The output format for vulnerability identification is \{Yes, No\}.
Moreover, you need to just directly output the \{CWE-ID\} without other information.

{$P_{guide}$: According to the relevant reasoning graph, the generated
guidance prompt contains the following steps. We will concatenate it with the $P_{identify}$: (Several steps STEP-1$\sim$STEP-$n$, where each step contains the instruction for how to analyze the target IR).}

\textit{- \textbf{Input:} The content of target IR, which is formatted as JSON.}

\textit{- \textbf{Input:} The textual description of all the selected graphs $\mathcal{G}^{Tar}_{r}$.}

\end{tcolorbox}
\end{center}
where {the prompt requires the output to explicitly output the classification results of vulnerability identification with two labels ``Yes (1), No (0)'', as well as the type of vulnerability ``CWE-ID''.
For vulnerability identification, we use the LLM's API to output the \textbf{probability} (e.g., ChatGPT utilizes the attribute in the response \texttt{logprobs.top\_logprobs} to calculate each output label) rather than the labels.
Referring to the previous works~\cite{DBLP:conf/sigsoft/PanZC0BHLH22,DBLP:conf/ccs/YamaguchiWGR13,DBLP:journals/tse/ScandariatoWHJ14,DBLP:journals/tse/ShinMWO11}, using the probability to determine the vulnerability-related IRs is a flexible method when the testing dataset is extremely imbalanced, and the output probability of positive label ``Yes'' to identify the vulnerability, i.e., $p(``Yes"|TarIR)\in[0,1]$. We utilize the threshold $\theta_{out}$ to decide the classification result, and $p(``Yes"|TarIR)\geq\theta_{out}$ means that the target IR is the vulnerability-related IR.
For CWE-ID prediction, we output the CWE-ID rather than using the classification probability, because the labels have similar distributions in the dataset.}
We feed this concatenated prompt target IR to the LLM and identify its vulnerability.
Since a vulnerability-related IR may have vulnerabilities with multiple CWE-IDs, we will refine the dataset by splitting them into multiple IRs (Section \ref{sec:dataset_preparation}).

\section{Experimental Design}\label{sec:exp}
To evaluate the performance of {\tool}, we investigate the following research questions (RQs):

\begin{itemize}[leftmargin=*]
    \item \textbf{RQ1: How does \tool\ perform on identifying the vulnerability-related IR and predicting the CWE-ID?}
    \item \textbf{RQ2: How does each component contribute to identifying vulnerability-related IRs and predicting CWE-IDs?}
    \item \textbf{RQ3: Can {\tool} identify emerging vulnerabilities from the open-source projects?}
\end{itemize}

\subsection{Dataset Preparation}\label{sec:dataset_preparation}

{In this section, we prepare our dataset in the following four steps: First, we select and collect the dataset after the rigorous review and selection process with security practitioners. 
{Second,
our manual analysis illustrates that some IRs may contain vulnerabilities that correspond to multiple CWE-IDs, so we refined the dataset by re-annotating the CWE-IDs to improve the data quality.}
Third, we collect the original IR to improve the dataset with its rich-text information.
Finally, we preprocess the dataset with token replacement.}

% Please add the following required packages to your document preamble:
% \usepackage{multirow}
% \usepackage[normalem]{ulem}
% \useunder{\uline}{\ul}{}
\begin{wraptable}{r}{0.5\textwidth}
    \centering
    \tiny
 \renewcommand{\arraystretch}{1}
 \vspace{-0.2cm}
\caption{The results of reviewing the data sources based on VA/evaluation dataset criteria.}
\vspace{-0.2cm}
\resizebox{0.5\columnwidth}{!}{
\begin{tabular}{l|ccc|ccc}
\toprule
\multirow{2}{*}{\textbf{Data Source}} & \multicolumn{3}{l|}{\textbf{VA-Dataset Criteria}} & \multicolumn{3}{l}{\textbf{Eval-Dataset Criteria}} \\
                             & \textbf{IRV}          & \textbf{NV}         & \textbf{USC}         & \textbf{TIR}          & \textbf{CPL}          & \textbf{NNS}         \\
\midrule
GHArchive~\cite{GHArchive}                    & \textcolor{green}{\ding{52}}            & \textcolor{green}{\ding{52}}          & \textcolor{green}{\ding{52}}           & \textcolor{green}{\ding{52}}            & \textcolor{green}{\ding{52}}            & \textcolor{green}{\ding{52}}           \\
D2A~\cite{DBLP:conf/icse/ZhengPLBEYLMS21}                          &      \textcolor{red}{\ding{56}}         &      \textcolor{red}{\ding{56}}       &     \textcolor{green}{\ding{52}}        &    \textcolor{green}{\ding{52}}           &      \textcolor{green}{\ding{52}}        &       \textcolor{red}{\ding{56}}      \\
ATT\&CK~\cite{ATTCK}                        &       \textcolor{red}{\ding{56}}        &     \textcolor{green}{\ding{52}}       &     \textcolor{green}{\ding{52}}       &       \textcolor{red}{\ding{56}}       &   \textcolor{green}{\ding{52}}         &     \textcolor{green}{\ding{52}}         \\
VulZoo~\cite{10.1145/3691620.3695345}                       &     \textcolor{green}{\ding{52}}           &        \textcolor{green}{\ding{52}}      &     \textcolor{red}{\ding{56}}         &     \textcolor{green}{\ding{52}}           &     \textcolor{green}{\ding{52}}           &       \textcolor{red}{\ding{56}}      \\
CAPEC~\cite{CAPEC1}                         &       \textcolor{green}{\ding{52}}        &       \textcolor{red}{\ding{56}}     &     \textcolor{green}{\ding{52}}        &      \textcolor{green}{\ding{52}}         &      \textcolor{red}{\ding{56}}        &      \textcolor{green}{\ding{52}}        \\
\midrule
KB~\cite{DBLP:conf/msr/PontaPSBD19}                           & \textcolor{green}{\ding{52}}          & \textcolor{green}{\ding{52}}           & \textcolor{green}{\ding{52}}             &    \textcolor{red}{\ding{56}}           &     \textcolor{green}{\ding{52}}         &     \textcolor{green}{\ding{52}}         \\
BigVul~\cite{DBLP:conf/msr/FanL0N20}                       & \textcolor{green}{\ding{52}}          & \textcolor{green}{\ding{52}}           & \textcolor{green}{\ding{52}}             &     \textcolor{red}{\ding{56}}          &       \textcolor{red}{\ding{56}}       &  \textcolor{green}{\ding{52}}            \\
OWASP~\cite{OWASP}                        & \textcolor{green}{\ding{52}}          & \textcolor{green}{\ding{52}}           & \textcolor{green}{\ding{52}}             &     \textcolor{red}{\ding{56}}          &      \textcolor{red}{\ding{56}}         &    \textcolor{green}{\ding{52}}          \\
Debian~\cite{debian_dataset}                       & \textcolor{green}{\ding{52}}          & \textcolor{green}{\ding{52}}           & \textcolor{green}{\ding{52}}             &     \textcolor{red}{\ding{56}}          &       \textcolor{red}{\ding{56}}       &   \textcolor{green}{\ding{52}}           \\
VDISC~\cite{DBLP:conf/icmla/RussellKHLHOEM18}                        & \textcolor{green}{\ding{52}}          & \textcolor{green}{\ding{52}}           & \textcolor{green}{\ding{52}}             &     \textcolor{red}{\ding{56}}          &        \textcolor{red}{\ding{56}}      &  \textcolor{green}{\ding{52}}  \\
\bottomrule
\end{tabular}}
% \vspace{-0.3cm}
\label{tab:eval_selection}
\end{wraptable}

\textbf{STEP-1: Selecting \& Collecting the Dataset.}
Before the dataset preprocessing, we search for five candidate data sources, i.e., \textbf{GHArchive}~\cite{GHArchive}, \textbf{D2A}~\cite{DBLP:conf/icse/ZhengPLBEYLMS21}, \textbf{VulZoo}~\cite{10.1145/3691620.3695345}, and MITRE's \textbf{ATT\&CK}~\cite{ATTCK} \& \textbf{CAPEC}~\cite{CAPEC1}.
Since the selected data sources for evaluating the {\tool} are large-scale and should contain traceable IRs,
the criteria for selecting the data source will not be limited to selecting VA datasets (Section \ref{sec:external_knowledge}), i.e., \textbf{Inclusion of Recent Vulnerabilities (IRV)}, \textbf{Number of Vulnerabilities (NV)}, and \textbf{Usage in Security Communities (USC)}.
We strictly follow Wu's work on vulnerability database reviewing~\cite{DBLP:conf/icse/WuSH0024} and add three additional criteria to review the evaluation dataset, and the selected data sources should satisfy all the criteria after reviewing. 

\begin{itemize}[leftmargin=*]
    \item \textbf{Criterion-1: Traceability of IRs (TIR).} The original vulnerability-related IRs will be traceable based on each sample's detail, and the CWE-IDs are also traceable in the data sources.
    \item \textbf{Criterion-2: Coverage of Programming Languages (CPL).} The vulnerabilities in the dataset include various programming languages, and {\tool} cannot be limited to identifying vulnerabilities in only a few programming languages.
    \item \textbf{Criterion-3: Number of Noise Samples (NNS).} The data sources should not contain too much noise, including the incorrect labels of vulnerability and CWE-ID.
\end{itemize}

From Table \ref{tab:eval_selection}, we can see that only the \textbf{GHArchive} satisfies all the criteria after dataset review and can be used as the data source for vulnerability identification. 
Note that, these data sources are different from the previous VA dataset
After we determine the data source we use in the experiment, we collect the dataset by strictly following Pan et al's work's basic process of constructing the dataset~\cite{DBLP:conf/sigsoft/PanZC0BHLH22}, so the labels of whether the IRs contain the vulnerability (i.e., \textbf{the first task:} identifying the vulnerability-related IRs, where the label is ``Yes or No'') are correct, and the dataset is considered as a benchmark dataset without noise data (i.e., IRs that are mislabeled with incorrect labels).
It is a comprehensive dataset that contains 3,884 officially disclosed vulnerabilities with their source links of GitHub IRs since 2015. This dataset is manually labeled by security practitioners in CVE and has been widely used in vulnerability identification tasks~\cite{DBLP:journals/pieee/LinWHZX20,DBLP:conf/icmla/RussellKHLHOEM18,DBLP:conf/ndss/LiZXO0WDZ18,DBLP:conf/csr2/OmarS23,DBLP:conf/raid/0001DACW23,DBLP:conf/sigsoft/PanZC0BHLH22}.
We collect the vulnerability information in this dataset, such as whether the IR is vulnerability-related, their disclosed CVE-ID, and relevant CWE-ID.

{\textbf{STEP-2: Refining the Dataset.} 
{Since the labels of vulnerability-related IR identification are reviewed and double-checked regularly by experienced security practitioners from well-known security organizations/institutions (e.g., CVE, NVD, OWASP, etc.), we believe that these labels are correct. 
However, for the labels of CWE-ID (i.e., \textbf{the second task:} predicting the CWE-IDs of the vulnerability) in our manual analysis, around 118 of 3,884 vulnerability-related IRs may contain vulnerabilities that correspond to multiple CWE-IDs.}
The original GHArchive dataset may miss these CWE-ID labels, which affects the reliability of evaluation results.
For example, the labeled CWE-ID in \textit{Fuel-CMS/issues/536} is the ``XSS Injection'' (CWE-79)~\cite{cwe_79}. Still, we find that the project may also encounter the ``Cross-Site Request Forgery (CSRF)'' (CWE-352)~\cite{cwe_352} based on the description of this IR, but the CWE-352 label is missed in the dataset.
To improve the quality of our dataset, we refine the dataset by adding new IRs with the missed labels.
To reduce the biases in the manual re-annotation, we have invited three security practitioners out of the authors, who have over five years of experience in software security, to determine whether the re-annotated dataset is correct. We ask them to independently check whether the new samples are accurate.
The average Cohen’s Kappa~\cite{DBLP:journals/jss/PerezDMT20} value is 0.9, which means the perfect agreement on the re-annotated dataset.}

\textbf{STEP-3: Collecting the Original Rich-text Information.}
For all the samples in the dataset, we retrieve their original rich-text information.
For GHArchive, each sample contains an external link to the original address of the GitHub IR.
We search the source of all the GitHub IRs and utilize Python's \texttt{BeautifulSoup} package to spider all the XML elements in the original pages, including the links to page screenshots (wrapped by \texttt{$<$a href=".jpg|.png"$>$}, \texttt{$<$/a$>$}) and code snippets (wrapped by \texttt{$<$code$>$}, \texttt{$<$/code$>$}).

\textbf{STEP-4: Preprocessing the Dataset.} 
The GitHub IRs collected from the web pages are in XML format, 
so we preprocess the IRs with the following steps: 
\ding{182} 
{We use \textit{\textbf{[SCR]}} to tag the screenshots and  \textit{\textbf{[CODE]}} to tag the code snippets. Then, we add new JSON fields to store the content of rich-text information in the input IRs so that each IR can be represented as {\{``Content'': ``\textbf{\textit{Text}}\&\textbf{\textit{[SCR]}}\&\textbf{\textit{[CODE]}}'', ``Rich-Text'': [``\textbf{\textit{[SCR]}}'': ``Link of Screenshot'', ``\textbf{\textit{[CODE]}}'': ``Detail of Code Snippet'']\}};
\ding{183} we merge similar code snippets and page screenshots, which may have few differences and describe similar vulnerability-related information; 
\ding{184} we remove other XML tags (e.g., \texttt{$<$td$>$}, \texttt{$<$tr$>$}, \texttt{$<$p$>$}, etc.) and retain the plain text inside these tags, then correct typos and lemmatize the texts~\cite{DBLP:conf/kbse/ShiJYCZMJW21}};
and \ding{185} we split the vulnerability-related IRs by sorting the IRs in time order, then choose the first 60\% of the IRs as the historical IR and the remaining 40\% as the target IR (proportion setting is decided by the hyper-parameter tuning, which is shown in Section \ref{sec:paramaeters}).

% Please add the following required packages to your document preamble:
% \usepackage{multirow}
% \ziyou{has rich-text/without rich-texts, split to target/historical IRs, not the D2A, this dataset is not convinced!}
\begin{table}[t]
\centering
\small
\caption{The statistic of the number of IRs in the original and refined datasets.}
\vspace{-0.2cm}
\resizebox{\columnwidth}{!}{
\begin{tabular}{cc|c|ccc|c|ccc}
\toprule
\multicolumn{2}{c|}{\multirow{2}{*}{\textbf{Dataset}}}&\multicolumn{4}{c|}{\textbf{Original}}&\multicolumn{4}{c}{\textbf{Refined}$^{*}$}
\\
& & \textbf{\#Total} & {\textbf{\#R-Text}} & {\textbf{\#SCR}} & {\textbf{\#CODE}}                    & \textbf{\#Total} & {\textbf{\#R-Text}} & {\textbf{\#SCR}} & {\textbf{\#CODE}} \\
\midrule
\multirow{2}{*}{\textbf{Vul-IR}} & {{Historical}} &2,306&683&495&619& 2,401 (+95)           & 720 (+37)   & 527 (+32)   & 630 (+11)                  \\
% & {\textbf{Non-Vul-IR}}  & -              & -   & -              & -                      \\
 & {{Target}} &1,578&956&664&760& 1,601 (+23)           & 966 (+10)   & 670 (+6) & 767 (+7)                       \\
\midrule
\multirow{2}{*}{\textbf{Non-Vul-IR}} & {{Historical}} & 714,790              & 466,179   & 352,675 & 330,753   & 714,790              & 466,179   & 352,675 & 330,753 \\
& {{Target}}& 476,528 & 201,391 & 139,578 & 276,639 & 476,528 & 201,391 & 139,578 & 276,639
\\
\bottomrule
\end{tabular}}
 \begin{tablenotes}
        \footnotesize
        \item[*] $^{*}$ The refined dataset for training and evaluating {\tool} and baselines. The value after each IR's number, i.e., (+\textit{value}), indicates the increased number of IRs after dataset refining (\textbf{STEP-2}). 
        Note that, the dataset refining step only adds the missing CWE-IDs for the vulnerability-related IRs, so the numbers in the \textbf{Non-Vul-IR} will not be changed.
\end{tablenotes}
\vspace{-0.2cm}
\label{tab:dataset}
\end{table}

{Table \ref{tab:dataset} shows the number of IRs in our dataset, where the column \textbf{\#Total} and \textbf{\#R-Text} indicate the total number of IRs, and IRs with rich-text, and the number \textbf{\#Total-\#R-Text} illustrates the IRs with only plain text.
Column \textbf{\#SCR}, and \textbf{\#CODE} indicate the number of IRs with page screenshots and code snippets in the \textbf{\#R-Text}.
The labels \textbf{Vul-IR} and \textbf{Non-Vul-IR} indicate whether the IR is vulnerability-related.
We obtained 4,002 vulnerability-related IRs with 1,686 rich-text IRs, where 2,401 (60\%) vulnerability-related IRs are split into historical IRs and 1,601 (40\%) are target IRs.}

\subsection{Baselines}

To evaluate the performance of our approach, 
we find that identifying vulnerabilities from IRs lacks baselines, so we manually select approaches that can be used in the vulnerability detection and natural language process (NLP) tasks, and then we select the baselines in the following criteria:

\begin{itemize}[leftmargin=*]
    \item \textbf{Criterion-1: Applicability of baselines.} The selected baselines can be applied to our tasks. For example, traditional vulnerability detectors mainly focus on finding vulnerabilities from source code in the repository, which cannot be applied to identify vulnerabilities from IRs. 
    \item \textbf{Criterion-2: Availability of baselines' artifacts.} Considering the reliability of the baseline's result, we select the baselines whose dataset and code are available. For the approaches that their artifacts are not available, we have asked the author of these approaches and conducted the experiments on the models if the authors respond to us.
    \item \textbf{Criterion-3: Performance\&Efficiency of baselines.} The selected baselines have better performances and lower time costs. After discussing with the security practitioners who participate in our dataset refining, as well as following the previous work~\cite{DBLP:journals/csur/HarzeviliBWWJN25}, we select the baselines whose performances are $\geq 30\%$ F1 on average, with the time cost $\leq 1$ min/per IR.
    The traditional LLM reasoning and RAG approaches, such as AutoRAG, MemoRAG, and ToolBench, cannot be selected based on this criterion.
\end{itemize}

After evaluating the novel approaches in vulnerability identification, LLM's reasoning, and RAG, we finally select three types of baselines that meet the previous three criteria, i.e., three deep learning (DL) and LLM baselines that are widely used in NLP tasks, with three LLM reasoning strategies that are applicable in identifying vulnerability-related IRs and predicting CWE-IDs.

\noindent\textbf{DL Baselines.}
We first retrain the Deep Learning (DL) baselines on our dataset to identify the vulnerability-related IRs and predict their CWE-IDs.
\textbf{LR}~\cite{liu2018application} is the linear regression method that predicts the vulnerability with the single-layer perception; \textbf{MLP}~\cite{wu2017vulnerability,li2019comparative} utilizes the multi-layer perceptions to classify the type of vulnerability.
In our experiment, these two DL baselines can outperform other baselines in identifying vulnerabilities from rich-text information.
Also, we introduce \textbf{{\toolcompare}}~\cite{DBLP:conf/sigsoft/PanZC0BHLH22}, which is the latest approach that utilizes the memory network to store the relationships between vulnerability types and reuse it to identify the new IRs.

\noindent
{\textbf{LLM Baselines.}
In addition to DL baselines, we also compare the {\tool} with the original LLMs, since they are widely used and achieve high performances on text classification and generation tasks.
These LLM baselines use the same prompt as {\tool}, i.e., $P_{identify}$, in Section \ref{sec:vul_identification} to identify the vulnerability by outputting the confidence scores.
\textbf{LLaMA}~\cite{DBLP:journals/corr/abs-2302-13971} is the LLM proposed by Meta and is trained on multiple language models with various inference budgets;
\textbf{GPT-3}~\cite{DBLP:conf/emnlp/YooPKLP21}
and \textbf{ChatGPT (i.e., GPT-3.5)}~\cite{LLMBackground} are two novel LLMs proposed by OpenAI, which use over 100B of parameters and are trained on over 10TB samples with multiple training strategies (few-shot, zero-shot, etc.).
We choose the following stable versions of LLMs: {LLaMA} (\textit{Llama-2-13b-chat-hf} (13B)~\cite{llama_version}), {GPT-3} (\textit{text-davinci-003} (175B)~\cite{gpt3_version}), and \textbf{ChatGPT} (\textit{gpt-3.5-turbo}~\cite{gpt35_version}).

\noindent\textbf{LLM Reasoning Strategies.}
Since LLMs are well-trained in these versions, we will not retrain them on our dataset.
Moreover, for each LLM, we also introduce three reasoning strategies to analyze the rich-text information, i.e., \textbf{+CoT-SC}~\cite{DBLP:conf/nips/Wei0SBIXCLZ22}, and 
\textbf{+ReAct}~\cite{DBLP:conf/iclr/YaoZYDSN023}, which are latest frameworks that control the LLM's reasoning with the highest performances on our tasks.
\textbf{+DS-Agent}~\cite{DBLP:journals/corr/abs-2402-17453} is the latest approach that firstly combines the RAG in the LLM reasoning process.}

To ensure a fair comparison, all the DL and LLM baselines use the same JSON-formatted IRs as input datasets with rich-text information, as is illustrated in \textbf{STEP-4} of Section \ref{sec:dataset_preparation}.

\subsection{Evaluation Metrics}

\noindent\textbf{Evaluation Metrics on Identifying Vulnerability-related IR.}
Since the labels of target IRs are imbalanced in Table \ref{tab:dataset}, we choose two sets of appropriate metrics to measure the performance of vulnerability-related IR identification.
The first set of metrics measures the performance of {\tool} on identifying the positive vulnerability-related IRs, i.e., \textbf{Precision}, \textbf{Recall}, and \textbf{F1-Score}.
These three metrics are useful in text classification tasks when the labels in the testing dataset are imbalanced. 
{Precision} calculates the ratio of correct positive predictions to the total positive predictions; {Recall} calculates the ratio of correct positive predictions to the ground-truth positive labels; and {F1-Score} is the harmony of Precision and Recall.

{Besides, since the number of negative samples is much larger than the positive ones ({\textit{imbalanced dataset}}), we choose threshold-independent metrics to measure the performances of {\tool} when the dataset is extremely imbalanced, i.e., \textbf{AUROC} (area under the Receiver Operating Characteristics curve) and \textbf{AUPRC} (area under the Precision-Recall curve). 
To intuitively illustrate the benefits of {\tool}, we plot the Precision-Recall curves by tuning the threshold $\theta_{out}$ within $[0,1]$ with 0.5 as the interval and measure the trade-off values.
Previous works illustrate that AUROC and AUPRC are indicative metrics when the dataset imbalance affects {\tool}'s performances~\cite{DBLP:conf/icml/DavisG06,DBLP:conf/icse/0001WCT20}.}

\noindent\textbf{Evaluation Metrics on Predicting CWE-IDs.}
{To measure the performance in CWE-ID prediction}, we apply the \textbf{Macro-P}, \textbf{Macro-R}, and \textbf{Macro-F1}.  
Macro-P and Macro-R are the macro averages of precision and recall on all the CWE labels (i.e., the equations are $\frac{1}{n}\sum^{n}_{cwe_i}pre(cwe_i)$ and $\frac{1}{n}\sum^{n}_{cwe_i}rec(cwe_i)$), and Macro-F1 is the harmony of Macro-P and Macro-R. 
Due to the equal importance among all the CWE-IDs, the macro average value can better reflect the prediction results than the micro average value. 
{Since we only evaluate CWE-ID prediction with positive labels in vulnerability identification, and different CWE-IDs have similar distributions in the vulnerability-related IRs, we will not utilize the AUROC and AUPRC in the CWE-ID prediction task.}

\noindent\textbf{Evaluation Metrics on Time Cost.}
To compare the time cost of the baselines and {\tool}, we utilize the average seconds of identifying a vulnerability-related IR and predicting its CWE-ID, which is denoted by s/IR. We calculate the average time cost after all the experiments are conducted.

\subsection{Experimental Settings}

\noindent\textbf{Experiment Details.}
Since the LLM's output may have randomness, we conducted \textbf{20-time experiments} on target IRs and calculated the mean value of these evaluation results, i.e., $\frac{1}{20}\sum_i^{20}res_i$, to reduce the biases, where $res_i$ indicates the $i_{th}$ metric result of experiments.
We use the LLM's API, such as \texttt{logprobs.top\_logprobs} to output the probability and \texttt{message.content} to output the response text.
To evaluate the advantages and practical application, we conduct the ablation study on four types of components and apply {\tool} on the newly proposed IRs to observe the proportion of disclosed vulnerabilities.

\noindent\textbf{Hyper-parameters.} 
We first set the proportion of the historical IRs as 60\% and the threshold $\theta_{sim}=0.7$ (parameter settings are shown in Section \ref{sec:paramaeters}).
{Then, we set the threshold for determining the output of vulnerability, i.e., $\theta_{out}$, within the range $[0.0, 1.0]$, and choose the optimal F1 value when $\theta_{out}=0.55$ (the tuning of the $\theta_{out}$ may affect the trade-off between Precision and Recall, so we illustrate the curve in Section \ref{sec:results}).
For each value in the parameter tuning, we also conduct 20-time experiments to determine whether the output of DL/LLM will not have biases and calculate the average score of the repeated experiments.}
For other parameters in {\tool}'s LLM module and the baselines, we set the value of them in DL/LLM, such as the number of layers, learning rate (i.e., $2e^{-5}$ for BERT encoder and $1e^{-4}$ for other modules), temperature (i.e., 0.3 for all the LLMs to ensure a stable probability distribution), etc., as default values.

\noindent\textbf{Hardware.} 
All the baselines and variants are retrained on these IRs in the refined dataset and run on a high-performance server with Ubuntu OS, four NVIDIA RTX A6000 GPUs, and 64GB RAM.
\section{Results}\label{sec:results}

\subsection{RQ1: Performances of Identifying Vulnerability-related IRs and Predicting CWE-IDs}

% Please add the following required packages to your document preamble:
% \usepackage{multirow}
% \usepackage[table,xcdraw]{xcolor}
% Beamer presentation requires \usepackage{colortbl} instead of \usepackage[table,xcdraw]{xcolor}
% \usepackage[normalem]{ulem}
% \useunder{\uline}{\ul}{}
\begin{table*}[t]
\caption{The results of baseline comparison of {\tool} on identifying vulnerability-related IRs and predicting CWE-IDs of these vulnerabilities on overall target IRs (\%).}
\vspace{-0.2cm}
\resizebox{\textwidth}{!}{\begin{tabular}{cl|r|lllll|lll}
\toprule
 &  & & \multicolumn{5}{c|}{\textbf{Vul-IR  Identification (VI)}}                                                             & \multicolumn{3}{c}{\textbf{CWE-ID Prediction (CP)}}                                                                           \\
\multirow{-2}{*}{\textbf{Category}} & \multicolumn{1}{c|}{\multirow{-2}{*}{\textbf{Methods}}}& \multicolumn{1}{c|}{\multirow{-2}{*}{\makecell[c]{\textbf{Time}\\\textbf{Cost Avg.}}}} & \textbf{Precision}                           & \textbf{Recall}                              & \textbf{F1-Score} & \textbf{AUROC}  & \textbf{AUPRC}                         & \textbf{Macro-P}                             & \textbf{Macro-R}                             & \textbf{Macro-F1}                        \\ \midrule
                   \multirow{3}{*}{\makecell[c]{\textbf{DL}\\\textbf{Baselines}}}  &              LR   &    1.0 s/IR                                      & 43.7                                & 24.0                                & 31.0   & 90.3 &	30.5 
                         & 34.1                                & 31.7                                & 32.9                            \\
                              &   MLP   &   1.7 s/IR                                        & 35.2                                & 34.7                                & 34.9    & 91.9  &	19.6 
                           & 42.7                                & 45.8                                & 44.2                           \\
& {\toolcompare}      &   2.0 s/IR                                  & 38.3                                & 71.6                                & 49.9     & 96.5 	& 33.2 
                          & 46.5                                & 52.4                                & 49.3                            \\ \midrule
                             \multirow{5}{*}{\makecell[c]{\textbf{LLaMA}\\ \textbf{Baselines \&} \\ \textbf{Our Approach}}}   &     Original    &     3.6 s/IR                                & 66.5                                & 74.1                                & 70.1     & 92.6  &	35.0 
                        & 57.5                                & 66.7                                & 61.8                         \\
                               &   +CoT-SC    &   11.7 s/IR                                   & 68.2                                & 70.3                                & 69.2  &  94.0  &	39.6 
                           & 59.4                                & 69.2                                & 63.9                           \\
                             &    +ReAct    &   15.0 s/IR                                    & 69.9                                & 70.1                                & 70.0   & 94.1  &	40.2 
                            & 60.2                                & 65.7                                & 62.8                          \\
& +DS-Agent &  20.0 s/IR & 69.3 & 69.5 & 69.4 & 93.6 &	37.9 
  & 62.3 & 61.7 & 62.0 \\
& \cellcolor[HTML]{EFEFEF}+{\tool}  &    \cellcolor[HTML]{EFEFEF} \textbf{7.9 s/IR}      & \cellcolor[HTML]{EFEFEF}\textbf{74.5} ($\textcolor{red}{\uparrow}$4.6) & \cellcolor[HTML]{EFEFEF}\textbf{75.3} ($\textcolor{red}{\uparrow}$1.2) & \cellcolor[HTML]{EFEFEF}\textbf{74.9} ($\textcolor{red}{\uparrow}$4.8) &
\cellcolor[HTML]{EFEFEF}\textbf{97.9} ($\textcolor{red}{\uparrow}$3.8) &	\cellcolor[HTML]{EFEFEF}\textbf{65.9} ($\textcolor{red}{\uparrow}$25.7)
 &
\cellcolor[HTML]{EFEFEF}\textbf{68.5} ($\textcolor{red}{\uparrow}$6.2) & \cellcolor[HTML]{EFEFEF}\textbf{72.4} ($\textcolor{red}{\uparrow}$3.2) & \cellcolor[HTML]{EFEFEF}\textbf{70.4} ($\textcolor{red}{\uparrow}$6.5) 
\\ \midrule
                            \multirow{5}{*}{\makecell[c]{\textbf{GPT-3}\\ \textbf{Baselines \&} \\ \textbf{Our Approach}}}    &         Original   &   4.9 s/IR                                    & 50.8                                & 72.0                                & 59.6   & 96.0  &	39.6 
                             & 66.8                                & 60.3                                & 63.4                                \\
                               &   +CoT-SC  &    19.5 s/IR                                 & 62.5                                & 76.4                                & 68.8 & 96.9  &	41.5 
                             & 67.2                                 & 61.5                                & 64.2                               \\
                            &      +ReAct  &   21.0 s/IR                                   & 63.0                                & 69.7                                & 66.2   & 96.7 	& 45.5 
                              & 67.5                                & 62.0                                & 64.6                              \\
& +DS-Agent &  25.5 s/IR & 64.6 &	98.5 &	47.7 
 & 96.1 &	39.9 &	63.9 
  &	67.9  &	65.8
\\
 & \cellcolor[HTML]{EFEFEF}+{\tool}   &     \cellcolor[HTML]{EFEFEF}     \textbf{9.2 s/IR}                         & \cellcolor[HTML]{EFEFEF}\textbf{69.5} ($\textcolor{red}{\uparrow}$4.9)                         & \cellcolor[HTML]{EFEFEF}\textbf{77.0} ($\textcolor{red}{\uparrow}$0.6)                         & \cellcolor[HTML]{EFEFEF}\textbf{73.1} ($\textcolor{red}{\uparrow}$4.3) 
 & \cellcolor[HTML]{EFEFEF}\textbf{98.2} ($\textcolor{red}{\uparrow}$1.3)	& \cellcolor[HTML]{EFEFEF}\textbf{77.2} ($\textcolor{red}{\uparrow}$31.7)

 & \cellcolor[HTML]{EFEFEF}\textbf{69.4} ($\textcolor{red}{\uparrow}$1.5)                         & \cellcolor[HTML]{EFEFEF}\textbf{75.3} ($\textcolor{red}{\uparrow}$7.4)                        & \cellcolor[HTML]{EFEFEF}\textbf{72.2} ($\textcolor{red}{\uparrow}$6.4)  
                      \\ \midrule
                           \multirow{5}{*}{\makecell[c]{\textbf{ChatGPT}\\ \textbf{Baselines \&} \\ \textbf{Our Approach}}}   &          Original    &   4.2 s/IR                                 & 80.7                                & 66.3                                & 72.8  & 97.3  &	45.9 
                              & 76.2                                & 63.9                                & 69.5                                \\
                              &    +CoT-SC   &    26.9 s/IR                                 & 85.3                                & 70.2                               & 77.0   & 98.1  &	46.8 
                            & 71.3                                & 70.2                                & 70.7                                 \\
                           &       +ReAct   &   30.0 s/IR                                   & 86.9                                & 67.4                                & 75.9   & 98.5 & 	53.6 
                            & 72.5                                & 70.7                                & 71.6                                \\
& +DS-Agent & 44.5 s/IR  & 84.2 &	72.5 &	77.9 & 70.2 &	60.2 &	75.3 &	73.2 &  66.7
\\
 & \cellcolor[HTML]{EFEFEF}+{\tool}       &   \cellcolor[HTML]{EFEFEF}       \textbf{11.7 s/IR}                     & \cellcolor[HTML]{EFEFEF}\textbf{90.2} ($\textcolor{red}{\uparrow}$3.3)                         & \cellcolor[HTML]{EFEFEF}\textbf{87.7} ($\textcolor{red}{\uparrow}$15.2)                        & \cellcolor[HTML]{EFEFEF}\textbf{88.9} ($\textcolor{red}{\uparrow}$11.0)
 & \cellcolor[HTML]{EFEFEF}\textbf{99.2} ($\textcolor{red}{\uparrow}$0.7) &	\cellcolor[HTML]{EFEFEF}\textbf{80.4} ($\textcolor{red}{\uparrow}$20.2)
& \cellcolor[HTML]{EFEFEF}\textbf{82.5} ($\textcolor{red}{\uparrow}$6.3)                         & \cellcolor[HTML]{EFEFEF}\textbf{85.0} ($\textcolor{red}{\uparrow}$9.7)                        & \cellcolor[HTML]{EFEFEF}\textbf{83.7} ($\textcolor{red}{\uparrow}$10.5)                    \\ \bottomrule
\end{tabular}}
\vspace{-0.2cm}
\label{tab:baseline_comparison}
\end{table*}

{In this experiment, we first evaluate the performances of {\tool} on the overall target IRs in our dataset, 
Then, we analyze the performances of {\tool} on target IRs with/without rich-text information and compare the performance of {\tool} with the best-performed DL/LLM baselines, i.e., {\tool} and ChatGPT. We also plot the Precision-Recall (PR) curves to intuitively illustrate the advantages of {\tool} to the baselines on different output thresholds $\theta_{out}$.}

\begin{figure}[t]
\begin{minipage}[H]{.63\textwidth}
\centering
% Please add the following required packages to your document preamble:
% \usepackage{multirow}
% \usepackage[table,xcdraw]{xcolor}
% Beamer presentation requires \usepackage{colortbl} instead of \usepackage[table,xcdraw]{xcolor}
% \usepackage[normalem]{ulem}
% \useunder{\uline}{\ul}{}
\begin{table}[H]
\caption{The results of baseline comparison on IR with/without rich-text information (\%).}
\vspace{-0.2cm}
\resizebox{\textwidth}{!}{\begin{tabular}{cl|lll|l}
\toprule
                                    & \multicolumn{1}{c}{}                         & \multicolumn{3}{|c}{\textbf{VI}}                                                                  & \multicolumn{1}{|c}{\textbf{CP}} \\
\multirow{-2}{*}{\textbf{Category}}          & \multicolumn{1}{c}{\multirow{-2}{*}{\textbf{Methods}}} & \multicolumn{1}{|c}{\textbf{F1-Score}}               & \multicolumn{1}{c}{\textbf{AUROC}}           & \multicolumn{1}{c}{\textbf{AUPRC}}            & \multicolumn{1}{|c}{\textbf{Macro-F1}}          \\
\midrule
                                    & {\toolcompare}                                       & 37.2                                 & 80.6                                & 30.9                                 & 36.8                                  \\
                                    & ChatGPT                                      & 65.3                                 & 86.9                                & 34.2                                 & 60.8                                  \\
                                    & +CoT-SC                                      & 68.9                                 & 89.2                                & 39.0                                 & 61.2                                  \\
                                    & +ReAct                                       & 66.2                                 & 89.9                                & 53.9                                 & 65.3                                  \\
                                    & +DS-Agent                                    & 67.0                                 & 87.5                                & 57.5                                 & 64.6                                  \\
\multirow{-6}{*}{\makecell[c]{\textbf{IR with} \\ \textbf{Rich-Text}}} & \cellcolor[HTML]{EFEFEF}+{\tool}             & \cellcolor[HTML]{EFEFEF}\textbf{85.6} ($\textcolor{red}{\uparrow}$16.7) & \cellcolor[HTML]{EFEFEF}\textbf{97.9} ($\textcolor{red}{\uparrow}$8.0) & \cellcolor[HTML]{EFEFEF}\textbf{76.6} ($\textcolor{red}{\uparrow}$19.1) & \cellcolor[HTML]{EFEFEF}\textbf{88.7} ($\textcolor{red}{\uparrow}$23.4)  \\
\midrule
                                    & {\toolcompare}                                       & 53.2                                 & 97.5                                & 36.6                                 & 53.7                                  \\
                                    & ChatGPT                                      & 79.6                                 & 98.4                                & 40.2                                 & 72.0                                  \\
                                    & +CoT-SC                                      & 80.1                                 & 98.9                                & 47.5                                 & 75.9                                  \\
                                    & +ReAct                                       & 81.7                                 & 99.3                                & 53.4                                 & 73.5                                  \\
                                    & +DS-Agent                                    & 82.9                                 & 99.1                                & 61.9                                 & 74.5                                  \\
\multirow{-6}{*}{\makecell[c]{\textbf{IR w/o} \\ \textbf{Rich-Text} \\ \textbf{(Plain-Text)}}}  & \cellcolor[HTML]{EFEFEF}+{\tool}             & \cellcolor[HTML]{EFEFEF}\textbf{92.2} ($\textcolor{red}{\uparrow}$9.3)  & \cellcolor[HTML]{EFEFEF}\textbf{99.8} ($\textcolor{red}{\uparrow}$0.5) & \cellcolor[HTML]{EFEFEF}\textbf{81.1} ($\textcolor{red}{\uparrow}$19.2) & \cellcolor[HTML]{EFEFEF}\textbf{80.2} ($\textcolor{red}{\uparrow}$4.3)  \\
\bottomrule
\end{tabular}}
\label{tab:split_comparison}
\end{table}
 \end{minipage}
 \hfill
\begin{minipage}[H]{.32\textwidth}
\centering
\caption{Precision-Recall (PR) curves of {\tool} and baselines.}
\vspace{0.1cm}
\includegraphics[width=\textwidth]{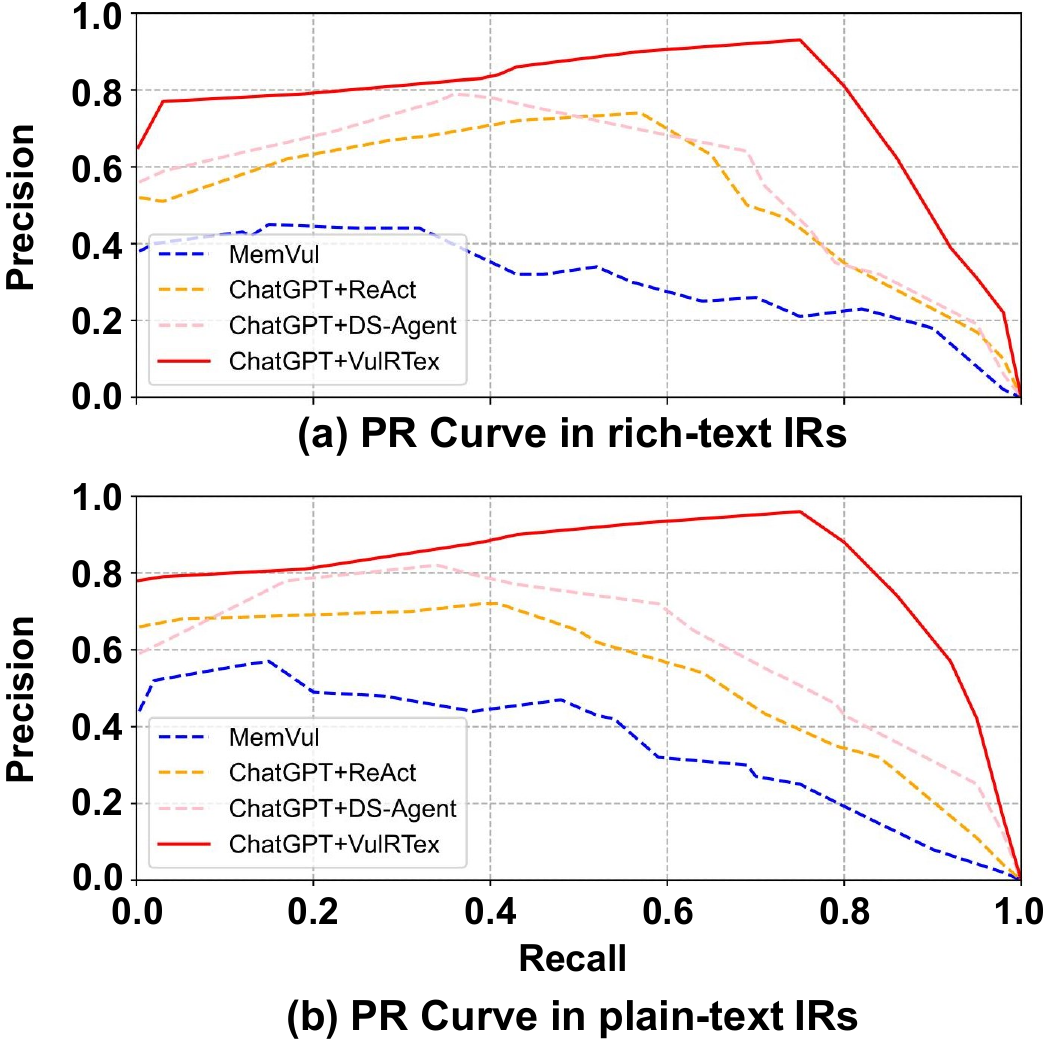}
\label{fig:pr_curve}
% \vspace{-0.8cm}
\end{minipage}
\vspace{-0.3cm}
\end{figure}

\begin{figure}[b]
\centering
\includegraphics[width=\columnwidth]{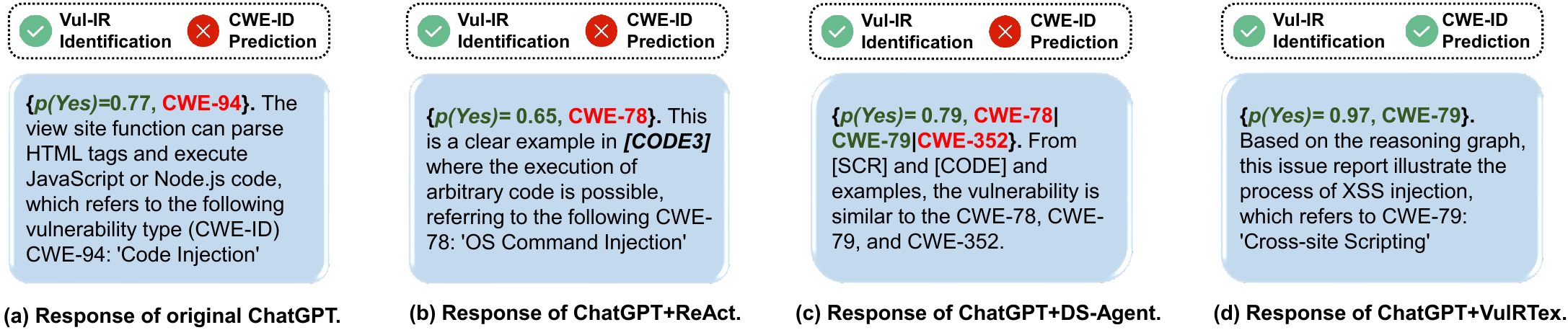}
\vspace{-0.7cm}
\caption{The case study of {\tool} on the motivation example (example in Fig. \ref{fig:motivation_example}).}
\vspace{-0.4cm}
\label{fig:case_study}
\end{figure}

\noindent
\textbf{Comparison Results on Overall IRs.}
Table \ref{tab:baseline_comparison}
illustrate the comparison results and the best performance of each column is highlighted with \textbf{bold face}. 
We can see that, the {\tool} can outperform all the DL, LLM, and LLM reasoning baselines. 
The ChatGPT+{\tool} obtains the highest performances with 88.9\% (F1), 99.2\% (AUROC), and 80.4\% (AUPRC) in vulnerability identification, and 83.7\% (Macro-F1) in CWE-ID Prediction,
improving the best baseline with +11.0\% (F1-Score), +0.7\% (AUROC), +20.2\% (AUPRC), and +10.5\% (Macro-F1).
Moreover, the time cost of each LLM+{\tool} is only higher than the original LLMs and around 2x lower than the baseline reasoning approaches. Especially for ChatGPT, the time cost of {\tool} reduces with over -15.2 s/IR.
Overall, the trade-off between performances and time cost illustrates the benefits of {\tool}.

\noindent\textbf{Case Study.}
{To qualitatively evaluate the {\tool}, we conduct the case study by analyzing the responses of ChatGPT on the motivation's IR.
We compare {\tool} with the original ChatGPT and two state-of-the-art (SOTA) reasoning approaches, i.e., ReAct and DS-Agent.
Fig. \ref{fig:case_study} shows the results of the case study. 
{We can see that the original ChatGPT/ChatGPT+ReAct cannot predict the accurate CWE-ID, and ChatGPT+DS-Agent fails to determine the CWE-ID based on the relevant information it retrieves from the historical IRs.
By comparison, {\tool} can accurately predict the CWE-ID based on the reasoning graphs and historical IRs.}
These results illustrate that {\tool} can intuitively outperform the performance of LLMs and other reasoning approaches.}

\noindent\textbf{Comparison Results on IRs with/without Rich-Text Information.}
{Table \ref{tab:split_comparison} shows the comparison results on target IR with/without rich-text information.
We can see that ChatGPT+{\tool} significantly outperforms the baselines on both rich-text and plain-text IRs.
For rich-text IRs, it significantly outperforms the best baseline by +19.1\% (AUPRC) in vulnerability identification and +23.4\% (Macro-F1) in CWE-ID prediction.
For plain-text IRs, it also outperforms the best baseline by +19.2\% (AUPRC) in vulnerability identification and +4.3\% (Macro-F1) in CWE-ID prediction.}
{To further evaluate the ability of {\tool} to adapt to the imbalanced dataset, we plot the \textbf{Precision-Recall (PR) curve} of {\tool} and representative baselines, i.e., {\toolcompare}, ChatGPT+ReAct, and ChatGPT+DS-Agent, which can reflect the performance of these models under different output threshold $\theta_{out}$.
In Fig. \ref{fig:pr_curve}, we can see that, the performances of {\tool} are higher than baselines under all the different thresholds, and the area (i.e., AUPRC) under its PR curve surrounds the other baselines' curves, which further illustrates {\tool}'s advantages.}
Overall, {\tool} outperforms the baselines on both IRs with or without rich-text information, which illustrates its practicality.

\begin{center}
\small
\begin{tcolorbox}[colback=gray!5,%gray background
                  colframe=black,% black frame color
                  width=\columnwidth,% Use 8cm total width,
                  arc=1mm, auto outer arc,
                  boxrule=0.5pt,
                  left=1pt,
                  right=1pt,
                  top=1pt,
                  bottom=1pt
                 ]
\textit{\textbf{Answering RQ1}: {\tool} improves all the baselines, and ChatGPT+{\tool} outperforms the best baseline with +19.2\% (AUPRC) in vulnerability identification and +4.3\% (Macro-F1) in CWE-ID prediction, with 2x lower time cost than the LLM reasoning baselines.
The PR curves further illustrate its advantages in identifying vulnerabilities in imbalanced datasets.}
\end{tcolorbox}
\end{center}

\subsection{RQ2: Ablation Study}

{We conduct the ablation study by comparing {\tool} with the following four types of variants:
\begin{itemize}[leftmargin=*]
    \item \textbf{Rich-text Information:} \textbf{w/o \textit{[SCR]}} \& \textbf{w/o \textit{[CODE]}} (remove page screenshots \& code snippets).
    \item \textbf{CoT-based Reasoning:} \textbf{w/o FECorr} (removing the LLM's factual error correction), as well as replacing our LLM reasoning approach with \textbf{CoT-SC} and \textbf{ReAct}.
    \item \textbf{Reasoning Graph Retrieval:} \textbf{w/o Pruning}, as well as replacing random-walking with \textbf{CART}~\cite{DBLP:journals/imcs/MathieuT23}, and \textbf{PageRank}~\cite{Page1999Page}, where CART and PageRank are representative graph pruning algorithms.
    \item \textbf{Text Matching Similarity:} Replacing TF-IDF with \textbf{Levenshtein Distance}~\cite{DBLP:conf/ntms/AfzalGLB18}, \textbf{Euclidean distance}, and \textbf{Cosine Distance}, where Euclidean and Cosine distance utilize the Word2Vec~\cite{DBLP:journals/corr/abs-1301-3781} to embed the two sentences $s_x$ and $s_y$ with the mean value of word embeddings, then calculate the similarity between the two sentence embeddings (i.e., $sim(w2v(s_x), w2v(s_y))$). 
\end{itemize}
where the symbol \textbf{w/o} means removing the component from the {\tool}, and the variant without \textbf{w/o} means replacing the component with the corresponding variants.}

\begin{figure}[t]
\centering
\includegraphics[width=\columnwidth]{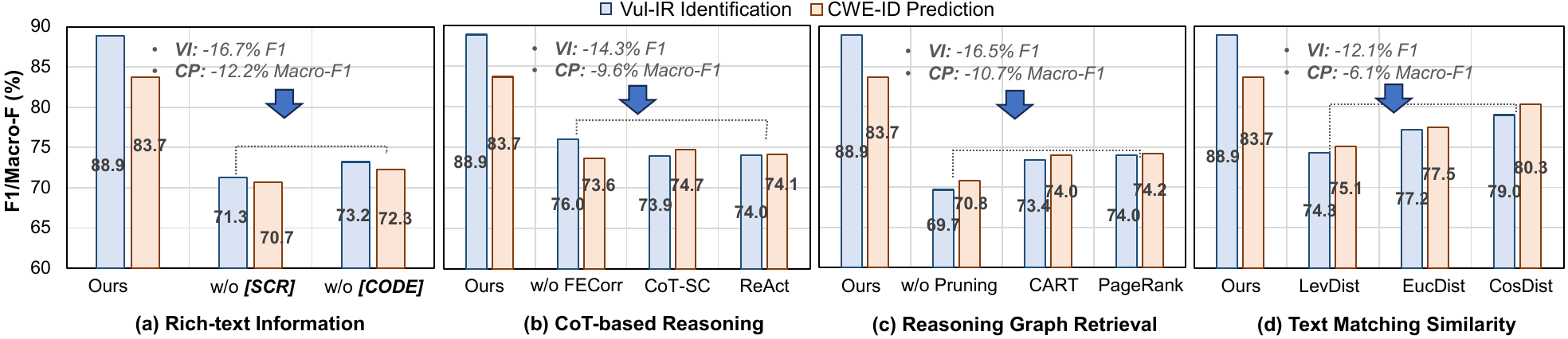}
\vspace{-0.6cm}
\caption{The results of ablation study.}
\vspace{-0.4cm}
\label{fig:ablation_study}
\end{figure}

From Fig. \ref{fig:ablation_study} (a) and (c), we can see that the variants in \textbf{rich-text information} (-16.7\% F1, -12.2\% Macro-F1) and \textbf{reasoning graph retrieval} (-16.5\% F1, -10.7\% Macro-F1) lead to a large decrease in vulnerability-related IR Identification and CWE-ID prediction.
Among the variants, removing the page screenshot analysis \textbf{w/o \textit{[SCR]}} (-17.6\% F1, -13.0\% Macro-F1) and removing the random-walking pruning (-19.2\% F1, -12.9\% Macro-F1) has the largest decrease.

From Fig. \ref{fig:ablation_study} (b) and (d), we can see that, compared with the other two types of variants in (a) and (c), the variants in \textbf{LLM reasoning approaches} (-14.3\% F1, -9.6\% Macro-F1) and \textbf{text matching similarity} (-12.1\% F1, -6.1\% Macro-F1) lead to a moderate decrease in vulnerability-related IR identification and CWE-ID prediction.
Among the variants, replacing TF-IDF similarity with Levenshtein distance (-14.6\% F1, -8.6\% Macro-F1) has the largest decrease.

\begin{center}
\small
\begin{tcolorbox}[colback=gray!5,%gray background
                  colframe=black,% black frame color
                  width=\columnwidth,% Use 8cm total width,
                  arc=1mm, auto outer arc,
                  boxrule=0.5pt,
                  left=1pt,
                  right=1pt,
                  top=1pt,
                  bottom=1pt
                 ]
\textit{\textbf{Answering RQ2}: {\tool} outperforms all the variants in the ablation study. 
The variants in rich-text information and reasoning graph retrieval have the largest contributions, and the CoT-based reasoning and text-matching similarities have moderate contributions on average.
}
\end{tcolorbox}
\end{center}

\subsection{RQ3: Performances on Identifying Emerging Vulnerabilities in Open-source Projects}

{To analyze the performances of {\tool} on identifying emerging vulnerabilities, 
We deploy {\tool} and continuously apply it to track the GitHub IRs within these projects proposed after \textit{Jan 1st, 2024}.
Since all the IRs in the previous experiments were collected before this date, these newly tracked GitHub IRs are \textbf{not included} in the collected dataset in Section \ref{sec:dataset_preparation}.
Then, we track the IRs with the issue-tracking system (e.g., Bugzilla~\cite{bugzilla} etc.) on these projects, and observe 30 IRs that may potentially describe vulnerabilities with {\tool}.
Then, we notify the OSS project's owners and IR authors through their emails or submit target IR comments if emails are not reserved.
We ask the IR authors and project owners the following two questions to verify whether our identified IRs really contain the vulnerability, and require assigning CVE-IDs for these vulnerabilities:}

\begin{itemize}[leftmargin=*]
    \item \textit{\textbf{Q1:} Can you manually identify whether the given IR we provide really contains the vulnerability with the corresponding CWE-ID? Please answer [Yes/No].}
    \item \textit{\textbf{Q2:} If this IR contains the vulnerability, Can you report it to the CVE for vulnerability disclosure?}
\end{itemize}

% Please add the following required packages to your document preamble:
% \usepackage{multirow}

\begin{table}[t]
\caption{The details of vulnerability-related IRs that are newly identified by {\tool} and assigned to CVE-IDs.}
\Huge
\vspace{-0.2cm}
\resizebox{\columnwidth}{!}{  
\begin{tabular}{lc|c|cll|lll|ll}
\toprule
\multicolumn{1}{c}{\multirow{2}{*}{\textbf{OSS Project}}} & \multirow{2}{*}{\textbf{Stars}} & \multirow{2}{*}{\makecell[c]{\textbf{\#Total-IR}\\ \textbf{in 2024}}} & \multicolumn{3}{c|}{\textbf{Statistics of Assigned CVE-IDs}} & \multicolumn{5}{c}{\textbf{Example of Assigned CVE-IDs}}  \\

& & & \multicolumn{1}{|c}{\textbf{\#Identified}} & \multicolumn{1}{c}{\textbf{\#CVE-Assigned}} & \multicolumn{1}{c}{\textbf{\#Potential}} & \multicolumn{1}{|c}{\textbf{Vul-IR}}                                      & \textbf{IR's Type} & \multicolumn{1}{c|}{\textbf{CWE-ID}} & \multicolumn{1}{c}{\textbf{CVE-Assigned}}  & \multicolumn{1}{c}{\textbf{Identified Date}} \\
\midrule
Rimedo-ts~\cite{rimendo}                 & 22             & 1                                    & 1                                         & 1/1 (100.0\%)                               & 0/1 (0.0\%)   & \href{https://github.com/onosproject/rimedo-ts/issues/16}{\#16}             & Plain-Text    & CWE-119                   & CVE-2024-34049                                                 & 02/20/2024 (\textbf{-69 Days})                             \\
Carla~\cite{carla}                     & 11.3K          & 300                                  & 6                                         & 1/6 (16.7\%)                                & 2/6 (33.3\%) & \href{https://github.com/carla-simulator/carla/issues/702}{\#7025}           & Rich-Text     & CWE-119                   & CVE-2024-33903                               & 01/10/2024 (\textbf{-110 Days})                            \\
Hyprland~\cite{hyprland}                  & 21.1K          & 720                                  & 2                                         & 1/2 (50.0\%)                                & 1/2 (50.0\%)          & \href{https://github.com/hyprwm/Hyprland/issues/5787}{\#5787}                 & Plain-Text    & CWE-362                   & CVE-2024-33904                              & 04/18/2024 (\textbf{-11 Days})                    \\
React~\cite{react}                     & 229K           & 197                                  & 4                                         & 0/4 (0.0\%)                                 & 1/4 (25.0\%)              & \href{https://github.com/facebook/react/issues/31174}{\#31174}                 & Rich-Text     & CWE-1333                  & \multicolumn{1}{c}{-}                               & 10/10/2024               \\
Python-jose~\cite{pythonjose}               & 1.5K           & 10                                   & 3                                         & 2/3 (66.7\%)                                & 0/3 (0.0\%)   & \href{https://github.com/mpdavis/python-jose/issues/344}{\#344}              & Rich-Text     & CWE-400                   & CVE-2024-33664                                & 03/13/2024 (\textbf{-43 Days})                           \\
Xxl-job~\cite{xxljob}                   & 27.5K          & 130                                  & 6                                         & 2/6 (33.3\%)                                & 2/6 (33.3\%)    & \href{https://github.com/xuxueli/xxl-job/issues/3375}{\#3375}                 & Rich-Text     & CWE-918                   & CVE-2024-24113                          & 01/14/2024 (\textbf{-25 Days})                         \\
Jerryscript~\cite{jerryscript}               & 6.9K           & 12                                   & 2                                         & 1/2 (50.0\%)                                & 1/2 (50.0\%)             & \href{https://github.com/jerryscript-project/jerryscript/issues/5135}{\#5135} & Rich-Text     & CWE-671                   & CVE-2024-33255                                                 & 03/29/2024 (\textbf{-28 Days})                \\
Node-server~\cite{nodeserver}               & 375            & 16                                   & 1                                         & 1/1 (100.0\%)                               & 0/1 (0.0\%)            & \href{https://github.com/honojs/node-server/issues/159}{\#159}               & Rich-Text     & CWE-20                    & CVE-2024-32652                                         & 04/18/2024 (\textbf{-1 Day})                       \\
Hugo~\cite{hugo}                      & 75.5K          & 115                                  & 2                                         & 1/2 (50.0\%)                                & 1/2 (50.0\%)              & \href{https://github.com/gohugoio/hugo/issues/12396}{\#12396}                  & Rich-Text     & CWE-20                    & CVE-2024-32875                                   & 04/20/2024 (\textbf{-3 Days})                \\
Kubernetes~\cite{kubernetes}                & 111K           & 597                                  & 3                                         & 1/3 (33.3\%)                                & 1/3 (33.3\%)          & \href{https://github.com/kubernetes/kubernetes/issues/124336}{\#124336}         & Rich-Text     & CWE-285                   & CVE-2024-3177                                                  & 04/20/2024 (\textbf{-2 Days})                   \\
\midrule
\multicolumn{2}{c|}{\textit{\textbf{Total}}}                                & 2,098                                 & 30                                        & 11/30 (36.7\%)                              & 9/30 (30.0\%) & \multicolumn{1}{c}{-} & \multicolumn{1}{c}{-} & \multicolumn{1}{c|}{-} & \multicolumn{1}{c}{-} & \multicolumn{1}{c}{-} \\
\bottomrule
\end{tabular}}
\vspace{-0.2cm}
\label{tab:application_emerging}
\end{table}

We pay attention to the popular OSS projects on GitHub with multiple participants and widely used in 2024 based on the \textit{Gitstar Ranking}~\cite{gitstar_ranking}, and remove some personal projects with fewer star ratings.
The column ``\textbf{Statistics of Assigned CVE-IDs}'' of Table \ref{tab:application_emerging} shows the number of identified IRs and the CVE-Assigned IRs. 
The newly identified vulnerability-related IRs belong to the 10 unique projects, from \textit{\textbf{Rimedo-ts}} to \textit{\textbf{Kubernetes}}. 
Among them, most of the projects obtain over 1K stars, and some projects (e.g., \textbf{\textit{React}}, \textbf{\textit{Hugo}}, and \textit{\textbf{Kubernetes}}) are among the top 100 large-scale projects with large user base.
We can see that 30 of 2,098 IRs are identified as vulnerability-related IRs.
Among these identified IRs, 11 of them (36.7\%) are assigned CVE-IDs after vulnerability identification, and 9 of them (30.0\%) are \textbf{potentially-assigned CVE}, which means that project owners admit that these IRs are vulnerability-related based on the feedback from the questionnaire, but have not been assigned CVE-IDs {yet}. These results indicate that {\tool} can identify emerging vulnerabilities from target IRs.

The column ``\textbf{Statistics of Assigned CVE-IDs}'' of Table \ref{tab:application_emerging} shows the details of CVE-Assigned IRs. We can see that, the date of vulnerability identification is earlier than the CVE-Assigned date.
For some open-source projects (i.e., \textbf{\textit{Rimedo-ts}}, \textbf{\textit{Carla}}, and \textbf{\textit{Python-jose}}), the identification time can even be 40 days ahead of the CVE-Assigned time.
{It is worth mentioning that {\tool} can identify only a few vulnerability-related IRs in large-scale projects, mainly because these projects are well-maintained and the proportion of these IRs is small.
Meanwhile, most vulnerability-related IRs we identified were disclosed in April because CVE usually disclosed the vulnerabilities within a certain period.}
In these cases, our method can successfully identify these vulnerabilities, which further illustrates the ability of {\tool} to identify emerging vulnerabilities in practice.

\begin{center}
\small
\begin{tcolorbox}[colback=gray!5,%gray background
                  colframe=black,% black frame color
                  width=\columnwidth,% Use 8cm total width,
                  arc=1mm, auto outer arc,
                  boxrule=0.5pt,
                  left=1pt,
                  right=1pt,
                  top=1pt,
                  bottom=1pt
                 ]
\textit{\textbf{Answering RQ3}: {\tool} can identify the emerging vulnerabilities from large-scale OSS projects.
Among these 30 newly identified vulnerability-related IRs from 10 representative projects, 11 IRs (36.7\%) are assigned CVE-IDs, and 9 IRs (30.0\%) will potentially be assigned CVE-IDs from the feedback from project owners.}
\end{tcolorbox}
\end{center}
\section{Discussion}\label{sec:discussion}
\subsection{Analysis of Hyper-Parameters}\label{sec:paramaeters}

{To analyze the effect of the historical IR's proportion and threshold $\theta_{sim}$ for TF-IDF similarity, we conduct the analysis of hyper-parameters on {\tool}. 
For the proportion of historical IRs, we choose the proportion from 30\% to 90\%, with intervals of 10\%.
For the $\theta_{sim}$, we choose the value from 0.50 to 0.90, with the intervals 0.05.}

\begin{wrapfigure}{r}{0.5\textwidth}
\centering
\vspace{-0.3cm}
\includegraphics[width=0.5\textwidth]{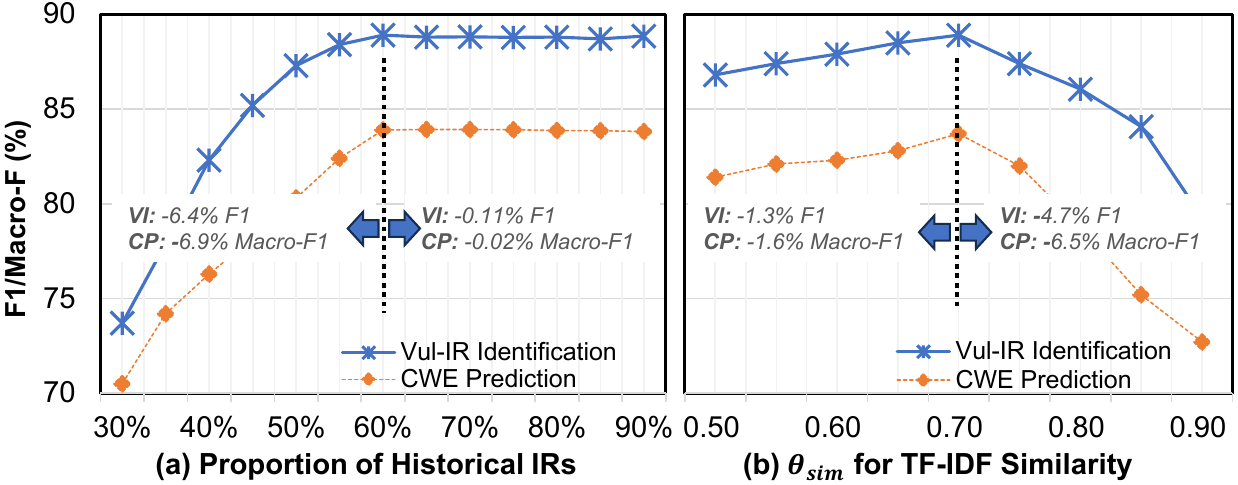}
\vspace{-0.6cm}
\caption{{Effects of two hyper-parameters in {\tool}.}}
\vspace{-0.3cm}
\label{fig:proportion_analysis}
\end{wrapfigure}

Fig. \ref{fig:proportion_analysis} shows the effect of hyper-parameters performances of {\tool} on average. We can see that, both parameters affect the performances of {\tool}.
For historical IR's proportion, from 30\% to 60\%, the performances of {\tool} have a large increase with +6.4\% F1 and +6.9\% Macro-F1; after 60\%, the fluctuations are small, with less than $\pm$0.1\% F1/Macro-F1.
For the $\theta_{sim}$, {\tool} achieves the optimal performances when $\theta_{sim}=0.7$, which are higher than other values with over +1\% F1/Macro-F1 performances on average.

In summary, the performance of {\tool} is affected by these two hyper-parameters. Choosing the proportion 60\% and $\theta_{sim}=0.7$ is {sufficient} to achieve optimal performance.

\subsection{Advantages of {\tool}}

The benefits of {\tool} come from three aspects, specifically, the ability to analyze rich-text information, the correction of factual errors, and the utilization of RAG to improve efficiency.
With these three advantages, {\tool} can not only accurately identify vulnerabilities in rich-text IRs but also outperform the baselines on plain-text IRs.

\begin{itemize}[leftmargin=*]
    \item \textbf{Advantage-1: The analysis of rich-text information.} Existing approaches, such as {\toolcompare}, cannot identify vulnerabilities in few-text IRs, and they need to track the source code in the repositories for further analysis. In comparison, {\tool} indicates that there will be rich-text elements, such as page screenshots and code snippets that can be utilized to complement the missing information in the few-text IRs and help identify the vulnerability triggering paths.   
    \item \textbf{Advantage-2: The correction of factual errors.} Since the IR may incorrectly describe how the vulnerability is triggered, and LLM itself may be trained on outdated and noisy datasets, the output may have some factual errors. With the help of factual error correction, the reasoning graphs of rich/plain-text IRs will be in line with the real-world situation, thus improving the performance of vulnerability identification.
    \item \textbf{Advantage-3: The utilization of RAG's thought.} Based on the reasoning graph retrieval, the LLMs will understand how the previous developers and users describe the triggering logic of similar vulnerabilities in their IRs, and LLMs will understand semantic information in the relevant texts and rich-text elements. Therefore, the thought of RAG not only improves the efficiency of {\tool}, but also enables LLM to understand the details of vulnerabilities and enhance the quality of its generated text in the security field.
\end{itemize}

\subsection{Incorrect Prediction of {\tool}}
Although {\tool} can accurately identify the vulnerabilities from rich-text IRs, there are still 15.0\% of IRs that are incorrectly predicted in our experiment. 
We manually inspect these incorrect predictions and find the following two reasons:

\begin{itemize}[leftmargin=*]
    \item \textbf{Case-1: Out-of-Scope Rich-Text Information (4.5\%).} The type of rich-text information is not included in our approach. For example, the IR \textit{TeamPass/issues/2688}~\cite{video_ir} utilizes the video stream to present the steps of XSS injection in the TeamPass project, but {\tool} cannot analyze the video stream, which leads to the incorrect prediction of CWE-ID.
    \item \textbf{Case-2: Missing the Original Pages of IR (10.5\%).} Some original pages of vulnerability-related IRs are deleted or hidden by authors, which leads to incorrect predictions. For example, the page of {coreruleset}'s issue \textit{security-tracker-private/issues/7}~\cite{hidden_ir} has been set as private, so we cannot retrieve the rich-text information.
\end{itemize}

\subsection{Threats to Validity}\label{sec:threat}

\noindent\textbf{Internal Threats.}
The first internal threat comes from dataset preparation. We only refer to the CVE to analyze whether this vulnerability is disclosed, but some vulnerabilities may only be disclosed in other security databases, such as CAPEC~\cite{CAPEC1}.
To alleviate this threat, we manually inspect 100 security GitHub IRs that are disclosed on CAPEC and find that 2/100 are not disclosed by CVE and are incorporated by our dataset simultaneously.
{Another threat comes from the TF-IDF that may introduce noisy data in correcting LLM factual errors.
The LLM itself can remove some noisy data with its reasoning ability, which alleviates this threat.
We have randomly sampled 100 vulnerability-related IRs from our dataset with target IRs, and manually inspected whether {\tool} can retrieve the matched golden knowledge in the knowledge $Know_t$. We find that 21/100 IRs may have LLM's factual errors, and all the IRs (21/21) are corrected with $Know_t$. Therefore, the impact of internal threats is small.}

\noindent\textbf{External Threat.}
The external threat comes from the potentially incorrect labels in the dataset. For example, CVE-2018-17566 indicates that \textit{top-think/think/issues/858} is the vulnerability-related IR with the ``SQL injection'' vulnerability, but in the following comments, other developers indicate that this vulnerability may not exist in practice. 
The impact of this threat is small due to the small proportion of samples (60/4,003), and we plan to report them to the CVE for further validation. 

\noindent\textbf{Constructive Threat.}
The constructive threat mainly comes from the metrics. We choose precision, recall, and F1-Score to evaluate the vulnerability-related IR identification, and choose Macro-P, Macro-R, and Macro-F1 to evaluate the performances on CWE-ID prediction.
We manually restore the rich-text information by retrieving the original IR from GitHub while calculating these metrics.
{This threat is mitigated by the fact that all the GitHub IRs are reviewed and discussed by our team members in the dataset preparation.}
\section{Related Works}\label{sec:rw}

% In this section, we will introduce the related works on automatic detection and identification of vulnerability, as well as the application of LLM-Agent in recent works.

\noindent{\textbf{Automatic Detection and Identification of Vulnerability.}
The automatic detection and identification of vulnerabilities have been investigated by researchers.
Automatic vulnerability detection aims to determine whether there are malicious codes that contain the vulnerabilities during the project development~\cite{DBLP:journals/nca/ZhuLSZ23,malhotra2015systematic,ghaffarian2017software,ji2018coming,jie2016survey,DBLP:journals/pieee/LinWHZX20,DBLP:journals/tse/ChakrabortyKDR22}.  
The researchers first proposed the statistic, dynamic, and hybrid techniques to detect the vulnerabilities with rules~\cite{DBLP:conf/sosp/EnglerCC01,DBLP:conf/snpd/LiangWWX16,DBLP:conf/sp/JangAB12}.
To improve detection accuracy and reduce the cost of manually designing rules, researchers have introduced machine learning (ML) approaches to detect the vulnerabilities, which combine the features extracted from codes and feed them into the basic ML models to predict the vulnerability types~\cite{DBLP:conf/ccs/YamaguchiWGR13,DBLP:journals/tse/ScandariatoWHJ14,DBLP:journals/tse/ShinMWO11}.
With the development of DL and LLM, researchers have introduced these novel models to automatically build code features and improve the efficiency of vulnerability detection tools~\cite{DBLP:conf/ndss/LiZXO0WDZ18,DBLP:conf/ndss/LiZXO0WDZ18,10.1145/3133956.3138840,DBLP:journals/corr/abs-2309-15324}.
Automatic vulnerability identification helps security practitioners identify whether artifacts (e.g., GitHub IRs, bug reports, etc.) submitted by developers actually contain vulnerabilities. 
Some researchers utilized text-mining methods to explore the security bug reports to identify the vulnerabilities~\cite{DBLP:conf/msr/GegickRX10,DBLP:conf/iecon/WijayasekaraMM14,DBLP:conf/hsi/WijayasekaraMWM12,DBLP:conf/kbse/WuZDYYCLJ20}, while other works focused on reducing the negative impact of vulnerability identification from class unbalancing~\cite{DBLP:journals/sqj/OyetoyanM21,DBLP:journals/tse/PetersTYN19,DBLP:journals/ese/ShuXCWM21,DBLP:conf/sigsoft/PanZC0BHLH22}.
% The GitHub IRs have been investigated by researchers to detect vulnerabilities. 
Our work focuses on identifying vulnerabilities in IRs.
Different from the previous works, we utilize rich-text information and {reasoning graph retrieval} to improve the accuracy of identifying vulnerability-related IRs and predict the CWE-IDs.}

% \ziyou{Some questions: what should we write in this section?}
\noindent\textbf{Application of LLM Reasoning and Agent.}
Recently, researchers have investigated the application of LLM's reasoning and agents on different research tasks.
Some of these works were based on textual description, and they utilized LLM agents to analyze the text-based reasoning logic.
Wang et al.~\cite{DBLP:journals/corr/abs-2310-01444} analyzed the logic of dialogues, and applied LLM agents in the communications.
Nan et al.~\cite{DBLP:journals/corr/abs-2311-09721} integrated the reasoning and action in LLM agents to describe the database question answering.
Other works combined visual information and proposed the embodied LLM agents to analyze more complex visual-related tasks.
Zheng et al.~\cite{DBLP:journals/corr/abs-2310-13255} proposed the Steve-Eye, an embodied agent that analyzes visual perception in open worlds.
Cherakara et al.~\cite{DBLP:conf/sigdial/CherakaraVSNKKFNMDRL23} proposed the FurChat, which is a conversational-based embodied agent that combines the open and close domain dialogues with facial expressions.
Schumann et al.~\cite{DBLP:journals/corr/abs-2307-06082} proposed the VELMA, which is an embodied agent that analyzes the language navigation and vision in street view.
Ma et al.~\cite{DBLP:journals/corr/abs-2309-08172} proposed the LASER, which utilizes the LLM agent to analyze the website navigation.
Different from these previous works, {\tool} incorporates the reasoning graph retrieval from the RAG's thought to improve the accuracy of LLM agents in analyzing rich-text information. 
\section{Conclusion and Future Work}\label{sec:conclusion}
In this paper, we propose the {\tool} to identify vulnerability-related IRs with rich-text information.
{\tool} first utilizes the reasoning ability of LLMs to prepare the {Vulnerability Reasoning Database} from historical IRs. Then, {\tool} retrieves the most relevant reasoning graphs from the prepared reasoning database to guide LLM in identifying vulnerabilities from target IRs. 
{Experiments conducted on 973,572 IRs show that {\tool} achieves the highest performance when the dataset is imbalanced outperforming the best baseline with +11.0\% F1 and +20.2\% AUPRC, with 2x lower time cost than the baseline reasoning approaches. 
{\tool} also has the highest performance on CWE-ID prediction, outperforming the best baseline with +10.5\% Macro-F1.}
Furthermore, {\tool} has been applied to identify the 30 emerging vulnerability-related IRs across 10 projects, and 11 of them are finally assigned CVE-IDs.

% To evaluate the performance of {\tool}, we conduct experiments on 973,572 IRs,
% % \lin{Only positive data, no negative IRs that are normal?}
% and the results show that {\tool} achieves the highest performance in identifying the vulnerability-related IRs and predicting CWE-IDs when the dataset is imbalanced, outperforming the best baseline with +11.0\% F1, +20.2\% AUPRC, and +10.5\% Macro-F1.
% Furthermore, {\tool} has been applied to identify 30 emerging vulnerabilities across 10 representative OSS projects outside the original dataset.
% Among them, 11 IRs are assigned CVE-IDs, which illustrates {\tool}'s practicability.

In the future, we plan to improve our approach with more vulnerability-related IRs from other collaborative platforms, such as GitLab, Gitee, etc. 
% We also plan to analyze the other rich-text information and the version-based changes in the projects.
{We also plan to introduce the insecure code commits and their patches that relate to the vulnerabilities to continuously improve the capability of {\tool} in more open-source projects.}
% After the GPT-4V is available without limitation, we plan to expand the new LLM and 
% GitLab, GPT-4V, Other multimodal information

% \section{Data Availability}
% To enhance {\tool}'s reproducibility, the source code and dataset of {\tool} are available in the replication package~\cite{model_data}. 
% % The source code and dataset of {\tool} are available on the replication package~\cite{model_data}.The source code and dataset of {\tool} are available on the replication package~\cite{model_data}.
\section*{Acknowledgments}

We sincerely appreciate the reviewers for their insightful suggestions. This work was supported by the National Key Research and Development Program of China (No. 2024YFF0618800), National Natural Science Foundation of China Grant No.62402484, No.62232016, Youth Innovation Promotion Association Chinese Academy of Sciences, Basic Research Program of ISCAS Grant No. ISCAS-JCZD-202405, and partially supported by the University of Queensland NSRSG Grant NS-2201 and Oracle Labs.
%%
%% The next two lines define the bibliography style to be used, and
%% the bibliography file.
\bibliographystyle{ACM-Reference-Format}
\bibliography{ref}

%%% -*-BibTeX-*-
%%% Do NOT edit. File created by BibTeX with style
%%% ACM-Reference-Format-Journals [18-Jan-2012].

\begin{thebibliography}{108}

%%% ====================================================================
%%% NOTE TO THE USER: you can override these defaults by providing
%%% customized versions of any of these macros before the \bibliography
%%% command.  Each of them MUST provide its own final punctuation,
%%% except for \shownote{}, \showDOI{}, and \showURL{}.  The latter two
%%% do not use final punctuation, in order to avoid confusing it with
%%% the Web address.
%%%
%%% To suppress output of a particular field, define its macro to expand
%%% to an empty string, or better, \unskip, like this:
%%%
%%% \newcommand{\showDOI}[1]{\unskip}   % LaTeX syntax
%%%
%%% \def \showDOI #1{\unskip}           % plain TeX syntax
%%%
%%% ====================================================================

\ifx \showCODEN    \undefined \def \showCODEN     #1{\unskip}     \fi
\ifx \showDOI      \undefined \def \showDOI       #1{#1}\fi
\ifx \showISBNx    \undefined \def \showISBNx     #1{\unskip}     \fi
\ifx \showISBNxiii \undefined \def \showISBNxiii  #1{\unskip}     \fi
\ifx \showISSN     \undefined \def \showISSN      #1{\unskip}     \fi
\ifx \showLCCN     \undefined \def \showLCCN      #1{\unskip}     \fi
\ifx \shownote     \undefined \def \shownote      #1{#1}          \fi
\ifx \showarticletitle \undefined \def \showarticletitle #1{#1}   \fi
\ifx \showURL      \undefined \def \showURL       {\relax}        \fi
% The following commands are used for tagged output and should be
% invisible to TeX
\providecommand\bibfield[2]{#2}
\providecommand\bibinfo[2]{#2}
\providecommand\natexlab[1]{#1}
\providecommand\showeprint[2][]{arXiv:#2}

\bibitem[iso(2018)]%
        {iso_disclosure}
 \bibinfo{year}{2018}\natexlab{}.
\newblock \bibinfo{title}{{ISO/IEC 29147:2018: Security techniques - Vulnerability disclosure.}}
\newblock \bibinfo{howpublished}{\url{https://www.iso.org/standard/72311.html.}}.
\newblock


\bibitem[ir_(2019)]%
        {ir_multimodal_compare}
 \bibinfo{year}{2019}\natexlab{}.
\newblock \bibinfo{title}{{XSS and CSRF in Blocks}}.
\newblock \bibinfo{howpublished}{\url{https://github.com/daylightstudio/fuel-cms/issues/536}}.
\newblock


\bibitem[ir_(2020)]%
        {ir_multimodal_example}
 \bibinfo{year}{2020}\natexlab{}.
\newblock \bibinfo{title}{{XSS in cmd.php for 1.2.5}}.
\newblock \bibinfo{howpublished}{\url{https://github.com/leenooks/phpldapadmin/issues/130}}.
\newblock


\bibitem[bug(2023)]%
        {bugzilla}
 \bibinfo{year}{2023}\natexlab{}.
\newblock \bibinfo{title}{{Bugzilla}}.
\newblock \bibinfo{howpublished}{\url{https://www.bugzilla.org/}}.
\newblock


\bibitem[CVE(2023)]%
        {CVE}
 \bibinfo{year}{2023}\natexlab{}.
\newblock \bibinfo{title}{{Common vulnerabilities and exposures}}.
\newblock \bibinfo{howpublished}{\url{https://cve.mitre.org/}}.
\newblock


\bibitem[CWE(2023)]%
        {CWE}
 \bibinfo{year}{2023}\natexlab{}.
\newblock \bibinfo{title}{{Common weakness enumeration}}.
\newblock \bibinfo{howpublished}{\url{https://cwe.mitre.org/}}.
\newblock


\bibitem[hid(2023)]%
        {hidden_ir}
 \bibinfo{year}{2023}\natexlab{}.
\newblock \bibinfo{title}{{Fix C9K-230327}}.
\newblock \bibinfo{howpublished}{\url{https://github.com/coreruleset/coreruleset/issues/3191}}.
\newblock


\bibitem[GHA(2023)]%
        {GHArchive}
 \bibinfo{year}{2023}\natexlab{}.
\newblock \bibinfo{title}{{GHArchive}}.
\newblock \bibinfo{howpublished}{\url{https://www.gharchive.org/}}.
\newblock


\bibitem[{\tool}(2024)]%
        {model_data}
\bibfield{author}{\bibinfo{person}{{\tool}}.} \bibinfo{year}{2024}\natexlab{}.
\newblock \bibinfo{title}{{Anonymized Repository}}.
\newblock \bibinfo{howpublished}{\url{https://anonymous.4open.science/r/VulRTex-0F94}}.
\newblock


\bibitem[rea(2024)]%
        {react}
 \bibinfo{year}{2024}\natexlab{}.
\newblock \bibinfo{title}{{Facebook/React}}.
\newblock \bibinfo{howpublished}{\url{https://github.com/facebook/react}}.
\newblock


\bibitem[deb(2024)]%
        {debian_dataset}
 \bibinfo{year}{2024}\natexlab{}.
\newblock \bibinfo{title}{{Gitstar Ranking}}.
\newblock \bibinfo{howpublished}{\url{https://www.akto.io/cves/vendor/debian}}.
\newblock


\bibitem[git(2024)]%
        {gitstar_ranking}
 \bibinfo{year}{2024}\natexlab{}.
\newblock \bibinfo{title}{{Gitstar Ranking}}.
\newblock \bibinfo{howpublished}{\url{https://gitstar-ranking.com/repositories}}.
\newblock


\bibitem[hug(2024)]%
        {hugo}
 \bibinfo{year}{2024}\natexlab{}.
\newblock \bibinfo{title}{{Gohugoio/Hugo}}.
\newblock \bibinfo{howpublished}{\url{https://github.com/gohugoio/hugo}}.
\newblock


\bibitem[nod(2024)]%
        {nodeserver}
 \bibinfo{year}{2024}\natexlab{}.
\newblock \bibinfo{title}{{Honojs/Node-server}}.
\newblock \bibinfo{howpublished}{\url{https://github.com/honojs/node-server}}.
\newblock


\bibitem[hyp(2024)]%
        {hyprland}
 \bibinfo{year}{2024}\natexlab{}.
\newblock \bibinfo{title}{{Hyprwm/Hyprland}}.
\newblock \bibinfo{howpublished}{\url{https://github.com/hyprwm/Hyprland}}.
\newblock


\bibitem[jer(2024)]%
        {jerryscript}
 \bibinfo{year}{2024}\natexlab{}.
\newblock \bibinfo{title}{{Jerryscript-project/Jerryscript}}.
\newblock \bibinfo{howpublished}{\url{https://github.com/jerryscript-project/jerryscript}}.
\newblock


\bibitem[kub(2024)]%
        {kubernetes}
 \bibinfo{year}{2024}\natexlab{}.
\newblock \bibinfo{title}{{Kubernetes/Kubernetes}}.
\newblock \bibinfo{howpublished}{\url{https://github.com/kubernetes/kubernetes}}.
\newblock


\bibitem[pyt(2024)]%
        {pythonjose}
 \bibinfo{year}{2024}\natexlab{}.
\newblock \bibinfo{title}{{Mpdavis/Python-jose}}.
\newblock \bibinfo{howpublished}{\url{https://github.com/mpdavis/python-jose}}.
\newblock


\bibitem[rim(2024)]%
        {rimendo}
 \bibinfo{year}{2024}\natexlab{}.
\newblock \bibinfo{title}{{Onosproject/Rimedo-ts}}.
\newblock \bibinfo{howpublished}{\url{https://github.com/onosproject/rimedo-ts}}.
\newblock


\bibitem[car(2024)]%
        {carla}
 \bibinfo{year}{2024}\natexlab{}.
\newblock \bibinfo{title}{{Onosproject/Rimedo-ts}}.
\newblock \bibinfo{howpublished}{\url{https://github.com/carla-simulator/carla}}.
\newblock


\bibitem[xxl(2024)]%
        {xxljob}
 \bibinfo{year}{2024}\natexlab{}.
\newblock \bibinfo{title}{{Xuxueli/Xxl-job}}.
\newblock \bibinfo{howpublished}{\url{https://github.com/xuxueli/xxl-job}}.
\newblock


\bibitem[Afzal et~al\mbox{.}(2018)]%
        {DBLP:conf/ntms/AfzalGLB18}
\bibfield{author}{\bibinfo{person}{Zeeshan Afzal}, \bibinfo{person}{Johan Garcia}, \bibinfo{person}{Stefan Lindskog}, {and} \bibinfo{person}{Anna Brunstr{\"{o}}m}.} \bibinfo{year}{2018}\natexlab{}.
\newblock \showarticletitle{Slice Distance: An Insert-Only Levenshtein Distance with a Focus on Security Applications}. In \bibinfo{booktitle}{\emph{9th {IFIP} International Conference on New Technologies, Mobility and Security, {NTMS} 2018}}. \bibinfo{publisher}{{IEEE}}, \bibinfo{pages}{1--5}.
\newblock


\bibitem[{Atlassian}(2023)]%
        {jira}
\bibfield{author}{\bibinfo{person}{{Atlassian}}.} \bibinfo{year}{2023}\natexlab{}.
\newblock \bibinfo{title}{{Jira, Issue \& Project Tracking Software}}.
\newblock \bibinfo{howpublished}{\url{https://www.atlassian.com/software/jira}}.
\newblock


\bibitem[Bilge and Dumitras(2012)]%
        {DBLP:conf/ccs/BilgeD12}
\bibfield{author}{\bibinfo{person}{Leyla Bilge} {and} \bibinfo{person}{Tudor Dumitras}.} \bibinfo{year}{2012}\natexlab{}.
\newblock \showarticletitle{Before we knew it: an empirical study of zero-day attacks in the real world}. In \bibinfo{booktitle}{\emph{the {ACM} Conference on Computer and Communications Security, CCS'12}}. \bibinfo{publisher}{{ACM}}, \bibinfo{pages}{833--844}.
\newblock


\bibitem[Chakraborty et~al\mbox{.}(2022)]%
        {DBLP:journals/tse/ChakrabortyKDR22}
\bibfield{author}{\bibinfo{person}{Saikat Chakraborty}, \bibinfo{person}{Rahul Krishna}, \bibinfo{person}{Yangruibo Ding}, {and} \bibinfo{person}{Baishakhi Ray}.} \bibinfo{year}{2022}\natexlab{}.
\newblock \showarticletitle{Deep Learning Based Vulnerability Detection: Are We There Yet?}
\newblock \bibinfo{journal}{\emph{{IEEE} Trans. Software Eng.}} \bibinfo{volume}{48}, \bibinfo{number}{9} (\bibinfo{year}{2022}), \bibinfo{pages}{3280--3296}.
\newblock


\bibitem[Chen et~al\mbox{.}(2023)]%
        {DBLP:conf/raid/0001DACW23}
\bibfield{author}{\bibinfo{person}{Yizheng Chen}, \bibinfo{person}{Zhoujie Ding}, \bibinfo{person}{Lamya Alowain}, \bibinfo{person}{Xinyun Chen}, {and} \bibinfo{person}{David~A. Wagner}.} \bibinfo{year}{2023}\natexlab{}.
\newblock \showarticletitle{DiverseVul: {A} New Vulnerable Source Code Dataset for Deep Learning Based Vulnerability Detection}. In \bibinfo{booktitle}{\emph{Proceedings of the 26th International Symposium on Research in Attacks, Intrusions and Defenses, {RAID} 2023}}. \bibinfo{publisher}{{ACM}}, \bibinfo{pages}{654--668}.
\newblock


\bibitem[Cherakara et~al\mbox{.}(2023)]%
        {DBLP:conf/sigdial/CherakaraVSNKKFNMDRL23}
\bibfield{author}{\bibinfo{person}{Neeraj Cherakara}, \bibinfo{person}{Finny Varghese}, \bibinfo{person}{Sheena Shabana}, \bibinfo{person}{Nivan Nelson}, \bibinfo{person}{Abhiram Karukayil}, \bibinfo{person}{Rohith Kulothungan}, \bibinfo{person}{Mohammed~Afil Farhan}, \bibinfo{person}{Birthe Nesset}, \bibinfo{person}{Meriam Moujahid}, \bibinfo{person}{Tanvi Dinkar}, \bibinfo{person}{Verena Rieser}, {and} \bibinfo{person}{Oliver Lemon}.} \bibinfo{year}{2023}\natexlab{}.
\newblock \showarticletitle{FurChat: An Embodied Conversational Agent using LLMs, Combining Open and Closed-Domain Dialogue with Facial Expressions}. In \bibinfo{booktitle}{\emph{Proceedings of the 24th Meeting of the Special Interest Group on Discourse and Dialogue, {SIGDIAL} 2023}}. \bibinfo{publisher}{Association for Computational Linguistics}, \bibinfo{pages}{588--592}.
\newblock


\bibitem[Corporation(2011)]%
        {CAPEC1}
\bibfield{author}{\bibinfo{person}{T.~M. Corporation}.} \bibinfo{year}{2011}\natexlab{}.
\newblock \bibinfo{title}{Common Attack Pattern Enumeration and Classification (CAPEC)}.
\newblock \bibinfo{howpublished}{\url{http://capec. mitre.org/}}.
\newblock


\bibitem[Davis and Goadrich(2006)]%
        {DBLP:conf/icml/DavisG06}
\bibfield{author}{\bibinfo{person}{Jesse Davis} {and} \bibinfo{person}{Mark Goadrich}.} \bibinfo{year}{2006}\natexlab{}.
\newblock \showarticletitle{The relationship between Precision-Recall and {ROC} curves}. In \bibinfo{booktitle}{\emph{Machine Learning, Proceedings of the Twenty-Third International Conference {(ICML} 2006)}} \emph{(\bibinfo{series}{{ACM} International Conference Proceeding Series}, Vol.~\bibinfo{volume}{148})}. \bibinfo{publisher}{{ACM}}, \bibinfo{pages}{233--240}.
\newblock


\bibitem[Engler et~al\mbox{.}(2001)]%
        {DBLP:conf/sosp/EnglerCC01}
\bibfield{author}{\bibinfo{person}{Dawson~R. Engler}, \bibinfo{person}{David~Yu Chen}, {and} \bibinfo{person}{Andy Chou}.} \bibinfo{year}{2001}\natexlab{}.
\newblock \showarticletitle{Bugs as Deviant Behavior: {A} General Approach to Inferring Errors in Systems Code}. In \bibinfo{booktitle}{\emph{Proceedings of the 18th {ACM} Symposium on Operating System Principles, {SOSP} 2001}}. \bibinfo{publisher}{{ACM}}, \bibinfo{pages}{57--72}.
\newblock


\bibitem[Fan et~al\mbox{.}(2020)]%
        {DBLP:conf/msr/FanL0N20}
\bibfield{author}{\bibinfo{person}{Jiahao Fan}, \bibinfo{person}{Yi Li}, \bibinfo{person}{Shaohua Wang}, {and} \bibinfo{person}{Tien~N. Nguyen}.} \bibinfo{year}{2020}\natexlab{}.
\newblock \showarticletitle{A {C/C++} Code Vulnerability Dataset with Code Changes and {CVE} Summaries}. In \bibinfo{booktitle}{\emph{{MSR} '20}}. \bibinfo{publisher}{{ACM}}, \bibinfo{pages}{508--512}.
\newblock


\bibitem[Gegick et~al\mbox{.}(2010)]%
        {DBLP:conf/msr/GegickRX10}
\bibfield{author}{\bibinfo{person}{Michael Gegick}, \bibinfo{person}{Pete Rotella}, {and} \bibinfo{person}{Tao Xie}.} \bibinfo{year}{2010}\natexlab{}.
\newblock \showarticletitle{Identifying security bug reports via text mining: An industrial case study}. In \bibinfo{booktitle}{\emph{Proceedings of the 7th International Working Conference on Mining Software Repositories, {MSR} 2010 (Co-located with ICSE)}}. \bibinfo{publisher}{{IEEE} Computer Society}, \bibinfo{pages}{11--20}.
\newblock


\bibitem[Ghaffarian and Shahriari(2017)]%
        {ghaffarian2017software}
\bibfield{author}{\bibinfo{person}{Seyed~Mohammad Ghaffarian} {and} \bibinfo{person}{Hamid~Reza Shahriari}.} \bibinfo{year}{2017}\natexlab{}.
\newblock \showarticletitle{Software vulnerability analysis and discovery using machine-learning and data-mining techniques: A survey}.
\newblock \bibinfo{journal}{\emph{Comput. Surveys}} \bibinfo{volume}{50}, \bibinfo{number}{4} (\bibinfo{year}{2017}), \bibinfo{pages}{1--36}.
\newblock


\bibitem[Guo et~al\mbox{.}(2024)]%
        {DBLP:journals/corr/abs-2402-17453}
\bibfield{author}{\bibinfo{person}{Siyuan Guo}, \bibinfo{person}{Cheng Deng}, \bibinfo{person}{Ying Wen}, \bibinfo{person}{Hechang Chen}, \bibinfo{person}{Yi Chang}, {and} \bibinfo{person}{Jun Wang}.} \bibinfo{year}{2024}\natexlab{}.
\newblock \showarticletitle{DS-Agent: Automated Data Science by Empowering Large Language Models with Case-Based Reasoning}.
\newblock \bibinfo{journal}{\emph{CoRR}}  \bibinfo{volume}{abs/2402.17453} (\bibinfo{year}{2024}).
\newblock
\showeprint[arXiv]{2402.17453}


\bibitem[Guu et~al\mbox{.}(2020)]%
        {DBLP:journals/corr/abs-2002-08909}
\bibfield{author}{\bibinfo{person}{Kelvin Guu}, \bibinfo{person}{Kenton Lee}, \bibinfo{person}{Zora Tung}, \bibinfo{person}{Panupong Pasupat}, {and} \bibinfo{person}{Ming{-}Wei Chang}.} \bibinfo{year}{2020}\natexlab{}.
\newblock \showarticletitle{{REALM:} Retrieval-Augmented Language Model Pre-Training}.
\newblock \bibinfo{journal}{\emph{CoRR}}  \bibinfo{volume}{abs/2002.08909} (\bibinfo{year}{2020}).
\newblock
\showeprint[arXiv]{2002.08909}


\bibitem[Harzevili et~al\mbox{.}(2025)]%
        {DBLP:journals/csur/HarzeviliBWWJN25}
\bibfield{author}{\bibinfo{person}{Nima~Shiri Harzevili}, \bibinfo{person}{Alvine~Boaye Belle}, \bibinfo{person}{Junjie Wang}, \bibinfo{person}{Song Wang}, \bibinfo{person}{Zhen Ming~(Jack) Jiang}, {and} \bibinfo{person}{Nachiappan Nagappan}.} \bibinfo{year}{2025}\natexlab{}.
\newblock \showarticletitle{A Systematic Literature Review on Automated Software Vulnerability Detection Using Machine Learning}.
\newblock \bibinfo{journal}{\emph{{ACM} Comput. Surv.}} \bibinfo{volume}{57}, \bibinfo{number}{3} (\bibinfo{year}{2025}), \bibinfo{pages}{55:1--55:36}.
\newblock


\bibitem[Householder et~al\mbox{.}(2017)]%
        {householder2017cert}
\bibfield{author}{\bibinfo{person}{Allen~D Householder}, \bibinfo{person}{Garret Wassermann}, \bibinfo{person}{Art Manion}, {and} \bibinfo{person}{Chris King}.} \bibinfo{year}{2017}\natexlab{}.
\newblock \showarticletitle{The cert guide to coordinated vulnerability disclosure}.
\newblock \bibinfo{journal}{\emph{Software Engineering Institute, Pittsburgh, PA}} (\bibinfo{year}{2017}).
\newblock


\bibitem[Huang et~al\mbox{.}(2023)]%
        {DBLP:journals/corr/abs-2311-05232}
\bibfield{author}{\bibinfo{person}{Lei Huang}, \bibinfo{person}{Weijiang Yu}, \bibinfo{person}{Weitao Ma}, \bibinfo{person}{Weihong Zhong}, \bibinfo{person}{Zhangyin Feng}, \bibinfo{person}{Haotian Wang}, \bibinfo{person}{Qianglong Chen}, \bibinfo{person}{Weihua Peng}, \bibinfo{person}{Xiaocheng Feng}, \bibinfo{person}{Bing Qin}, {and} \bibinfo{person}{Ting Liu}.} \bibinfo{year}{2023}\natexlab{}.
\newblock \showarticletitle{A Survey on Hallucination in Large Language Models: Principles, Taxonomy, Challenges, and Open Questions}.
\newblock \bibinfo{journal}{\emph{CoRR}}  \bibinfo{volume}{abs/2311.05232} (\bibinfo{year}{2023}).
\newblock


\bibitem[{Hugging Face}(2023)]%
        {llama_version}
\bibfield{author}{\bibinfo{person}{{Hugging Face}}.} \bibinfo{year}{2023}\natexlab{}.
\newblock \bibinfo{title}{{meta-llama/Llama-2-13b-chat-hfd}}.
\newblock \bibinfo{howpublished}{\url{https://huggingface.co/meta-llama/Llama-2-13b-chat-hf}}.
\newblock


\bibitem[Jang et~al\mbox{.}(2012)]%
        {DBLP:conf/sp/JangAB12}
\bibfield{author}{\bibinfo{person}{Jiyong Jang}, \bibinfo{person}{Abeer Agrawal}, {and} \bibinfo{person}{David Brumley}.} \bibinfo{year}{2012}\natexlab{}.
\newblock \showarticletitle{ReDeBug: Finding Unpatched Code Clones in Entire {OS} Distributions}. In \bibinfo{booktitle}{\emph{{IEEE} Symposium on Security and Privacy, {SP} 2012, 21-23 May 2012, San Francisco, California, {USA}}}. \bibinfo{publisher}{{IEEE} Computer Society}, \bibinfo{pages}{48--62}.
\newblock


\bibitem[Ji et~al\mbox{.}(2018)]%
        {ji2018coming}
\bibfield{author}{\bibinfo{person}{Tiantian Ji}, \bibinfo{person}{Yue Wu}, \bibinfo{person}{Chang Wang}, \bibinfo{person}{Xi Zhang}, {and} \bibinfo{person}{Zhongru Wang}.} \bibinfo{year}{2018}\natexlab{}.
\newblock \showarticletitle{The coming era of alphahacking?: A survey of automatic software vulnerability detection, exploitation and patching techniques}. In \bibinfo{booktitle}{\emph{2018 IEEE third international conference on data science in cyberspace (DSC)}}. IEEE, \bibinfo{pages}{53--60}.
\newblock


\bibitem[Jiang et~al\mbox{.}(2023a)]%
        {DBLP:conf/kbse/JiangSYW23}
\bibfield{author}{\bibinfo{person}{Ziyou Jiang}, \bibinfo{person}{Lin Shi}, \bibinfo{person}{Guowei Yang}, {and} \bibinfo{person}{Qing Wang}.} \bibinfo{year}{2023}\natexlab{a}.
\newblock \showarticletitle{SCPatcher: Mining Crowd Security Discussions to Enrich Secure Coding Practices}. In \bibinfo{booktitle}{\emph{38th {IEEE/ACM} International Conference on Automated Software Engineering, {ASE} 2023}}. \bibinfo{publisher}{{IEEE}}, \bibinfo{pages}{358--370}.
\newblock


\bibitem[Jiang et~al\mbox{.}(2023b)]%
        {DBLP:conf/emnlp/JiangXGSLDYCN23}
\bibfield{author}{\bibinfo{person}{Zhengbao Jiang}, \bibinfo{person}{Frank~F. Xu}, \bibinfo{person}{Luyu Gao}, \bibinfo{person}{Zhiqing Sun}, \bibinfo{person}{Qian Liu}, \bibinfo{person}{Jane Dwivedi{-}Yu}, \bibinfo{person}{Yiming Yang}, \bibinfo{person}{Jamie Callan}, {and} \bibinfo{person}{Graham Neubig}.} \bibinfo{year}{2023}\natexlab{b}.
\newblock \showarticletitle{Active Retrieval Augmented Generation}. In \bibinfo{booktitle}{\emph{{EMNLP}'23}}. \bibinfo{publisher}{Association for Computational Linguistics}, \bibinfo{pages}{7969--7992}.
\newblock


\bibitem[Jie et~al\mbox{.}(2016)]%
        {jie2016survey}
\bibfield{author}{\bibinfo{person}{Gong Jie}, \bibinfo{person}{Kuang Xiao-Hui}, {and} \bibinfo{person}{Liu Qiang}.} \bibinfo{year}{2016}\natexlab{}.
\newblock \showarticletitle{Survey on software vulnerability analysis method based on machine learning}. In \bibinfo{booktitle}{\emph{2016 IEEE first international conference on data science in cyberspace (DSC)}}. IEEE, \bibinfo{pages}{642--647}.
\newblock


\bibitem[Kim et~al\mbox{.}(2024)]%
        {DBLP:journals/corr/abs-2410-20878}
\bibfield{author}{\bibinfo{person}{Dongkyu Kim}, \bibinfo{person}{Byoungwook Kim}, \bibinfo{person}{Donggeon Han}, {and} \bibinfo{person}{Matous Eibich}.} \bibinfo{year}{2024}\natexlab{}.
\newblock \showarticletitle{AutoRAG: Automated Framework for optimization of Retrieval Augmented Generation Pipeline}.
\newblock \bibinfo{journal}{\emph{CoRR}}  \bibinfo{volume}{abs/2410.20878} (\bibinfo{year}{2024}).
\newblock
\showeprint[arXiv]{2410.20878}


\bibitem[Lewis et~al\mbox{.}(2020)]%
        {DBLP:conf/nips/LewisPPPKGKLYR020}
\bibfield{author}{\bibinfo{person}{Patrick S.~H. Lewis}, \bibinfo{person}{Ethan Perez}, \bibinfo{person}{Aleksandra Piktus}, \bibinfo{person}{Fabio Petroni}, \bibinfo{person}{Vladimir Karpukhin}, \bibinfo{person}{Naman Goyal}, \bibinfo{person}{Heinrich K{\"{u}}ttler}, \bibinfo{person}{Mike Lewis}, \bibinfo{person}{Wen{-}tau Yih}, \bibinfo{person}{Tim Rockt{\"{a}}schel}, \bibinfo{person}{Sebastian Riedel}, {and} \bibinfo{person}{Douwe Kiela}.} \bibinfo{year}{2020}\natexlab{}.
\newblock \showarticletitle{Retrieval-Augmented Generation for Knowledge-Intensive {NLP} Tasks}. In \bibinfo{booktitle}{\emph{NeurIPS'20}}.
\newblock


\bibitem[Li et~al\mbox{.}(2019)]%
        {li2019comparative}
\bibfield{author}{\bibinfo{person}{Zhen Li}, \bibinfo{person}{Deqing Zou}, \bibinfo{person}{Jing Tang}, \bibinfo{person}{Zhihao Zhang}, \bibinfo{person}{Mingqian Sun}, {and} \bibinfo{person}{Hai Jin}.} \bibinfo{year}{2019}\natexlab{}.
\newblock \showarticletitle{A comparative study of deep learning-based vulnerability detection system}.
\newblock \bibinfo{journal}{\emph{IEEE Access}}  \bibinfo{volume}{7} (\bibinfo{year}{2019}), \bibinfo{pages}{103184--103197}.
\newblock


\bibitem[Li et~al\mbox{.}(2018)]%
        {DBLP:conf/ndss/LiZXO0WDZ18}
\bibfield{author}{\bibinfo{person}{Zhen Li}, \bibinfo{person}{Deqing Zou}, \bibinfo{person}{Shouhuai Xu}, \bibinfo{person}{Xinyu Ou}, \bibinfo{person}{Hai Jin}, \bibinfo{person}{Sujuan Wang}, \bibinfo{person}{Zhijun Deng}, {and} \bibinfo{person}{Yuyi Zhong}.} \bibinfo{year}{2018}\natexlab{}.
\newblock \showarticletitle{VulDeePecker: {A} Deep Learning-Based System for Vulnerability Detection}. In \bibinfo{booktitle}{\emph{25th Annual Network and Distributed System Security Symposium, {NDSS} 2018}}. \bibinfo{publisher}{The Internet Society}.
\newblock


\bibitem[Liang et~al\mbox{.}(2016)]%
        {DBLP:conf/snpd/LiangWWX16}
\bibfield{author}{\bibinfo{person}{Hongliang Liang}, \bibinfo{person}{Lei Wang}, \bibinfo{person}{Dongyang Wu}, {and} \bibinfo{person}{Jiuyun Xu}.} \bibinfo{year}{2016}\natexlab{}.
\newblock \showarticletitle{{MLSA:} {A} static bugs analysis tool based on {LLVM} {IR}}. In \bibinfo{booktitle}{\emph{17th {IEEE/ACIS} International Conference on Software Engineering, Artificial Intelligence, Networking and Parallel/Distributed Computing, {SNPD} 2016}}. \bibinfo{publisher}{{IEEE} Computer Society}, \bibinfo{pages}{407--412}.
\newblock


\bibitem[Lin et~al\mbox{.}(2020)]%
        {DBLP:journals/pieee/LinWHZX20}
\bibfield{author}{\bibinfo{person}{Guanjun Lin}, \bibinfo{person}{Sheng Wen}, \bibinfo{person}{Qing{-}Long Han}, \bibinfo{person}{Jun Zhang}, {and} \bibinfo{person}{Yang Xiang}.} \bibinfo{year}{2020}\natexlab{}.
\newblock \showarticletitle{Software Vulnerability Detection Using Deep Neural Networks: {A} Survey}.
\newblock \bibinfo{journal}{\emph{Proc. {IEEE}}} \bibinfo{volume}{108}, \bibinfo{number}{10} (\bibinfo{year}{2020}), \bibinfo{pages}{1825--1848}.
\newblock


\bibitem[Lin et~al\mbox{.}(2017)]%
        {10.1145/3133956.3138840}
\bibfield{author}{\bibinfo{person}{Guanjun Lin}, \bibinfo{person}{Jun Zhang}, \bibinfo{person}{Wei Luo}, \bibinfo{person}{Lei Pan}, {and} \bibinfo{person}{Yang Xiang}.} \bibinfo{year}{2017}\natexlab{}.
\newblock \showarticletitle{POSTER: Vulnerability Discovery with Function Representation Learning from Unlabeled Projects}. In \bibinfo{booktitle}{\emph{Proceedings of the 2017 ACM SIGSAC Conference on Computer and Communications Security}} (Dallas, Texas, USA) \emph{(\bibinfo{series}{CCS '17})}. \bibinfo{publisher}{Association for Computing Machinery}, \bibinfo{pages}{2539–2541}.
\newblock
\showISBNx{9781450349468}


\bibitem[Liu and Zhang(2018)]%
        {liu2018application}
\bibfield{author}{\bibinfo{person}{Tao Liu} {and} \bibinfo{person}{Longtao Zhang}.} \bibinfo{year}{2018}\natexlab{}.
\newblock \showarticletitle{Application of logistic regression in web vulnerability scanning}. In \bibinfo{booktitle}{\emph{2018 International Conference on Sensor Networks and Signal Processing (SNSP)}}. IEEE, \bibinfo{pages}{486--490}.
\newblock


\bibitem[Ma et~al\mbox{.}(2023)]%
        {DBLP:journals/corr/abs-2309-08172}
\bibfield{author}{\bibinfo{person}{Kaixin Ma}, \bibinfo{person}{Hongming Zhang}, \bibinfo{person}{Hongwei Wang}, \bibinfo{person}{Xiaoman Pan}, {and} \bibinfo{person}{Dong Yu}.} \bibinfo{year}{2023}\natexlab{}.
\newblock \showarticletitle{{LASER:} {LLM} Agent with State-Space Exploration for Web Navigation}.
\newblock \bibinfo{journal}{\emph{CoRR}}  \bibinfo{volume}{abs/2309.08172} (\bibinfo{year}{2023}).
\newblock


\bibitem[Malhotra(2015)]%
        {malhotra2015systematic}
\bibfield{author}{\bibinfo{person}{Ruchika Malhotra}.} \bibinfo{year}{2015}\natexlab{}.
\newblock \showarticletitle{A systematic review of machine learning techniques for software fault prediction}.
\newblock \bibinfo{journal}{\emph{Applied Soft Computing}}  \bibinfo{volume}{27} (\bibinfo{year}{2015}), \bibinfo{pages}{504--518}.
\newblock


\bibitem[Mathieu and Turovlin(2023)]%
        {DBLP:journals/imcs/MathieuT23}
\bibfield{author}{\bibinfo{person}{Richard~G. Mathieu} {and} \bibinfo{person}{Alan~E. Turovlin}.} \bibinfo{year}{2023}\natexlab{}.
\newblock \showarticletitle{Lost in the middle - a pragmatic approach for {ERP} managers to prioritize known vulnerabilities by applying classification and regression trees {(CART)}}.
\newblock \bibinfo{journal}{\emph{Inf. Comput. Secur.}} \bibinfo{volume}{31}, \bibinfo{number}{5} (\bibinfo{year}{2023}), \bibinfo{pages}{655--674}.
\newblock


\bibitem[Mikolov et~al\mbox{.}(2013)]%
        {DBLP:journals/corr/abs-1301-3781}
\bibfield{author}{\bibinfo{person}{Tom{\'{a}}s Mikolov}, \bibinfo{person}{Kai Chen}, \bibinfo{person}{Greg Corrado}, {and} \bibinfo{person}{Jeffrey Dean}.} \bibinfo{year}{2013}\natexlab{}.
\newblock \showarticletitle{Efficient Estimation of Word Representations in Vector Space}. In \bibinfo{booktitle}{\emph{1st International Conference on Learning Representations, {ICLR} 2013}}.
\newblock


\bibitem[MITRE(2014)]%
        {ATTCK}
\bibfield{author}{\bibinfo{person}{MITRE}.} \bibinfo{year}{2014}\natexlab{}.
\newblock \bibinfo{title}{{Adversarial Tactics, Techniques \& Common Knowledge (ATT\&CK)}}.
\newblock \bibinfo{howpublished}{\url{https://attack.mitre.org}}.
\newblock


\bibitem[MITRE(2023a)]%
        {cwe_352}
\bibfield{author}{\bibinfo{person}{MITRE}.} \bibinfo{year}{2023}\natexlab{a}.
\newblock \bibinfo{title}{{CWE-352: Cross-Site Request Forgery (CSRF) }}.
\newblock \bibinfo{howpublished}{\url{https://cwe.mitre.org/data/definitions/352.html}}.
\newblock


\bibitem[MITRE(2023b)]%
        {cwe_79}
\bibfield{author}{\bibinfo{person}{MITRE}.} \bibinfo{year}{2023}\natexlab{b}.
\newblock \bibinfo{title}{{CWE-79: Improper Neutralization of Input During Web Page Generation ('Cross-site Scripting') (4.13)}}.
\newblock \bibinfo{howpublished}{\url{https://cwe.mitre.org/data/definitions/79.html}}.
\newblock


\bibitem[MITRE(2023c)]%
        {cwe_94}
\bibfield{author}{\bibinfo{person}{MITRE}.} \bibinfo{year}{2023}\natexlab{c}.
\newblock \bibinfo{title}{{CWE-94: Improper Control of Generation of Code ('Code Injection')}}.
\newblock \bibinfo{howpublished}{\url{https://cwe.mitre.org/data/definitions/79.html}}.
\newblock


\bibitem[Nan et~al\mbox{.}(2023)]%
        {DBLP:journals/corr/abs-2311-09721}
\bibfield{author}{\bibinfo{person}{Linyong Nan}, \bibinfo{person}{Ellen Zhang}, \bibinfo{person}{Weijin Zou}, \bibinfo{person}{Yilun Zhao}, \bibinfo{person}{Wenfei Zhou}, {and} \bibinfo{person}{Arman Cohan}.} \bibinfo{year}{2023}\natexlab{}.
\newblock \showarticletitle{On Evaluating the Integration of Reasoning and Action in {LLM} Agents with Database Question Answering}.
\newblock \bibinfo{journal}{\emph{CoRR}}  \bibinfo{volume}{abs/2311.09721} (\bibinfo{year}{2023}).
\newblock


\bibitem[{nilsteampassnet}(2019)]%
        {video_ir}
\bibfield{author}{\bibinfo{person}{{nilsteampassnet}}.} \bibinfo{year}{2019}\natexlab{}.
\newblock \bibinfo{title}{{Stored XSS in log of Failed Logins}}.
\newblock \bibinfo{howpublished}{\url{https://github.com/nilsteampassnet/TeamPass/issues/2688}}.
\newblock


\bibitem[Omar and Shiaeles(2023)]%
        {DBLP:conf/csr2/OmarS23}
\bibfield{author}{\bibinfo{person}{Marwan Omar} {and} \bibinfo{person}{Stavros Shiaeles}.} \bibinfo{year}{2023}\natexlab{}.
\newblock \showarticletitle{VulDetect: {A} novel technique for detecting software vulnerabilities using Language Models}. In \bibinfo{booktitle}{\emph{{IEEE} International Conference on Cyber Security and Resilience, {CSR} 2023}}. \bibinfo{publisher}{{IEEE}}, \bibinfo{pages}{105--110}.
\newblock
\urldef\tempurl%
\url{https://doi.org/10.1109/CSR57506.2023.10224924}
\showDOI{\tempurl}


\bibitem[OpenAI(2023)]%
        {LLMBackground}
\bibfield{author}{\bibinfo{person}{OpenAI}.} \bibinfo{year}{2023}\natexlab{}.
\newblock \bibinfo{title}{{Chatgpt: A language model for conversational AI}}.
\newblock \bibinfo{howpublished}{\url{https://www.openai.com/research/chatgpt/}}.
\newblock


\bibitem[{OpenAI}(2023a)]%
        {gpt3_version}
\bibfield{author}{\bibinfo{person}{{OpenAI}}.} \bibinfo{year}{2023}\natexlab{a}.
\newblock \bibinfo{title}{{GPT-3-Models}}.
\newblock \bibinfo{howpublished}{\url{https://platform.openai.com/docs/models/gpt-3}}.
\newblock


\bibitem[{OpenAI}(2023b)]%
        {gpt35_version}
\bibfield{author}{\bibinfo{person}{{OpenAI}}.} \bibinfo{year}{2023}\natexlab{b}.
\newblock \bibinfo{title}{{GPT-3.5-Models}}.
\newblock \bibinfo{howpublished}{\url{https://platform.openai.com/docs/models/gpt-3-5}}.
\newblock


\bibitem[OWASP(2023)]%
        {OWASP}
\bibfield{author}{\bibinfo{person}{OWASP}.} \bibinfo{year}{2023}\natexlab{}.
\newblock \bibinfo{title}{{Open web application security project}}.
\newblock \bibinfo{howpublished}{\url{https://www.owasp.org/index.php/Main Page}}.
\newblock


\bibitem[Oyetoyan and Morrison(2021)]%
        {DBLP:journals/sqj/OyetoyanM21}
\bibfield{author}{\bibinfo{person}{Tosin~Daniel Oyetoyan} {and} \bibinfo{person}{Patrick Morrison}.} \bibinfo{year}{2021}\natexlab{}.
\newblock \showarticletitle{An improved text classification modelling approach to identify security messages in heterogeneous projects}.
\newblock \bibinfo{journal}{\emph{Softw. Qual. J.}} \bibinfo{volume}{29}, \bibinfo{number}{2} (\bibinfo{year}{2021}), \bibinfo{pages}{509--553}.
\newblock


\bibitem[Page et~al\mbox{.}(1999)]%
        {Page1999Page}
\bibfield{author}{\bibinfo{person}{Page}, \bibinfo{person}{Lawrence}, \bibinfo{person}{Brin}, \bibinfo{person}{Sergey}, {and} \bibinfo{person}{Terry}.} \bibinfo{year}{1999}\natexlab{}.
\newblock \showarticletitle{The PageRank citation ranking: Bringing order to the web}.
\newblock \bibinfo{journal}{\emph{stanford digital libraries working paper}} (\bibinfo{year}{1999}).
\newblock


\bibitem[Pan and Tomlinson(2016)]%
        {article}
\bibfield{author}{\bibinfo{person}{Liuxuan Pan} {and} \bibinfo{person}{Allan Tomlinson}.} \bibinfo{year}{2016}\natexlab{}.
\newblock \showarticletitle{{A Systematic Review of Information Security Risk Assessment}}.
\newblock \bibinfo{journal}{\emph{International Journal of Safety and Security Engineering}}  \bibinfo{volume}{6} (\bibinfo{date}{06} \bibinfo{year}{2016}), \bibinfo{pages}{270--281}.
\newblock


\bibitem[Pan et~al\mbox{.}(2022)]%
        {DBLP:conf/sigsoft/PanZC0BHLH22}
\bibfield{author}{\bibinfo{person}{Shengyi Pan}, \bibinfo{person}{Jiayuan Zhou}, \bibinfo{person}{Filipe~Roseiro C{\^{o}}go}, \bibinfo{person}{Xin Xia}, \bibinfo{person}{Lingfeng Bao}, \bibinfo{person}{Xing Hu}, \bibinfo{person}{Shanping Li}, {and} \bibinfo{person}{Ahmed~E. Hassan}.} \bibinfo{year}{2022}\natexlab{}.
\newblock \showarticletitle{Automated unearthing of dangerous issue reports}. In \bibinfo{booktitle}{\emph{Proceedings of the 30th {ACM} Joint European Software Engineering Conference and Symposium on the Foundations of Software Engineering, {ESEC/FSE} 2022}}. \bibinfo{publisher}{{ACM}}, \bibinfo{pages}{834--846}.
\newblock


\bibitem[P{\'{e}}rez et~al\mbox{.}(2020)]%
        {DBLP:journals/jss/PerezDMT20}
\bibfield{author}{\bibinfo{person}{Jorge~E. P{\'{e}}rez}, \bibinfo{person}{Jessica D{\'{\i}}az}, \bibinfo{person}{Javier~Garc{\'{\i}}a Martin}, {and} \bibinfo{person}{Bernardo Tabuenca}.} \bibinfo{year}{2020}\natexlab{}.
\newblock \showarticletitle{{Systematic Literature Reviews in Software Engineering - Enhancement of the Study Selection Process Using Cohen's Kappa Statistic}}.
\newblock \bibinfo{journal}{\emph{J. Syst. Softw.}}  \bibinfo{volume}{168} (\bibinfo{year}{2020}), \bibinfo{pages}{110657}.
\newblock
\urldef\tempurl%
\url{https://doi.org/10.1016/j.jss.2020.110657}
\showDOI{\tempurl}


\bibitem[Perozzi et~al\mbox{.}(2014)]%
        {DBLP:conf/kdd/PerozziAS14}
\bibfield{author}{\bibinfo{person}{Bryan Perozzi}, \bibinfo{person}{Rami Al{-}Rfou}, {and} \bibinfo{person}{Steven Skiena}.} \bibinfo{year}{2014}\natexlab{}.
\newblock \showarticletitle{DeepWalk: online learning of social representations}. In \bibinfo{booktitle}{\emph{The 20th {ACM} {SIGKDD} International Conference on Knowledge Discovery and Data Mining, {KDD} '14}}. \bibinfo{publisher}{{ACM}}, \bibinfo{pages}{701--710}.
\newblock


\bibitem[Peters et~al\mbox{.}(2019)]%
        {DBLP:journals/tse/PetersTYN19}
\bibfield{author}{\bibinfo{person}{Fayola Peters}, \bibinfo{person}{Thein~Than Tun}, \bibinfo{person}{Yijun Yu}, {and} \bibinfo{person}{Bashar Nuseibeh}.} \bibinfo{year}{2019}\natexlab{}.
\newblock \showarticletitle{Text Filtering and Ranking for Security Bug Report Prediction}.
\newblock \bibinfo{journal}{\emph{{IEEE} Trans. Software Eng.}} \bibinfo{volume}{45}, \bibinfo{number}{6} (\bibinfo{year}{2019}), \bibinfo{pages}{615--631}.
\newblock


\bibitem[Ponta et~al\mbox{.}(2019)]%
        {DBLP:conf/msr/PontaPSBD19}
\bibfield{author}{\bibinfo{person}{Serena~Elisa Ponta}, \bibinfo{person}{Henrik Plate}, \bibinfo{person}{Antonino Sabetta}, \bibinfo{person}{Michele Bezzi}, {and} \bibinfo{person}{C{\'{e}}dric Dangremont}.} \bibinfo{year}{2019}\natexlab{}.
\newblock \showarticletitle{A manually-curated dataset of fixes to vulnerabilities of open-source software}. In \bibinfo{booktitle}{\emph{Proceedings of the 16th International Conference on Mining Software Repositories, {MSR} 2019}}. \bibinfo{publisher}{{IEEE} / {ACM}}, \bibinfo{pages}{383--387}.
\newblock


\bibitem[Qian et~al\mbox{.}(2024)]%
        {DBLP:journals/corr/abs-2409-05591}
\bibfield{author}{\bibinfo{person}{Hongjin Qian}, \bibinfo{person}{Peitian Zhang}, \bibinfo{person}{Zheng Liu}, \bibinfo{person}{Kelong Mao}, {and} \bibinfo{person}{Zhicheng Dou}.} \bibinfo{year}{2024}\natexlab{}.
\newblock \showarticletitle{MemoRAG: Moving towards Next-Gen {RAG} Via Memory-Inspired Knowledge Discovery}.
\newblock \bibinfo{journal}{\emph{CoRR}}  \bibinfo{volume}{abs/2409.05591} (\bibinfo{year}{2024}).
\newblock
\showeprint[arXiv]{2409.05591}


\bibitem[Qin et~al\mbox{.}(2023)]%
        {DBLP:journals/corr/abs-2304-08354}
\bibfield{author}{\bibinfo{person}{Yujia Qin}, \bibinfo{person}{Shengding Hu}, \bibinfo{person}{Yankai Lin}, \bibinfo{person}{Weize Chen}, \bibinfo{person}{Ning Ding}, \bibinfo{person}{Ganqu Cui}, \bibinfo{person}{Zheni Zeng}, \bibinfo{person}{Yufei Huang}, \bibinfo{person}{Chaojun Xiao}, \bibinfo{person}{Chi Han}, \bibinfo{person}{Yi~Ren Fung}, \bibinfo{person}{Yusheng Su}, \bibinfo{person}{Huadong Wang}, \bibinfo{person}{Cheng Qian}, \bibinfo{person}{Runchu Tian}, \bibinfo{person}{Kunlun Zhu}, \bibinfo{person}{Shihao Liang}, \bibinfo{person}{Xingyu Shen}, \bibinfo{person}{Bokai Xu}, \bibinfo{person}{Zhen Zhang}, \bibinfo{person}{Yining Ye}, \bibinfo{person}{Bowen Li}, \bibinfo{person}{Ziwei Tang}, \bibinfo{person}{Jing Yi}, \bibinfo{person}{Yuzhang Zhu}, \bibinfo{person}{Zhenning Dai}, \bibinfo{person}{Lan Yan}, \bibinfo{person}{Xin Cong}, \bibinfo{person}{Yaxi Lu}, \bibinfo{person}{Weilin Zhao}, \bibinfo{person}{Yuxiang Huang}, \bibinfo{person}{Junxi Yan}, \bibinfo{person}{Xu Han}, \bibinfo{person}{Xian
  Sun}, \bibinfo{person}{Dahai Li}, \bibinfo{person}{Jason Phang}, \bibinfo{person}{Cheng Yang}, \bibinfo{person}{Tongshuang Wu}, \bibinfo{person}{Heng Ji}, \bibinfo{person}{Zhiyuan Liu}, {and} \bibinfo{person}{Maosong Sun}.} \bibinfo{year}{2023}\natexlab{}.
\newblock \showarticletitle{Tool Learning with Foundation Models}.
\newblock \bibinfo{journal}{\emph{CoRR}}  \bibinfo{volume}{abs/2304.08354} (\bibinfo{year}{2023}).
\newblock
\showeprint[arXiv]{2304.08354}


\bibitem[Ruan et~al\mbox{.}(2024)]%
        {10.1145/3691620.3695345}
\bibfield{author}{\bibinfo{person}{Bonan Ruan}, \bibinfo{person}{Jiahao Liu}, \bibinfo{person}{Weibo Zhao}, {and} \bibinfo{person}{Zhenkai Liang}.} \bibinfo{year}{2024}\natexlab{}.
\newblock \showarticletitle{VulZoo: A Comprehensive Vulnerability Intelligence Dataset}. In \bibinfo{booktitle}{\emph{Proceedings of the 39th IEEE/ACM International Conference on Automated Software Engineering}} \emph{(\bibinfo{series}{ASE '24})}. \bibinfo{publisher}{Association for Computing Machinery}, \bibinfo{pages}{2334–2337}.
\newblock
\showISBNx{9798400712487}


\bibitem[Russell et~al\mbox{.}(2018)]%
        {DBLP:conf/icmla/RussellKHLHOEM18}
\bibfield{author}{\bibinfo{person}{Rebecca~L. Russell}, \bibinfo{person}{Louis~Y. Kim}, \bibinfo{person}{Lei~H. Hamilton}, \bibinfo{person}{Tomo Lazovich}, \bibinfo{person}{Jacob Harer}, \bibinfo{person}{Onur Ozdemir}, \bibinfo{person}{Paul~M. Ellingwood}, {and} \bibinfo{person}{Marc~W. McConley}.} \bibinfo{year}{2018}\natexlab{}.
\newblock \showarticletitle{Automated Vulnerability Detection in Source Code Using Deep Representation Learning}. In \bibinfo{booktitle}{\emph{17th {IEEE} International Conference on Machine Learning and Applications, {ICMLA} 2018}}. \bibinfo{publisher}{{IEEE}}, \bibinfo{pages}{757--762}.
\newblock


\bibitem[Salton and Buckley(1988)]%
        {tfidf}
\bibfield{author}{\bibinfo{person}{Gerard Salton} {and} \bibinfo{person}{Christopher Buckley}.} \bibinfo{year}{1988}\natexlab{}.
\newblock \showarticletitle{{Term-weighting Approaches in Automatic Text Retrieval}}.
\newblock \bibinfo{journal}{\emph{Information Processing \& Management}} \bibinfo{volume}{24}, \bibinfo{number}{5} (\bibinfo{year}{1988}), \bibinfo{pages}{513--523}.
\newblock
\showISSN{0306-4573}


\bibitem[Scandariato et~al\mbox{.}(2014)]%
        {DBLP:journals/tse/ScandariatoWHJ14}
\bibfield{author}{\bibinfo{person}{Riccardo Scandariato}, \bibinfo{person}{James Walden}, \bibinfo{person}{Aram Hovsepyan}, {and} \bibinfo{person}{Wouter Joosen}.} \bibinfo{year}{2014}\natexlab{}.
\newblock \showarticletitle{Predicting Vulnerable Software Components via Text Mining}.
\newblock \bibinfo{journal}{\emph{{IEEE} Trans. Software Eng.}} \bibinfo{volume}{40}, \bibinfo{number}{10} (\bibinfo{year}{2014}), \bibinfo{pages}{993--1006}.
\newblock


\bibitem[Schumann et~al\mbox{.}(2023)]%
        {DBLP:journals/corr/abs-2307-06082}
\bibfield{author}{\bibinfo{person}{Raphael Schumann}, \bibinfo{person}{Wanrong Zhu}, \bibinfo{person}{Weixi Feng}, \bibinfo{person}{Tsu{-}Jui Fu}, \bibinfo{person}{Stefan Riezler}, {and} \bibinfo{person}{William~Yang Wang}.} \bibinfo{year}{2023}\natexlab{}.
\newblock \showarticletitle{{VELMA:} Verbalization Embodiment of {LLM} Agents for Vision and Language Navigation in Street View}.
\newblock \bibinfo{journal}{\emph{CoRR}}  \bibinfo{volume}{abs/2307.06082} (\bibinfo{year}{2023}).
\newblock


\bibitem[Shi et~al\mbox{.}(2021b)]%
        {DBLP:conf/emnlp/Shi0D0HZS21}
\bibfield{author}{\bibinfo{person}{Ensheng Shi}, \bibinfo{person}{Yanlin Wang}, \bibinfo{person}{Lun Du}, \bibinfo{person}{Hongyu Zhang}, \bibinfo{person}{Shi Han}, \bibinfo{person}{Dongmei Zhang}, {and} \bibinfo{person}{Hongbin Sun}.} \bibinfo{year}{2021}\natexlab{b}.
\newblock \showarticletitle{{CAST:} Enhancing Code Summarization with Hierarchical Splitting and Reconstruction of Abstract Syntax Trees}. In \bibinfo{booktitle}{\emph{Proceedings of the 2021 Conference on Empirical Methods in Natural Language Processing, {EMNLP} 2021}}. \bibinfo{publisher}{Association for Computational Linguistics}, \bibinfo{pages}{4053--4062}.
\newblock


\bibitem[Shi et~al\mbox{.}(2021a)]%
        {DBLP:conf/kbse/ShiJYCZMJW21}
\bibfield{author}{\bibinfo{person}{Lin Shi}, \bibinfo{person}{Ziyou Jiang}, \bibinfo{person}{Ye Yang}, \bibinfo{person}{Xiao Chen}, \bibinfo{person}{Yumin Zhang}, \bibinfo{person}{Fangwen Mu}, \bibinfo{person}{Hanzhi Jiang}, {and} \bibinfo{person}{Qing Wang}.} \bibinfo{year}{2021}\natexlab{a}.
\newblock \showarticletitle{{ISPY:} Automatic Issue-Solution Pair Extraction from Community Live Chats}. In \bibinfo{booktitle}{\emph{36th {IEEE/ACM} International Conference on Automated Software Engineering, {ASE} 2021}}. \bibinfo{publisher}{{IEEE}}, \bibinfo{pages}{142--154}.
\newblock


\bibitem[Shin et~al\mbox{.}(2011)]%
        {DBLP:journals/tse/ShinMWO11}
\bibfield{author}{\bibinfo{person}{Yonghee Shin}, \bibinfo{person}{Andrew Meneely}, \bibinfo{person}{Laurie~A. Williams}, {and} \bibinfo{person}{Jason~A. Osborne}.} \bibinfo{year}{2011}\natexlab{}.
\newblock \showarticletitle{Evaluating Complexity, Code Churn, and Developer Activity Metrics as Indicators of Software Vulnerabilities}.
\newblock \bibinfo{journal}{\emph{{IEEE} Trans. Software Eng.}} \bibinfo{volume}{37}, \bibinfo{number}{6} (\bibinfo{year}{2011}), \bibinfo{pages}{772--787}.
\newblock


\bibitem[Shu et~al\mbox{.}(2021)]%
        {DBLP:journals/ese/ShuXCWM21}
\bibfield{author}{\bibinfo{person}{Rui Shu}, \bibinfo{person}{Tianpei Xia}, \bibinfo{person}{Jianfeng Chen}, \bibinfo{person}{Laurie~A. Williams}, {and} \bibinfo{person}{Tim Menzies}.} \bibinfo{year}{2021}\natexlab{}.
\newblock \showarticletitle{How to Better Distinguish Security Bug Reports (Using Dual Hyperparameter Optimization)}.
\newblock \bibinfo{journal}{\emph{Empir. Softw. Eng.}} \bibinfo{volume}{26}, \bibinfo{number}{3} (\bibinfo{year}{2021}), \bibinfo{pages}{53}.
\newblock


\bibitem[Stocco et~al\mbox{.}(2020)]%
        {DBLP:conf/icse/0001WCT20}
\bibfield{author}{\bibinfo{person}{Andrea Stocco}, \bibinfo{person}{Michael Weiss}, \bibinfo{person}{Marco Calzana}, {and} \bibinfo{person}{Paolo Tonella}.} \bibinfo{year}{2020}\natexlab{}.
\newblock \showarticletitle{Misbehaviour prediction for autonomous driving systems}. In \bibinfo{booktitle}{\emph{{ICSE} '20: 42nd International Conference on Software Engineering}}. \bibinfo{publisher}{{ACM}}, \bibinfo{pages}{359--371}.
\newblock


\bibitem[{Tencent}(2023)]%
        {tecentOCR}
\bibfield{author}{\bibinfo{person}{{Tencent}}.} \bibinfo{year}{2023}\natexlab{}.
\newblock \bibinfo{title}{{OCR, Tencent Cloud}}.
\newblock \bibinfo{howpublished}{\url{https://www.tencentcloud.com/document/product/1045/49147}}.
\newblock


\bibitem[Touvron et~al\mbox{.}(2023)]%
        {DBLP:journals/corr/abs-2302-13971}
\bibfield{author}{\bibinfo{person}{Hugo Touvron}, \bibinfo{person}{Thibaut Lavril}, \bibinfo{person}{Gautier Izacard}, \bibinfo{person}{Xavier Martinet}, \bibinfo{person}{Marie{-}Anne Lachaux}, \bibinfo{person}{Timoth{\'{e}}e Lacroix}, \bibinfo{person}{Baptiste Rozi{\`{e}}re}, \bibinfo{person}{Naman Goyal}, \bibinfo{person}{Eric Hambro}, \bibinfo{person}{Faisal Azhar}, \bibinfo{person}{Aur{\'{e}}lien Rodriguez}, \bibinfo{person}{Armand Joulin}, \bibinfo{person}{Edouard Grave}, {and} \bibinfo{person}{Guillaume Lample}.} \bibinfo{year}{2023}\natexlab{}.
\newblock \showarticletitle{LLaMA: Open and Efficient Foundation Language Models}.
\newblock \bibinfo{journal}{\emph{CoRR}}  \bibinfo{volume}{abs/2302.13971} (\bibinfo{year}{2023}).
\newblock


\bibitem[Wang et~al\mbox{.}(2023a)]%
        {DBLP:journals/corr/abs-2309-15324}
\bibfield{author}{\bibinfo{person}{Jin Wang}, \bibinfo{person}{Zishan Huang}, \bibinfo{person}{Hengli Liu}, \bibinfo{person}{Nianyi Yang}, {and} \bibinfo{person}{Yinhao Xiao}.} \bibinfo{year}{2023}\natexlab{a}.
\newblock \showarticletitle{DefectHunter: {A} Novel LLM-Driven Boosted-Conformer-based Code Vulnerability Detection Mechanism}.
\newblock \bibinfo{journal}{\emph{CoRR}}  \bibinfo{volume}{abs/2309.15324} (\bibinfo{year}{2023}).
\newblock


\bibitem[Wang et~al\mbox{.}(2023b)]%
        {DBLP:journals/corr/abs-2310-01444}
\bibfield{author}{\bibinfo{person}{Kuan Wang}, \bibinfo{person}{Yadong Lu}, \bibinfo{person}{Michael Santacroce}, \bibinfo{person}{Yeyun Gong}, \bibinfo{person}{Chao Zhang}, {and} \bibinfo{person}{Yelong Shen}.} \bibinfo{year}{2023}\natexlab{b}.
\newblock \showarticletitle{Adapting {LLM} Agents Through Communication}.
\newblock \bibinfo{journal}{\emph{CoRR}}  \bibinfo{volume}{abs/2310.01444} (\bibinfo{year}{2023}).
\newblock


\bibitem[Wang et~al\mbox{.}(2022)]%
        {DBLP:journals/corr/abs-2207-00747}
\bibfield{author}{\bibinfo{person}{Xuezhi Wang}, \bibinfo{person}{Jason Wei}, \bibinfo{person}{Dale Schuurmans}, \bibinfo{person}{Quoc~V. Le}, \bibinfo{person}{Ed~H. Chi}, {and} \bibinfo{person}{Denny Zhou}.} \bibinfo{year}{2022}\natexlab{}.
\newblock \showarticletitle{Rationale-Augmented Ensembles in Language Models}.
\newblock \bibinfo{journal}{\emph{CoRR}}  \bibinfo{volume}{abs/2207.00747} (\bibinfo{year}{2022}).
\newblock


\bibitem[Wei et~al\mbox{.}(2022)]%
        {DBLP:conf/nips/Wei0SBIXCLZ22}
\bibfield{author}{\bibinfo{person}{Jason Wei}, \bibinfo{person}{Xuezhi Wang}, \bibinfo{person}{Dale Schuurmans}, \bibinfo{person}{Maarten Bosma}, \bibinfo{person}{Brian Ichter}, \bibinfo{person}{Fei Xia}, \bibinfo{person}{Ed~H. Chi}, \bibinfo{person}{Quoc~V. Le}, {and} \bibinfo{person}{Denny Zhou}.} \bibinfo{year}{2022}\natexlab{}.
\newblock \showarticletitle{Chain-of-Thought Prompting Elicits Reasoning in Large Language Models}. In \bibinfo{booktitle}{\emph{NeurIPS}}.
\newblock


\bibitem[Wijayasekara et~al\mbox{.}(2014)]%
        {DBLP:conf/iecon/WijayasekaraMM14}
\bibfield{author}{\bibinfo{person}{Dumidu Wijayasekara}, \bibinfo{person}{Milos Manic}, {and} \bibinfo{person}{Miles McQueen}.} \bibinfo{year}{2014}\natexlab{}.
\newblock \showarticletitle{Vulnerability identification and classification via text mining bug databases}. In \bibinfo{booktitle}{\emph{{IECON} 2014 - 40th Annual Conference of the {IEEE} Industrial Electronics Society}}. \bibinfo{publisher}{{IEEE}}, \bibinfo{pages}{3612--3618}.
\newblock


\bibitem[Wijayasekara et~al\mbox{.}(2012)]%
        {DBLP:conf/hsi/WijayasekaraMWM12}
\bibfield{author}{\bibinfo{person}{Dumidu Wijayasekara}, \bibinfo{person}{Milos Manic}, \bibinfo{person}{Jason~L. Wright}, {and} \bibinfo{person}{Miles McQueen}.} \bibinfo{year}{2012}\natexlab{}.
\newblock \showarticletitle{Mining Bug Databases for Unidentified Software Vulnerabilities}. In \bibinfo{booktitle}{\emph{2012 5th International Conference on Human System Interactions}}. \bibinfo{publisher}{{IEEE}}, \bibinfo{pages}{89--96}.
\newblock


\bibitem[Wu et~al\mbox{.}(2017)]%
        {wu2017vulnerability}
\bibfield{author}{\bibinfo{person}{Fang Wu}, \bibinfo{person}{Jigang Wang}, \bibinfo{person}{Jiqiang Liu}, {and} \bibinfo{person}{Wei Wang}.} \bibinfo{year}{2017}\natexlab{}.
\newblock \showarticletitle{Vulnerability detection with deep learning}. In \bibinfo{booktitle}{\emph{2017 3rd IEEE international conference on computer and communications (ICCC)}}. IEEE, \bibinfo{pages}{1298--1302}.
\newblock


\bibitem[Wu et~al\mbox{.}(2024)]%
        {DBLP:conf/icse/WuSH0024}
\bibfield{author}{\bibinfo{person}{Susheng Wu}, \bibinfo{person}{Wenyan Song}, \bibinfo{person}{Kaifeng Huang}, \bibinfo{person}{Bihuan Chen}, {and} \bibinfo{person}{Xin Peng}.} \bibinfo{year}{2024}\natexlab{}.
\newblock \showarticletitle{Identifying Affected Libraries and Their Ecosystems for Open Source Software Vulnerabilities}. In \bibinfo{booktitle}{\emph{{ICSE}'24}}. \bibinfo{publisher}{{ACM}}, \bibinfo{pages}{162:1--162:12}.
\newblock


\bibitem[Wu et~al\mbox{.}(2020)]%
        {DBLP:conf/kbse/WuZDYYCLJ20}
\bibfield{author}{\bibinfo{person}{Yueming Wu}, \bibinfo{person}{Deqing Zou}, \bibinfo{person}{Shihan Dou}, \bibinfo{person}{Siru Yang}, \bibinfo{person}{Wei Yang}, \bibinfo{person}{Feng Cheng}, \bibinfo{person}{Hong Liang}, {and} \bibinfo{person}{Hai Jin}.} \bibinfo{year}{2020}\natexlab{}.
\newblock \showarticletitle{SCDetector: Software Functional Clone Detection Based on Semantic Tokens Analysis}. In \bibinfo{booktitle}{\emph{35th {IEEE/ACM} International Conference on Automated Software Engineering, {ASE} 2020}}. \bibinfo{publisher}{{IEEE}}, \bibinfo{pages}{821--833}.
\newblock


\bibitem[Yamaguchi et~al\mbox{.}(2013)]%
        {DBLP:conf/ccs/YamaguchiWGR13}
\bibfield{author}{\bibinfo{person}{Fabian Yamaguchi}, \bibinfo{person}{Christian Wressnegger}, \bibinfo{person}{Hugo Gascon}, {and} \bibinfo{person}{Konrad Rieck}.} \bibinfo{year}{2013}\natexlab{}.
\newblock \showarticletitle{Chucky: exposing missing checks in source code for vulnerability discovery}. In \bibinfo{booktitle}{\emph{2013 {ACM} {SIGSAC} Conference on Computer and Communications Security, CCS'13}}. \bibinfo{publisher}{{ACM}}, \bibinfo{pages}{499--510}.
\newblock


\bibitem[Yao et~al\mbox{.}(2023)]%
        {DBLP:conf/iclr/YaoZYDSN023}
\bibfield{author}{\bibinfo{person}{Shunyu Yao}, \bibinfo{person}{Jeffrey Zhao}, \bibinfo{person}{Dian Yu}, \bibinfo{person}{Nan Du}, \bibinfo{person}{Izhak Shafran}, \bibinfo{person}{Karthik~R. Narasimhan}, {and} \bibinfo{person}{Yuan Cao}.} \bibinfo{year}{2023}\natexlab{}.
\newblock \showarticletitle{ReAct: Synergizing Reasoning and Acting in Language Models}. In \bibinfo{booktitle}{\emph{The Eleventh International Conference on Learning Representations, {ICLR} 2023}}. \bibinfo{publisher}{OpenReview.net}.
\newblock


\bibitem[Yoo et~al\mbox{.}(2021)]%
        {DBLP:conf/emnlp/YooPKLP21}
\bibfield{author}{\bibinfo{person}{Kang~Min Yoo}, \bibinfo{person}{Dongju Park}, \bibinfo{person}{Jaewook Kang}, \bibinfo{person}{Sang{-}Woo Lee}, {and} \bibinfo{person}{Woo{-}Myoung Park}.} \bibinfo{year}{2021}\natexlab{}.
\newblock \showarticletitle{GPT3Mix: Leveraging Large-scale Language Models for Text Augmentation}. In \bibinfo{booktitle}{\emph{Findings of the Association for Computational Linguistics: {EMNLP} 2021}}. \bibinfo{publisher}{Association for Computational Linguistics}, \bibinfo{pages}{2225--2239}.
\newblock


\bibitem[Yu et~al\mbox{.}(2022)]%
        {DBLP:journals/tosem/YuHLLWX22}
\bibfield{author}{\bibinfo{person}{Hao Yu}, \bibinfo{person}{Xing Hu}, \bibinfo{person}{Ge Li}, \bibinfo{person}{Ying Li}, \bibinfo{person}{Qianxiang Wang}, {and} \bibinfo{person}{Tao Xie}.} \bibinfo{year}{2022}\natexlab{}.
\newblock \showarticletitle{Assessing and Improving an Evaluation Dataset for Detecting Semantic Code Clones via Deep Learning}.
\newblock \bibinfo{journal}{\emph{{ACM} Trans. Softw. Eng. Methodol.}} \bibinfo{volume}{31}, \bibinfo{number}{4} (\bibinfo{year}{2022}), \bibinfo{pages}{62:1--62:25}.
\newblock


\bibitem[Zeng et~al\mbox{.}(2020)]%
        {DBLP:conf/iclr/ZengZSKP20}
\bibfield{author}{\bibinfo{person}{Hanqing Zeng}, \bibinfo{person}{Hongkuan Zhou}, \bibinfo{person}{Ajitesh Srivastava}, \bibinfo{person}{Rajgopal Kannan}, {and} \bibinfo{person}{Viktor~K. Prasanna}.} \bibinfo{year}{2020}\natexlab{}.
\newblock \showarticletitle{GraphSAINT: Graph Sampling Based Inductive Learning Method}. In \bibinfo{booktitle}{\emph{8th International Conference on Learning Representations, {ICLR} 2020}}. \bibinfo{publisher}{OpenReview.net}.
\newblock


\bibitem[Zhang et~al\mbox{.}(2019)]%
        {DBLP:conf/icse/ZhangWZ0WL19}
\bibfield{author}{\bibinfo{person}{Jian Zhang}, \bibinfo{person}{Xu Wang}, \bibinfo{person}{Hongyu Zhang}, \bibinfo{person}{Hailong Sun}, \bibinfo{person}{Kaixuan Wang}, {and} \bibinfo{person}{Xudong Liu}.} \bibinfo{year}{2019}\natexlab{}.
\newblock \showarticletitle{A novel neural source code representation based on abstract syntax tree}. In \bibinfo{booktitle}{\emph{Proceedings of the 41st International Conference on Software Engineering, {ICSE} 2019}}. \bibinfo{publisher}{{IEEE} / {ACM}}, \bibinfo{pages}{783--794}.
\newblock


\bibitem[Zheng et~al\mbox{.}(2023)]%
        {DBLP:journals/corr/abs-2310-13255}
\bibfield{author}{\bibinfo{person}{Sipeng Zheng}, \bibinfo{person}{Jiazheng Liu}, \bibinfo{person}{Yicheng Feng}, {and} \bibinfo{person}{Zongqing Lu}.} \bibinfo{year}{2023}\natexlab{}.
\newblock \showarticletitle{Steve-Eye: Equipping LLM-based Embodied Agents with Visual Perception in Open Worlds}.
\newblock \bibinfo{journal}{\emph{CoRR}}  \bibinfo{volume}{abs/2310.13255} (\bibinfo{year}{2023}).
\newblock


\bibitem[Zheng et~al\mbox{.}(2021)]%
        {DBLP:conf/icse/ZhengPLBEYLMS21}
\bibfield{author}{\bibinfo{person}{Yunhui Zheng}, \bibinfo{person}{Saurabh Pujar}, \bibinfo{person}{Burn~L. Lewis}, \bibinfo{person}{Luca Buratti}, \bibinfo{person}{Edward~A. Epstein}, \bibinfo{person}{Bo Yang}, \bibinfo{person}{Jim Laredo}, \bibinfo{person}{Alessandro Morari}, {and} \bibinfo{person}{Zhong Su}.} \bibinfo{year}{2021}\natexlab{}.
\newblock \showarticletitle{{D2A:} {A} Dataset Built for AI-Based Vulnerability Detection Methods Using Differential Analysis}. In \bibinfo{booktitle}{\emph{{ICSE} {(SEIP)}'21}}. \bibinfo{publisher}{{IEEE}}, \bibinfo{pages}{111--120}.
\newblock


\bibitem[Zhou et~al\mbox{.}(2023)]%
        {DBLP:conf/iclr/ZhouMHPPCB23}
\bibfield{author}{\bibinfo{person}{Yongchao Zhou}, \bibinfo{person}{Andrei~Ioan Muresanu}, \bibinfo{person}{Ziwen Han}, \bibinfo{person}{Keiran Paster}, \bibinfo{person}{Silviu Pitis}, \bibinfo{person}{Harris Chan}, {and} \bibinfo{person}{Jimmy Ba}.} \bibinfo{year}{2023}\natexlab{}.
\newblock \showarticletitle{Large Language Models are Human-Level Prompt Engineers}. In \bibinfo{booktitle}{\emph{The Eleventh International Conference on Learning Representations, {ICLR} 2023}}. \bibinfo{publisher}{OpenReview.net}.
\newblock


\bibitem[Zhu et~al\mbox{.}(2023)]%
        {DBLP:journals/nca/ZhuLSZ23}
\bibfield{author}{\bibinfo{person}{Yuhui Zhu}, \bibinfo{person}{Guanjun Lin}, \bibinfo{person}{Lipeng Song}, {and} \bibinfo{person}{Jun Zhang}.} \bibinfo{year}{2023}\natexlab{}.
\newblock \showarticletitle{The application of neural network for software vulnerability detection: a review}.
\newblock \bibinfo{journal}{\emph{Neural Comput. Appl.}} \bibinfo{volume}{35}, \bibinfo{number}{2} (\bibinfo{year}{2023}), \bibinfo{pages}{1279--1301}.
\newblock


\end{thebibliography}

\end{document}